\documentclass[11pt]{article}

\usepackage[T1]{fontenc}
\usepackage[utf8]{inputenc}

\usepackage[tt=false,type1=true]{libertine}
\usepackage[varqu]{zi4}
\usepackage[libertine]{newtxmath}
\usepackage{microtype}

\usepackage[
  a4paper,
  margin=1in,
  headheight=14pt,
  headsep=24pt,
  footskip=30pt
]{geometry}

\usepackage{fancyhdr}
\newcommand{\keywords}[1]{%
	\par\noindent\textbf{Keywords:} #1
}

\usepackage{xspace}
\usepackage{tabularx}
\usepackage{multirow}
\usepackage{enumitem}
\usepackage{siunitx}
\usepackage{longtable}
\usepackage{tikz}
\usetikzlibrary{
	arrows.meta,
	positioning,
	fit,
	calc,
	shapes.misc,
	backgrounds
}
\usepackage{pgfplots}
\usepackage{pgfplotstable}
\usepackage{amsmath}
\usepackage{booktabs}
\usepackage{graphicx}
\usepackage{url}
\usepackage{hyperref}
\usepackage[capitalize,noabbrev]{cleveref}
\usepackage{xcolor}
\usepackage{lscape}
\usepackage{float}
\usepackage{makecell}
\usepackage{seqsplit}
\usepackage{array}

\usepackage{booktabs}
\usepackage{array}
\usepackage{ragged2e}
\usepackage{xltabular}
\usepackage{xurl}

\usepackage{caption}
\keepXColumns
\newcolumntype{Y}{>{\RaggedRight\arraybackslash}X}

\newcommand{\reqtab}[1]{{\ttfamily\scriptsize\nolinkurl{#1}}}

\newcolumntype{L}[1]{>{\raggedright\arraybackslash}p{#1}}

\newcommand{\code}[1]{\nolinkurl{#1}}

\usepackage[backend=biber,style=numeric]{biblatex}
\newcommand{\mlkem}{ML-KEM\xspace}
\newcommand{\mldsa}{ML-DSA\xspace}
\newcommand{\pkixcore}{\texttt{pkix-core}\xspace}
\newcommand{\strictmode}{\texttt{strict}\xspace}
\newcommand{\deployablemode}{\texttt{deployable}\xspace}
\newcommand{\gatepack}[1]{\textsf{#1}}
\newcommand{\tool}[1]{\textsc{#1}}
\newcommand{\req}[1]{\texttt{#1}}
\newcommand{\json}[1]{\texttt{#1}}

\newcommand{\etal}{et~al.\xspace}

\makeatletter

\newcommand{\@toptitlebar}{%
  \hrule height 2pt
  \vskip 0.25in
  \vskip -\parskip
}

\newcommand{\@bottomtitlebar}{%
  \vskip 0.29in
  \vskip -\parskip
  \hrule height 2pt
  \vskip 0.09in
}

\renewcommand{\maketitle}{%
  \par
  \begingroup
    \thispagestyle{plain}
    \begin{center}
      \vspace*{0.1in}
      \@toptitlebar
      {\LARGE\scshape \@title\par}
      \@bottomtitlebar
      \vspace{0.15in}
      {\normalsize\bfseries \@author\par}
      \vspace{0.2in}
    \end{center}
  \endgroup
}

\renewcommand{\section}{%
	\@startsection{section}{1}{\z@}%
	{-2.0ex \@plus -0.5ex \@minus -0.2ex}%
	{1.2ex \@plus 0.2ex}%
	{\large\bfseries\raggedright}}

\renewcommand{\subsection}{%
	\@startsection{subsection}{2}{\z@}%
	{-1.8ex \@plus -0.5ex \@minus -0.2ex}%
	{0.8ex \@plus 0.2ex}%
	{\normalsize\bfseries\raggedright}}

\renewcommand{\subsubsection}{%
	\@startsection{subsubsection}{3}{\z@}%
	{-1.5ex \@plus -0.5ex \@minus -0.2ex}%
	{0.5ex \@plus 0.2ex}%
	{\normalsize\bfseries\raggedright}}

\makeatother

\newcommand{\minorhead}[1]{%
  \par\smallskip\noindent\textbf{#1}\par\nobreak\smallskip
}

\providecommand{\phantomsection}{}

\makeatletter
\renewcommand{\thepart}{\Roman{part}}

\newcommand{\partline}[1]{%
	\clearpage
	\refstepcounter{part}%
	\phantomsection
	\addcontentsline{toc}{part}{Part \thepart.\ #1}%
	\begin{center}
		\vspace*{0.08\textheight}
		{\Huge\bfseries Part \thepart\par}
		\vspace{0.4em}
		{\Large\bfseries #1\par}
	\end{center}
	\vspace{1.5em}
}

\renewcommand*\l@part[2]{%
	\ifnum\c@tocdepth>-2\relax
	\addpenalty{-\@highpenalty}%
	\vskip 0.4em
	\@dottedtocline{0}{0pt}{5.5em}{#1}{#2}%
	\fi
}
\makeatother

\title{From Public-Key Linting to Operational Post-Quantum X.509 Assurance for ML-KEM and ML-DSA: Registry-Driven Policy, Mutation-Based Evaluation, and Import Validation}

\author{%
José Luis Delgado \\
\small Universitat Oberta de Catalunya \\
\texttt{jdelgado13@uoc.edu}
}

\begin{document}

\maketitle

\begin{abstract}
	The final FIPS and PKIX standards for ML-KEM and ML-DSA establish the
	normative floor, but they do not by themselves provide deployment assurance.
	In practical post-quantum X.509 deployments, failures arise in
	certificate-profile semantics, SubjectPublicKeyInfo representation, and
	private-key-container import. Current PQ public-key linting also lacks a
	reproducible workflow that assigns checks to the certification authority or the
	artifact importer and specifies their behavior under deployment-facing policy.
	We present an operational post-quantum X.509 assurance framework for ML-KEM
	and ML-DSA in the narrow executable profile \pkixcore. The framework maps 17
	final-standards requirements into an assurance registry indexed by owner,
	stage, detector kind, normative strength, and mode-specific action; groups
	those requirements into three operator gate packs; covers
	certificate/profile, SPKI/public-key, and private-key-container/import
	surfaces; and evaluates them through a frozen mutation-based corpus supported
	by bounded public-appendix and cross-tool evidence.
	
	Across a controlled corpus of 48 artifacts, comprising 21 valid and 27 invalid
	cases, the artifact detects all expected invalid artifacts in both strict and
	deployable modes with zero false positives. Strict mode blocks all 17 active
	requirements; deployable mode preserves the same underlying detection coverage
	and downgrades exactly one exercised ML-KEM canonicality condition from block
	to warning. On the importer-owned private-key surface, all 7 active
	requirements are covered, with 7/7 expected invalid detections and no open
	detector gaps. On a comparable certificate subset, a frozen JZLint baseline
	meets 5/10 expected invalid detections and fatally rejects 3 valid ML-KEM
	certificates, whereas the local artifact meets 10/10 with no fatal valid
	rejections. A bounded public appendix of 26 artifacts across two providers and
	all six ML-KEM/ML-DSA private-key parameter sets, together with a cross-tool
	matrix over 57 artifacts, also records a material divergence between parse
	acceptance and policy conformance without treating the evidence as a
	prevalence study or benchmark shootout. The results support an operational
	X.509 assurance workflow for CA pre-issuance and private-key import that
	extends prior PQ public-key linting work.
\end{abstract}


\newpage
\tableofcontents

\partline{Foundations}
\section{Introduction}
\label{sec:introduction}

\minorhead{Why final standards do not remove operational assurance risk}
\label{subsec:intro-why-standards-not-enough}

The standardization path for lattice-based post-quantum cryptography has reached final specifications. At the algorithmic layer, \mlkem and \mldsa are defined by FIPS~203 and FIPS~204 \cite{fips203,fips204}. At the X.509/PKIX layer, RFC~9881 specifies the conventions for \mldsa signatures, subject public keys, and private-key encodings, and RFC~9935 specifies the corresponding conventions for \mlkem \cite{rfc9881,rfc9935}. This closure removes much of the ambiguity that surrounded draft-era engineering and gives implementation work a final normative baseline.

The remaining problem is operational assurance. Standards specify the shape of conforming algorithms and artifacts, yet they do not assign enforcement to a particular stage, identify the owner of each enforcement decision, or define how a violation should propagate through a deployment-facing workflow. A certification authority, an issuance pipeline, and a private-key importer occupy different operational positions. They inspect different artifact surfaces, carry different responsibilities, and require different failure semantics. Final requirements therefore have to be translated into accountable gates before issuance and before import.

The PQ setting sharpens this problem because the relevant failure modes span more than one syntactic layer. An artifact may identify the correct algorithm family and still fail at certificate-profile semantics, SubjectPublicKeyInfo representation, or private-key import consistency. Conversely, an implementation may parse an artifact successfully while accepting a semantically invalid or misprofiled object. Standardization fixes the normative baseline; assurance must translate that baseline into executable operational decisions.

\minorhead{Failure surfaces across certificate, SPKI, and import boundaries}
\label{subsec:intro-failure-surfaces}

In the scope of this paper, operational assurance spans three distinct surfaces. The first is the \emph{certificate/profile} surface, where failures arise from PKIX semantics rather than from raw encoding alone. Examples include \texttt{keyUsage} requirements, constraints on the signature \texttt{AlgorithmIdentifier}, and PKIX-specific prohibitions such as the exclusion of HashML-DSA from the covered certificate profile \cite{rfc9881,rfc9935}. These are issuance-time policy conditions whose failure should be attributed to the CA-side workflow.

The second surface is \emph{SPKI/public-key}. The relevant defects include algorithm-identifier family mismatches, parameters that must be absent, exact public-key lengths tied to the parameter set, and canonicality conditions on represented public-key material. A public key may therefore be attached to the right certificate skeleton while remaining structurally or semantically wrong at the representation boundary. Such defects also belong to the pre-issuance path, but they are distinct from certificate-profile defects and should be reported separately.

The third surface is \emph{private-key-container/import}. This surface is easy to miss when assurance is reduced to certificate linting, yet it is necessary for deployment-facing validation. Private-key containers can be malformed, use the wrong CHOICE form, encode the wrong seed or expanded-key length, or violate seed-to-expanded consistency conditions. Some of these checks depend on import-side validation logic rather than certificate parsing. They should therefore be modeled as importer-owned assurance requirements within the import path itself.

These three surfaces are operationally adjacent but not reducible to one another. A certificate can be well formed even when the associated public-key payload is not canonical. A public key can parse even when the private key supplied for import is internally inconsistent. A tool can accept an object at parse time without enforcing a deployment-relevant PKIX constraint. Parse acceptance, structural well-formedness, profile conformance, and import validity are related but non-identical judgments.

\minorhead{From public-key linting to operational assurance}
\label{subsec:intro-from-linting-to-assurance}

Recent work by Karatsiolis \etal opened the line of public-key linting for \mlkem and \mldsa and showed that post-quantum X.509 requires dedicated checks rather than a naive transplantation of classical certificate logic \cite{karatsiolis2026pqlinting}. That contribution is the closest prior step. This paper extends that line into operational assurance.

With final standards in place, the problem is to organize their clauses into a reproducible assurance workflow that answers four operator questions together: \emph{what} is checked, \emph{where} it is checked, \emph{who} owns the check, and \emph{how} the result should act under policy. Accordingly, this paper presents a workflow-centric assurance artifact for X.509/PKIX around \mlkem and \mldsa, not an expanded lint catalogue.

This framing also determines the use of comparison evidence. A certificate-level baseline remains supporting evidence, especially when it reveals incompleteness or runtime fragility, but it is not the main contribution. Ecosystem behavior observations are also supporting evidence, because they help separate parse acceptance from policy conformance. The central claim is architectural: the field needs accountable post-quantum X.509 assurance, not isolated PQ public-key lints.

\minorhead{Thesis}
\label{subsec:intro-thesis-identity}

The central claim is the following.

\begin{quote}
	\emph{Once final PQ standards exist, the central assurance problem is the execution of accountable decisions across certificate, SPKI, and private-key import boundaries.}
\end{quote}

This paper addresses that problem with \pkixcore, a narrow, executable profile for \mlkem and \mldsa. The profile reifies final-standards requirements into a registry whose records are assigned by owner and stage, grouped into operator gate packs, evaluated under two policy modes, and tested against a frozen mutation-based corpus. The artifact is designed for operator use and reproducible evaluation. Reference implementation, frozen corpus, and reproduction materials are available at \href{https://github.com/hypergalois/pqc-x509-assurance}{\texttt{github.com/hypergalois/pqc-x509-assurance}}.

\section{Related Work}
\label{sec:related-work}

\minorhead{Public-key linting for \mlkem and \mldsa}
\label{subsec:related-pq-linting}

The nearest prior work is the public-key linting line opened by Karatsiolis
\etal \cite{karatsiolis2026pqlinting}. It shows that
\mlkem and \mldsa require dedicated PQ-aware lint logic and cannot be handled by
an unmodified inheritance of classical X.509 checking. This paper builds on that work, but changes the unit of contribution. The focus here is a workflow-centric assurance artifact with owner/stage assignment, gate packs, policy modes, import validation, and frozen replay outputs.

\minorhead{Final standards as normative foundations}
\label{subsec:related-final-standards-foundations}

FIPS~203 and FIPS~204 provide the algorithmic baseline for \mlkem and \mldsa
respectively \cite{fips203,fips204}, and RFC~9881 and RFC~9935 provide the
final PKIX/X.509 conventions that close the draft-era uncertainty for the
certificate, SPKI, and private-key encodings in scope \cite{rfc9881,rfc9935}.

\minorhead{\tool{JZLint} as frozen baseline}
\label{subsec:related-jzlint-baseline}

The frozen \tool{JZLint} snapshot is used as a certificate-level implementation
baseline \cite{jzlint_snapshot}. Its function is to provide
supporting comparative evidence about current certificate-level capability and
runtime fragility under a fixed local replay environment.

\minorhead{\tool{libcrux} as narrow import substrate}
\label{subsec:related-libcrux-substrate}

The libcrux snapshot
\cite{libcrux_snapshot} is the narrow cryptographic substrate used by
the import-validation bridge, especially for the private-key container checks
whose validity depends on seed, expanded-key, and consistency semantics. The
paper therefore cites libcrux as implementation substrate.

Prior work establishes the need for PQ-aware linting; final standards establish the normative baseline; and the frozen baseline and import substrate provide bounded implementation anchors. This paper builds on that base to study operational post-quantum X.509 assurance.

\minorhead{Contributions}
\label{subsec:intro-contributions}

\begin{itemize}
	\item We define an \emph{operational post-quantum X.509 assurance model} for \mlkem and \mldsa in \pkixcore, organized by owner, stage, and gate pack rather than by lints alone.
	
	\item We realize that model as a \emph{registry-driven policy artifact} with 17 active requirements distributed across three artifact surfaces and three operator-facing gate packs, together with explicit \strictmode and \deployablemode semantics.
	
	\item We provide a \emph{controlled mutation-based evaluation} over a frozen corpus of 48 artifacts, comprising 21 valid and 27 invalid cases, and show full expected invalid detection with zero false positives in the declared scope.
	
	\item We extend assurance beyond certificate material by including \emph{private-key-container/import validation} as a first-class surface, covering all 7 importer-owned requirements in the current profile.
	
	\item We present \emph{disciplined supporting evidence} through a certificate-level frozen baseline comparison and a bounded public appendix, while keeping both subordinate to the workflow-centric claim of the paper.
\end{itemize}

\section{Problem Setting and Scope}
\label{sec:problem-setting}

\minorhead{The precise assurance question}
\label{subsec:scope-precise-question}

\emph{Given final FIPS and PKIX standards for \mlkem and \mldsa, how can their executable requirements be translated into a reproducible, owner-assigned, mode-aware assurance workflow across certificate issuance and private-key import, without claiming runtime coverage that the artifact does not execute?}

First, the paper concerns \emph{executable requirements} grounded in standards text. Second, it concerns \emph{workflow assignment}, where responsibility is attached to concrete owners and stages. Third, it does not let certificate-side evidence stand in for runtime behavior.

\minorhead{The \pkixcore profile}
\label{subsec:scope-pkix-core}

The \pkixcore profile is the paper's executable profile boundary. It captures the intersection of final algorithmic standards and final PKIX/X.509 conventions that the artifact can instantiate, mutate, and evaluate reproducibly for \mlkem and \mldsa.

\minorhead{Artifact surfaces in scope}
\label{subsec:scope-artifact-surfaces}

The active profile spans three artifact surfaces.

\begin{enumerate}
	\item \textbf{Certificate/profile}. This surface covers X.509 certificate semantics whose assurance meaning is determined by PKIX policy, including \texttt{keyUsage} obligations and signature \texttt{AlgorithmIdentifier} requirements.
	\item \textbf{SPKI/public-key}. This surface covers SubjectPublicKeyInfo and the embedded public-key payload, including algorithm-identifier encoding, absent-parameters rules, length constraints, and canonicality conditions.
	\item \textbf{Private-key-container/import}. This surface covers private-key encodings and import-side validity conditions, including CHOICE form, exact lengths, and seed-to-expanded consistency.
\end{enumerate}

\Cref{tab:pkixcore-scope} summarizes how these surfaces are mapped to owners, gate packs, and operator actions in the current profile.

\begin{figure}[H]
	\caption{\pkixcore assurance scope by surface, owner, and operational role. The profile contains 17 active requirements: 10 owned by CA pre-issuance and 7 owned by the artifact importer.} 
	\vspace{.5cm}
	\label{tab:pkixcore-scope}
	\centering
	\footnotesize
	\begin{tabularx}{\textwidth}{@{}>{\raggedright\arraybackslash}p{2.7cm}>{\raggedright\arraybackslash}p{2.3cm}>{\raggedright\arraybackslash}p{2.8cm}c>{\raggedright\arraybackslash}X>{\raggedright\arraybackslash}p{2.1cm}@{}}
		\toprule
		Surface / gate pack & Owner & Artifact unit & Active reqs. & Representative obligations & Default role in workflow \\
		\midrule
		Certificate/profile\\ \gatepack{ca-certificate-profile}
		& \texttt{ca-preissuance}
		& X.509 certificate profile
		& 5
		& \mlkem \texttt{keyUsage} must be \texttt{keyEncipherment}-only when present; \mldsa certificates must carry at least one signing-related bit; signature \texttt{AlgorithmIdentifier} parameters must be absent; HashML-DSA is forbidden in the covered PKIX profile
		& CA issuance gate \\
		
		SPKI/public-key\\ \gatepack{ca-spki-public-key}
		& \texttt{ca-preissuance}
		& SubjectPublicKeyInfo and public-key payload
		& 5
		& OID family and absent-parameters rules; exact public-key lengths for parameter sets; \mlkem encode/decode identity as a canonicality condition
		& CA issuance gate \\
		
		Private-key-container/import\\ \gatepack{import-private-key}
		& \texttt{artifact-importer}
		& PKCS\#8-style private-key container and import substrate
		& 7
		& Seed and expanded-key lengths; CHOICE form; seed/expanded consistency; \mlkem expanded-key hash check
		& Import acceptance gate \\
		
		Runtime-consumer boundary
		& \texttt{runtime-consumer}
		& Consumer-time behavior
		& 0
		& Explicitly modeled as a future boundary rather than silently implied by certificate-side evidence
		& Out of executable scope \\
		\bottomrule
	\end{tabularx}
\end{figure}

\minorhead{Owners and stages}
\label{subsec:scope-owners-stages}

The owner-stage split is the organizing abstraction of \pkixcore. In \pkixcore, \texttt{ca-preissuance} owns two stages: \texttt{certificate/profile} and \texttt{SPKI/public-key}. The \texttt{artifact-importer} owns the \texttt{private-key-container/import} stage. The \texttt{runtime-consumer} is kept explicit as a boundary role, but it is outside executable scope until active requirements and evidence support it. This structure answers a practical question that many standards-conformant tools still leave unclear: who is responsible for running which gate before an artifact crosses into the next stage of the workflow?

\begin{figure}[H]
	\centering
	\begin{tikzpicture}[x=1cm,y=1cm,>=Latex,
		band/.style={draw, rounded corners=2pt, fill=black!6, minimum height=0.9cm, align=center, font=\small\bfseries},
		lane/.style={draw, rounded corners=2pt, minimum width=4.1cm, minimum height=0.85cm, align=center, font=\small\bfseries},
		gate/.style={draw, rounded corners=2pt, minimum width=4.1cm, minimum height=1.0cm, align=center, font=\small},
		note/.style={draw, dashed, rounded corners=2pt, minimum width=4.1cm, minimum height=1.0cm, align=center, font=\small},
		flow/.style={-{Latex[length=2.3mm]}, thick}
		]
		
		\def\xL{2.0}
		\def\xC{7.0}
		\def\xR{12.0}
		
		\node[band, minimum width=14.2cm] (floor) at (7.0,6.2) {Final normative floor: FIPS~203, FIPS~204, RFC~9881, RFC~9935};
		
		\node[lane] (ca)   at (\xL,4.8) {CA pre-issuance};
		\node[lane] (imp)  at (\xC,4.8) {Artifact importer};
		\node[lane] (run)  at (\xR,4.8) {Runtime consumer};
		
		\node[gate] (cert)  at (\xL,3.2) {\gatepack{ca-certificate-profile}\\5 requirements};
		\node[gate] (spki)  at (\xL,1.7) {\gatepack{ca-spki-public-key}\\5 requirements};
		\node[gate] (priv)  at (\xC,2.45) {\gatepack{import-private-key}\\7 requirements};
		\node[note] (bound) at (\xR,2.45) {Explicit boundary\\0 active requirements};
		
		\draw[flow] (floor.south west) ++(1.8,0) .. controls +(0,-0.7) and +(0,0.7) .. (ca.north);
		\draw[flow] (floor.south) .. controls +(0,-0.8) and +(0,0.8) .. (imp.north);
		\draw[flow] (floor.south east) ++(-1.8,0) .. controls +(0,-0.7) and +(0,0.7) .. (run.north);
		
		\draw[flow] (ca) -- (cert);
		\draw[flow] (cert) -- (spki);
		\draw[flow] (imp) -- (priv);
		\draw[flow] (run) -- (bound);
		
		\node[draw, dotted, rounded corners=2pt, fit=(cert)(spki), inner sep=3.5mm,
		label={[font=\footnotesize]below:10 CA-owned requirements across two issuance stages}] {};
		\node[draw, dotted, rounded corners=2pt, fit=(priv), inner sep=3.5mm,
		label={[font=\footnotesize]below:7 importer-owned requirements}] {};
		
	\end{tikzpicture}
	\caption{Operational assurance boundaries in \pkixcore. Final standards provide the normative floor, but the actionable workflow emerges only after requirements are assigned by owner and stage.}
	\label{fig:assurance-boundaries}
\end{figure}
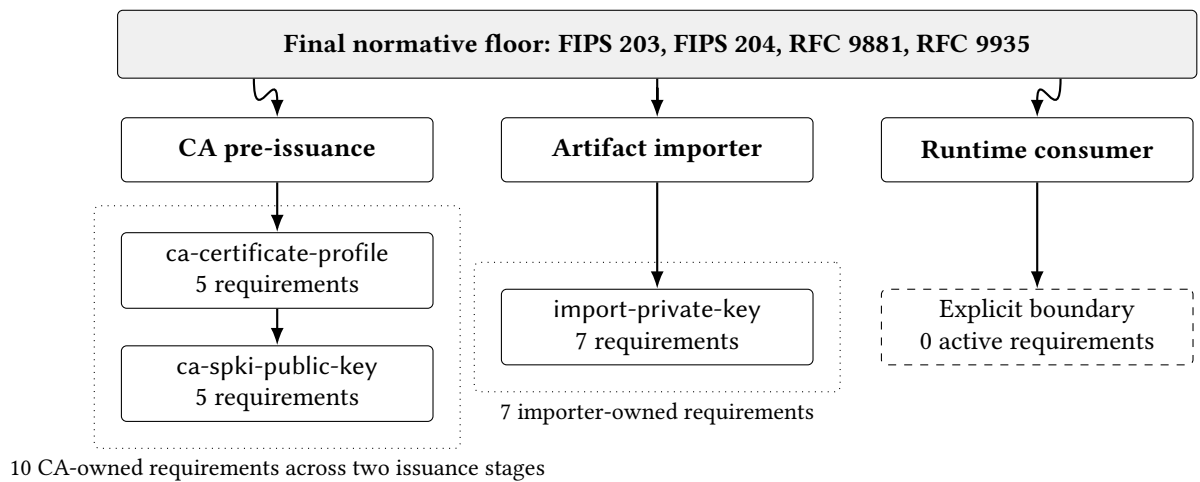

The notion of \emph{gate pack} makes this split usable. A gate pack is the smallest operator-facing assurance bundle that specifies what should be run at a given stage by a given owner. In the current profile, there are exactly three packs: one for certificate/profile checks, one for SPKI/public-key checks, and one for private-key/import checks. This packaging lets the artifact produce playbooks, pack-level summaries, and mode-aware outputs that can be consumed without reading the implementation first.

\minorhead{Non-goals and explicit exclusions}
\label{subsec:scope-non-goals}

\begin{itemize}
	\item \textbf{Benchmark breadth.} The paper does not claim to be a complete benchmark across all ecosystem tools or all possible PQ X.509 implementations.
	\item \textbf{Prevalence claims.} The bounded public appendix is external-validity support, not a census of Internet-deployed PQ artifacts.
	\item \textbf{Runtime-consumer execution.} Runtime-consumer behavior remains explicit but out of executable scope in \pkixcore.
	\item \textbf{Algorithm-family breadth.} The artifact is limited to \mlkem and \mldsa, and does not attempt to cover every PQ family or hybrid issuance profile.
	\item \textbf{Baseline-centered framing.} The certificate-level baseline and cross-tool observations are supporting evidence, not the center of the contribution.
\end{itemize}

\minorhead{Threat model and operational assumptions}
\label{subsec:scope-threat-model}

The operational threat model is the threat of malformed, misprofiled, or semantically inconsistent artifacts crossing the wrong boundary unchecked. The artifact is designed to intercept such failures before certificate issuance and before private-key import. It therefore assumes a deployment pipeline in which the relevant objects can be inspected at those stages and in which the owners of those stages can act on block-or-warn outputs.

The work does \emph{not} model every conceivable security failure. It does not address side channels, key-generation entropy failures, runtime protocol misuse, or consumer behavior after artifact acceptance. It also does not claim that every external toolchain will behave identically. The paper assumes a final normative baseline, a frozen set of third-party snapshots, a controlled corpus with labeled mutations, and a bounded public appendix for external support.

\section{Normative Basis and Translation Method}
\label{sec:normative-basis}

\minorhead{Final normative baseline}
\label{subsec:normative-floor}

At the algorithmic layer, the normative baseline consists of FIPS~203 for \mlkem and FIPS~204 for \mldsa \cite{fips203,fips204}. At the PKIX/X.509 layer, it consists of RFC~9881 for \mldsa and RFC~9935 for \mlkem \cite{rfc9881,rfc9935}. These documents are treated as the authoritative source of algorithm identifiers, absent-parameter rules, public-key sizes, certificate \texttt{keyUsage} semantics, and private-key encoding conventions in the covered profile.

Draft-era PQ engineering accumulated provisional logic, partial implementations, and outdated assumptions. A workflow-centric assurance artifact cannot base operator-facing policy on that sediment. It needs fixed clauses, stable identifiers, and a basis on which disagreements can be localized to a requirement rather than to a shifting standards process.

\minorhead{Requirement extraction procedure}
\label{subsec:requirement-extraction}

The procedure consists of five steps.

\begin{enumerate}
	\item \textbf{Locate an executable clause.} We identify a final-standards clause that imposes a concrete condition on an artifact representation, profile semantic, or import boundary.
	\item \textbf{Normalize it into a requirement sentence.} The clause is rewritten as a single explicit requirement whose truth value can be checked against an artifact.
	\item \textbf{Assign operational placement.} The requirement is mapped to an artifact type, stage, owner, and gate pack.
	\item \textbf{Bind it to evaluation design.} The record is linked to mutation families, an expected detector path, and a baseline-status annotation describing whether a frozen external baseline appears to cover, partially cover, or miss the requirement.
	\item \textbf{Bind it to policy.} The record receives a normative-strength label, a constructibility status, mode-specific actions, and a short operational justification.
\end{enumerate}

The result is a registry record that is normative, operational, and evaluable. It preserves traceability to the standards text through explicit source locators, and records enough structure to generate policy matrices, owner-stage summaries, coverage reports, and operator gate packs.

\minorhead{Requirement schema and fields}
\label{subsec:requirement-schema}

The registry schema is designed to prevent a requirement from floating free of provenance, workflow, or evidence. \Cref{tab:requirement-schema} groups the fields by their manuscript function.

\begin{table}[H]
	\caption{Registry field groups and their role in the assurance workflow. The schema keeps each requirement traceable to final standards, placeable in the workflow, and testable against frozen evidence.}
	\vspace{.5cm}
	\label{tab:requirement-schema}
	\centering
	\footnotesize
	\begin{tabularx}{\textwidth}{@{}>{\raggedright\arraybackslash}p{3.1cm}>{\raggedright\arraybackslash}p{5.2cm}>{\raggedright\arraybackslash}X@{}}
		\toprule
		Field group & Representative fields & Purpose in the artifact \\
		\midrule
		Normative provenance
		& \texttt{source}, \texttt{source\_locators}, \texttt{requirement}, \texttt{normative\_strength}
		& Anchors each record in the final standards corpus and preserves the exact clause that justifies the check. \\
		
		Artifact placement
		& \texttt{algorithm}, \texttt{artifact\_type}, \texttt{stage}, \texttt{profile}, \texttt{fault\_family}
		& States where the requirement lives: which algorithm family it concerns, which artifact surface it targets, and which defect family it represents. \\
		
		Operational ownership
		& \texttt{owner}, \texttt{gate\_pack}, \texttt{mode\_action}, \texttt{justification}
		& Assigns responsibility, groups the requirement into an operator-facing bundle, and determines whether a finding blocks or warns under each mode. \\
		
		Evaluation linkage
		& \texttt{mutation\_family}, \texttt{expected\_detector}, \texttt{baseline\_status}, \texttt{constructibility}
		& Connects the normative claim to the controlled corpus, to detector expectations, to comparative evidence, and to whether the requirement is currently executable in the profile. \\
		
		Identity and reporting
		& \texttt{id}, \texttt{severity}
		& Gives each requirement a stable manuscript-facing handle and a report-facing severity label. \\
		\bottomrule
	\end{tabularx}
\end{table}

This schema avoids a common failure in security-engineering papers: a detector exists, but its justification, execution point, and evidentiary basis are unclear. In the present artifact, a requirement that lacks one of these dimensions is incomplete by design.

\minorhead{Constructibility and activation criteria}
\label{subsec:constructibility-activation}

A requirement becomes active only when it satisfies the constructibility discipline of the profile. In the current release line, active requirements must belong to \pkixcore, be marked \texttt{covered}, map consistently to the owner-stage expectations of their declared gate pack, carry valid actions for both \strictmode and \deployablemode, and include a non-empty operational justification.

The schema can represent future states such as planned or externally dependent coverage, but the current profile activates only the 17 requirements whose execution path is present and evidenced. The registry limits claims to requirements that are both implemented and evidenced.

\minorhead{From normative text to executable requirement}
\label{subsec:clause-to-requirement}

\begin{figure}[H]
	\centering
	\begin{tikzpicture}[x=1cm,y=1cm,>=Latex,
		box/.style={
			draw, rounded corners=2pt,
			minimum height=1.4cm,
			text width=2.55cm,
			inner sep=3pt,
			align=center,
			font=\small
		},
		smallbox/.style={
			draw, rounded corners=2pt,
			minimum height=1.15cm,
			text width=2.55cm,
			inner sep=2pt,
			align=center,
			font=\scriptsize
		},
		flow/.style={-{Latex[length=2.2mm]}, thick}
		]
		
		\def\xA{0.0}
		\def\xB{3.1}
		\def\xC{6.2}
		\def\xD{9.3}
		\def\xE{12.4}
		
		\node[box, fill=black!5] (clause) at (\xA,0)
		{Final standards clause\\RFC/FIPS rule};
		
		\node[box] (norm) at (\xB,0)
		{Normalized requirement\\single checkable sentence};
		
		\node[box] (reg) at (\xC,0)
		{Registry record\\owner, stage,\\gate pack, mode};
		
		\node[box] (eval) at (\xD,0)
		{Evaluation link\\mutation family\\+ detector};
		
		\node[box, fill=black!5] (gate) at (\xE,0)
		{Operator-facing outcome\\block /\\warn / pass};
		
		\draw[flow] (clause) -- (norm);
		\draw[flow] (norm) -- (reg);
		\draw[flow] (reg) -- (eval);
		\draw[flow] (eval) -- (gate);
		
		\node[smallbox] (ex1) at (\xA,-2.0)
		{Example clause:\\ML-KEM cert.\\keyUsage must be\\keyEncipherment-only};
		
		\node[smallbox] (ex2) at (\xB,-2.0)
		{\begin{tabular}{@{}c@{}}
				\ttfamily MLKEM-CERT-KU-\\
				KEYENCIPHERMENT\\
				-ONLY
		\end{tabular}};
		
		\node[smallbox] (ex3) at (\xC,-2.0)
		{\begin{tabular}{@{}c@{}}
				\ttfamily ca-preissuance\\
				certificate/\\profile\\
				ca-certificate-\\profile
		\end{tabular}};
		
		\node[smallbox] (ex4) at (\xD,-2.0)
		{Mutations:\\missing bit,\\extra bit,\\empty KU};
		
		\node[smallbox] (ex5) at (\xE,-2.0)
		{\begin{tabular}{@{}c@{}}
				\strictmode: block\\
				\deployablemode: block
		\end{tabular}};
		
		\draw[flow] (clause) -- (ex1.north);
		\draw[flow] (norm) -- (ex2.north);
		\draw[flow] (reg) -- (ex3.north);
		\draw[flow] (eval) -- (ex4.north);
		\draw[flow] (gate) -- (ex5.north);
		
	\end{tikzpicture}
	\caption{From normative text to executable gate. The registry is the mechanism that turns standards prose into an operator-owned, corpus-exercised assurance requirement.}
	\label{fig:translation-pipeline}
\end{figure}
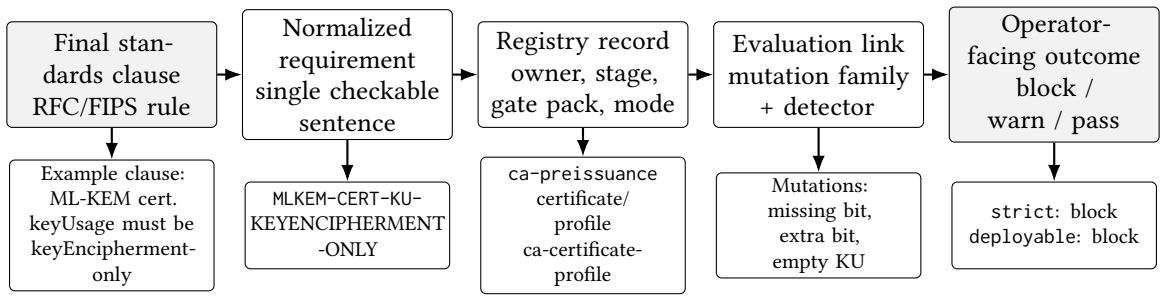

\Cref{fig:translation-pipeline} illustrates the translation pipeline from standards clause to operator-facing gate. Consider the \mlkem certificate rule that, when \texttt{keyUsage} is present, \texttt{keyEncipherment} must be the only active bit \cite{rfc9935}. In the registry, this becomes \req{MLKEM-CERT-KU-KEYENCIPHERMENT-ONLY}. The record is assigned to \texttt{artifact\_type=certificate}, \texttt{stage=certificate/profile}, \texttt{owner=ca-preissuance}, \texttt{gate\_pack=ca-certificate-profile}, and \texttt{detector\_kind=policy}. It is then linked to mutation families such as missing \texttt{keyEncipherment}, extra prohibited bits, and empty \texttt{keyUsage}; finally, it receives \texttt{block} as its action in both modes.

The operational context around the detector is part of the requirement. It has a normative anchor, a workflow owner, a corpus realization, and an output policy. The same discipline applies to structural constraints such as absent parameters and exact public-key lengths, and to import-side conditions such as seed/expanded consistency checks.

\minorhead{Registry-driven policy}
\label{subsec:why-registry-driven-policy}

A registry-driven policy model resolves three issues that ordinary detector lists leave open. First, it provides \emph{traceability}: every active check can be traced back to final standards text and forward to evidence. Second, it provides \emph{accountability}: every check belongs to an owner, a stage, and a gate pack rather than floating in a generic lint namespace. Third, it provides \emph{operational differentiation}: the same underlying detector evidence can be mapped to distinct actions under \strictmode and \deployablemode without duplicating the normative record.

The registry also constrains the manuscript to claims backed by registry records, policy outputs, corpus mutations, and frozen summaries. In that sense, the registry binds standards, workflow, policy, and evaluation within a single assurance artifact.

\partline{Realization}
\section{Operational Assurance Model}
\label{sec:operational-model}

\minorhead{Owner-stage workflow}
\label{subsec:owner-stage-workflow}

Part~I fixed the normative baseline, the executable scope, and the requirement
schema. The next step is to translate those records into a workflow that an
operator can run. The present artifact places every active requirement at the
intersection of an \emph{owner} and a \emph{stage}. In the current \pkixcore
release line, two owners carry executable responsibility: \emph{CA pre-issuance}
and \emph{artifact importer}. The CA owns the two issuance-side stages,
\emph{certificate/profile} and \emph{SPKI/public-key}, which together account
for 10 active requirements. The importer owns the
\emph{private-key-container/import} stage, which accounts for the remaining 7.
The \emph{runtime-consumer} role is retained explicitly in the workflow model,
but has zero active requirements, as anticipated by the scope boundary in
\Cref{fig:assurance-boundaries,tab:pkixcore-scope}.

This split answers the operator question that standards text alone does not
settle: who must run which gate before the artifact is allowed to cross into the
next stage of the pipeline. CA-side defects and importer-side defects are both
assurance defects, but they arise under different authority, different failure
semantics, and different remediation paths. A certificate-profile violation must
be corrected before issuance. A malformed private-key container must be rejected
before import.

The owner-stage assignment is frozen into the workflow package emitted by the
artifact. The machine-readable workflow view records the active owners, their
stages, their commands, and their mode-specific requirement actions. The
human-readable workflow view translates the same information into an operator
recipe. Because both views are generated from the registry, the deployment
guidance cannot drift from the executable artifact through separate manual
maintenance.

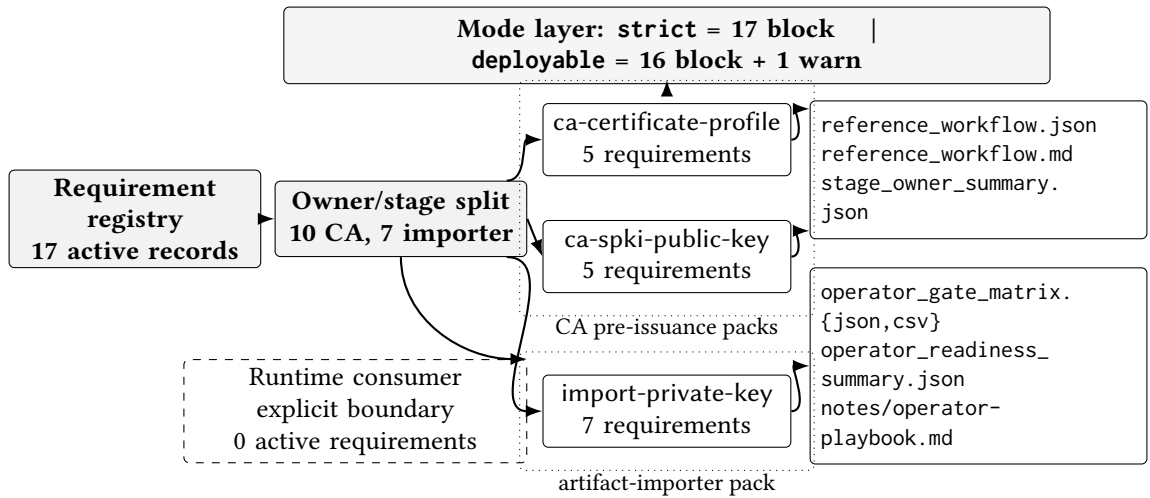
\begin{figure}[H]
	\centering
	\begin{tikzpicture}[x=1cm,y=1cm,>=Latex,
		block/.style={
			draw, rounded corners=2pt,
			minimum height=1.0cm,
			text width=3.10cm,
			inner sep=3pt,
			align=center,
			font=\small\bfseries,
			fill=black!5
		},
		gate/.style={
			draw, rounded corners=2pt,
			minimum height=0.95cm,
			text width=3.05cm,
			inner sep=3pt,
			align=center,
			font=\small
		},
		outbox/.style={
			draw, rounded corners=2pt,
			minimum height=1.55cm,
			text width=4.15cm,
			inner sep=4pt,
			align=left,
			font=\footnotesize
		},
		band/.style={
			draw, rounded corners=2pt,
			minimum height=0.85cm,
			text width=9.8cm,
			inner sep=4pt,
			align=center,
			font=\small\bfseries,
			fill=black!4
		},
		boundary/.style={
			draw, dashed, rounded corners=2pt,
			minimum height=1.0cm,
			text width=4.30cm,
			inner sep=3pt,
			align=center,
			font=\small
		},
		flow/.style={-{Latex[length=2.2mm]}, thick}
		]
		
		\node[block] (reg) at (1.8,3.7)
		{Requirement registry\\17 active records};
		
		\node[block] (split) at (5.3,3.7)
		{Owner/stage split\\10 CA, 7 importer};
		
		\node[gate] (cert) at (8.8,4.75)
		{\gatepack{ca-certificate-profile}\\5 requirements};
		
		\node[gate] (spki) at (8.8,3.20)
		{\gatepack{ca-spki-public-key}\\5 requirements};
		
		\node[gate] (priv) at (8.8,1.15)
		{\gatepack{import-private-key}\\7 requirements};
		
		\node[band] (mode) at (8.8,6.0)
		{Mode layer: \strictmode = 17 block \quad|\quad \deployablemode = 16 block + 1 warn};
		
		\node[outbox] (wf) at (12.9,4.35)
		{\begin{tabular}{@{}l@{}}
				\texttt{reference\_workflow.json}\\
				\texttt{reference\_workflow.md}\\
				\texttt{stage\_owner\_summary.}\\
				\texttt{json}
		\end{tabular}};
		
		\node[outbox] (ops) at (12.9,1.75)
		{\begin{tabular}{@{}l@{}}
				\texttt{operator\_gate\_matrix.}\\
				\texttt{\{json,csv\}}\\
				\texttt{operator\_readiness\_}\\
				\texttt{summary.json}\\
				\texttt{notes/operator-}\\
				\texttt{playbook.md}
		\end{tabular}};
		
		\node[boundary] (run) at (4.7,1.15)
		{Runtime consumer\\explicit boundary\\0 active requirements};
		
		\node[draw, dotted, rounded corners=2pt, fit=(cert)(spki), inner sep=3mm] (capacks) {};
		\node[draw, dotted, rounded corners=2pt, fit=(priv), inner sep=3mm] (imppack) {};
		
		\node[font=\footnotesize] at (8.8,2.25) {CA pre-issuance packs};
		\node[font=\footnotesize] at (8.8,0.18) {artifact-importer pack};
		
		\draw[flow] (reg) -- (split);
		
		\draw[flow] (split.20)  to[out=0,in=180] (cert.west);
		\draw[flow] (split.east) -- (spki.west);
		\draw[flow] (split.-20) to[out=0,in=180] (priv.west);
		
		\draw[flow] (split.south) to[out=-90,in=180] (run.north east);
		
		\draw[flow] (mode.south) -- (capacks.north);
		
		\draw[flow] (cert.east) to[out=0,in=180] (wf.160);
		\draw[flow] (spki.east) to[out=0,in=180] (wf.200);
		\draw[flow] (priv.east) to[out=0,in=180] (ops.180);
		
	\end{tikzpicture}
	\caption{From registry records to operator workflow. Requirements are assigned to owners and stages, grouped into gate packs, and emitted as machine-readable and human-readable workflow outputs.}
	\label{fig:workflow-package}
\end{figure}

\minorhead{Gate packs and accountability}
\label{subsec:gate-packs-accountability}

Gate packs operationalize the owner-stage model. Each pack is the smallest
operator-facing assurance bundle that can be invoked, reviewed, and assigned
without reading the source tree. The current profile defines three packs, each
mapped one-to-one to an owner-stage row:
\gatepack{ca-certificate-profile},
\gatepack{ca-spki-public-key}, and
\gatepack{import-private-key}. The policy layer enforces this mapping, so a
requirement cannot be assigned to the wrong owner or stage.

Because each pack carries explicit responsibility, failures map directly to the
relevant workflow: \gatepack{ca-certificate-profile} to issuance and
\gatepack{import-private-key} to the importer. \Cref{tab:gate-pack-summary}
captures the current view: three packs, three active owner-stage rows,
17 requirements, and one mode-dependent non-blocking path.

\begin{table}[H]
	\caption{Operator-facing gate packs in the current \pkixcore release line. The table records the one-to-one mapping between pack, owner, and stage, and identifies the outputs that a CA or importer inspects after running the corresponding gate.}
	\vspace{.5cm}
	\label{tab:gate-pack-summary}
	\centering
	\scriptsize
	\begin{tabularx}{\textwidth}{@{}>{\raggedright\arraybackslash}p{2.55cm}>{\raggedright\arraybackslash}p{2.0cm}>{\raggedright\arraybackslash}p{2.35cm}c c c >{\raggedright\arraybackslash}X@{}}
		\toprule
		Gate pack & Owner & Stage & Req. & \strictmode & \deployablemode & Principal operator outputs \\
		\midrule
		\gatepack{ca-certificate-profile}
		& CA pre-issuance
		& certificate/profile
		& 5
		& 5 block
		& 5 block
		& \texttt{extended\_registry\_summary*}, \texttt{policy\_summary*}, \texttt{certificate\_spki\_coverage*}, \texttt{operator\_readiness\_summary.json} \\
		
		\gatepack{ca-spki-public-key}
		& CA pre-issuance
		& SPKI/public-key
		& 5
		& 5 block
		& 4 block, 1 warn
		& \texttt{extended\_registry\_summary*}, \texttt{policy\_summary*}, \texttt{certificate\_spki\_coverage*}, \texttt{operator\_gate\_matrix.json} \\
		
		\gatepack{import-private-key}
		& Artifact importer
		& private-key-container/import
		& 7
		& 7 block
		& 7 block
		& \texttt{extended\_registry\_summary*}, \texttt{policy\_summary*}, \texttt{private\_key\_coverage*}, \texttt{operator\_gate\_matrix.json} \\
		\midrule
		Total & --- & --- & 17 & 17 block & 16 block, 1 warn & \texttt{reference\_workflow.\{json,md\}} and \texttt{notes/operator-playbook.md} provide the workflow-wide view. \\
		\bottomrule
	\end{tabularx}
\end{table}

\newpage

\minorhead{Mode semantics: \strictmode and \deployablemode}
\label{subsec:mode-semantics}

The policy model separates \emph{detection} from \emph{operator action}.
Detectors produce findings against artifact content; policy then interprets
those findings under a selected mode. In the current profile, both modes operate
over the same 17 active requirements and the same implemented detector set.
\strictmode is the assurance-maximizing interpretation, under which every active
requirement blocks. \deployablemode is the operator-facing issuance posture: it
preserves the same detector evidence, but the policy layer can downgrade selected
conditions from block to warning when that yields a lower-noise deployment-facing
surface without discarding the underlying signal.

The present release line exercises that split exactly once. The single warning
path is \\ \req{MLKEM-SPKI-ENCODE-DECODE-IDENTITY}, a canonicality condition on
\mlkem public-key material that is implemented locally and exercised by an
unreduced ByteEncode12 mutation in the controlled corpus. In \strictmode the
condition blocks; in \deployablemode it warns. Every other active requirement
blocks in both modes. The modes distinguish the assurance-maximizing posture from
the deployment-facing posture while preserving the raw evidence.

\begin{figure}[H]
	\centering
	\begin{tikzpicture}[x=1cm,y=1cm,>=Latex,
		modebox/.style={
			draw, rounded corners=2pt,
			text width=4.35cm,
			minimum height=3.65cm,
			inner sep=5pt,
			align=left,
			font=\small
		},
		warnbox/.style={
			draw, fill=black!7, rounded corners=2pt,
			text width=4.10cm,
			inner sep=4pt,
			align=left,
			font=\footnotesize
		},
		flow/.style={-{Latex[length=2mm]}, thick}
		]
		
		\node[modebox] (strict) at (0,0) {%
			\textbf{\strictmode}\\[1mm]
			17 blocking requirements\\[1.5mm]
			certificate/profile: 5 block\\
			SPKI/public-key: 5 block\\
			private-key/import: 7 block\\[2mm]
			{\footnotesize No warning path}
		};
		
		\node[modebox] (deploy) at (8.0,0) {%
			\textbf{\deployablemode}\\[1mm]
			16 blocking requirements\\[1.5mm]
			certificate/profile: 5 block\\
			SPKI/public-key: 4 block, 1 warn\\
			private-key/import: 7 block
		};
		
		\node[warnbox, anchor=north west] (warn)
		at ([xshift=2mm,yshift=-2mm]deploy.south west) {%
			Warning path:\\
			\texttt{MLKEM-SPKI-ENCODE-}\\
			\texttt{DECODE-IDENTITY}
		};
		
		\draw[flow]
		([yshift=3mm]strict.north east) --
		([yshift=3mm]deploy.north west)
		node[midway, above, font=\footnotesize, align=center]
		{same detector evidence\\different operator action};
		
	\end{tikzpicture}
	\caption{Mode-aware policy in the current artifact.x.}
	\label{fig:policy-mode-split}
\end{figure}
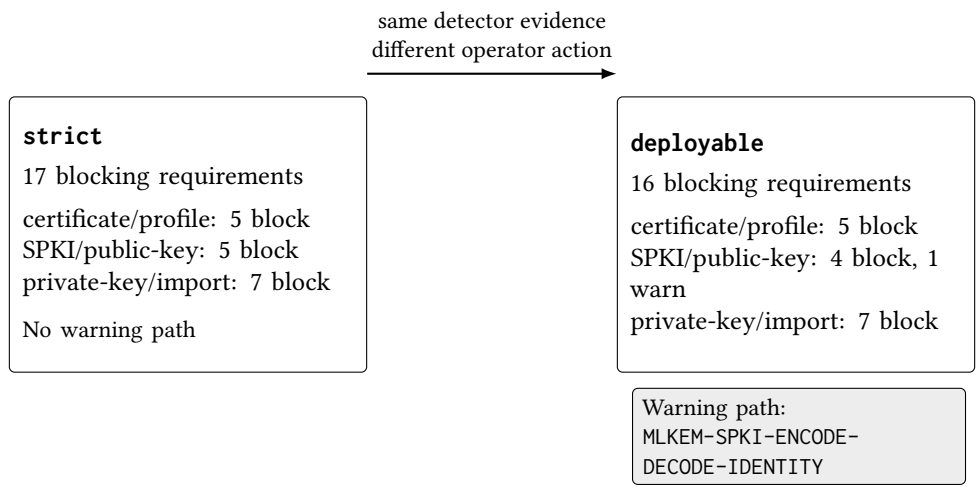

\minorhead{Runtime-consumer as explicit boundary}
\label{subsec:runtime-boundary}

The runtime-consumer role is modeled and kept outside the executable profile.
Issuance-side or importer-side evidence is not treated as a proxy for end-to-end
runtime consumer behavior. The artifact keeps runtime visible as a role in the
workflow package, records that it has zero active executable requirements, and
does not derive pseudo-coverage from certificate-only or importer-only results.

\minorhead{Outputs produced for operators}
\label{subsec:operator-outputs}

The artifact emits outputs in two strata. The first stratum is
\emph{policy- and workflow-facing}. This includes
\json{policy\_summary\_strict.json},
\json{policy\_summary\_deployable.json},
\json{stage\_owner\_summary.json},
\json{reference\_workflow.json},
\json{reference\_workflow.md},
\json{operator\_gate\_matrix.json}, \\
\json{operator\_gate\_matrix.csv}, and
\json{operator\_readiness\_summary.json}. These files answer which requirements
belong to which owner, what each mode does, which pack a requirement inhabits,
and what the default operator posture should be.

The second stratum is \emph{execution-facing}. Here the primary outputs are the
mode-specific evaluation summaries and coverage reports, including
\json{extended\_registry\_summary\_*.json},
\\ \json{certificate\_spki\_coverage\_*.json}, and
\json{private\_key\_coverage\_*.json}. These outputs are inspected after running
the relevant gate-pack commands. The human-readable playbook in \\
\texttt{notes/operator-playbook.md} spans both strata: it names the commands to
execute and the files to inspect.

The artifact can therefore be used as a paper companion and as an operational
recipe. A reviewer can read the workflow files without running the code. An
operator can run the commands and inspect the matching summaries. An author can
write the paper against outputs that are already frozen and hashable. These are
three users of the same package, not three divergent documentation paths.

\minorhead{Operational interpretation}
\label{subsec:operational-interpretation}

The artifact is an assurance workflow package that can be delegated, audited,
and replayed. In that form, it extends public-key linting into an assurance
workflow \cite{karatsiolis2026pqlinting}. The relevant question is whether a
defect is assigned to the correct owner, rendered under a defensible action
policy, and presented in a form that an operator can use before issuance or
before import.

The workflow also clarifies why \deployablemode defaults to the CA-facing path
while \strictmode is the reference assurance posture. Issuance pipelines benefit
from a low-noise default gate that still preserves policy evidence. Audit and
pre-release review benefit from the maximal blocking interpretation. The
importer path blocks all currently modeled private-key defects in both modes
because the current profile gives no operational reason to soften those checks.

\section{Artifact Architecture and Implementation}
\label{sec:artifact-architecture}

\minorhead{Repository design}
\label{subsec:repository-design}

The repository is organized according to an evidence-first discipline. Normative
sources are frozen and documented under \texttt{reference/}; the machine-readable
assurance registry lives in \texttt{requirements.json}; local implementation code
lives in \texttt{src/}; valid and mutated artifacts plus their ledgers live in
\texttt{corpus/}; reproducible runner entrypoints live in \texttt{experiments/};
raw machine-readable outputs accumulate under \texttt{results/}; and playbooks
and manuscript-lock materials live under \texttt{notes/}.

The layout addresses three recurrent risks in reproducible assurance work:
undocumented provenance, undocumented evaluation, and manuscript drift. The
repository design counters these risks directly. Provenance is captured through
reference manifests, corpus hashes, and package manifests. Evaluation is captured
through replayable scripts and machine-readable summaries. Manuscript drift is
controlled by locating claim maps, visual plans, related-work positioning, and
release-bundle plans in the same tree as the executable artifact.

\begin{figure}[H]
	\centering
	\resizebox{\linewidth}{!}{%
		\begin{tikzpicture}[x=1cm,y=1cm,>=Latex,
			block/.style={
				draw, rounded corners=2pt,
				minimum height=1.05cm,
				text width=2.75cm,
				inner sep=3pt,
				align=center,
				font=\small\bfseries,
				fill=black!5
			},
			upper/.style={
				draw, rounded corners=2pt,
				minimum height=0.95cm,
				text width=3.00cm,
				inner sep=3pt,
				align=center,
				font=\scriptsize
			},
			small/.style={
				draw, rounded corners=2pt,
				minimum height=0.95cm,
				text width=2.45cm,
				inner sep=3pt,
				align=center,
				font=\scriptsize
			},
			note/.style={
				draw, dashed, rounded corners=2pt,
				minimum height=0.95cm,
				text width=3.30cm,
				inner sep=3pt,
				align=center,
				font=\scriptsize
			},
			flow/.style={-{Latex[length=2.1mm]}, thick}
			]
			
			\node[block] (refs) at (1.6,3.9)
			{\texttt{reference/}\\frozen norms};
			
			\node[block, text width=2.95cm] (reg) at (4.8,3.9)
			{\texttt{requirements.}\\\texttt{json}\\registry};
			
			\node[block] (src) at (8.0,3.9)
			{\texttt{src/}\\assurance engine};
			
			\node[block] (corpus) at (11.2,3.9)
			{\texttt{corpus/}\\valid + mutated};
			
			\node[block] (exp) at (14.4,3.9)
			{\texttt{experiments/}\\replay + runners};
			
			\node[upper] (tp) at (4.0,5.95)
			{\texttt{third\_party/}\\frozen upstream\\snapshots};
			
			\node[upper, text width=3.15cm] (bridge) at (8.3,5.95)
			{\texttt{tools/}\\\texttt{libcrux\_import\_check}\\narrow import substrate};
			
			\node[block, text width=3.65cm] (res) at (8.0,1.00)
			{\texttt{results/}\\machine-readable evidence};
			
			\node[small] (notes) at (11.8,1.00)
			{\texttt{notes/}\\playbooks + locks};
			
			\node[note] (pkg) at (15.35,1.00)
			{\texttt{artifact\_release\_}\\\texttt{package.json}};
			
			\draw[flow] (refs) -- (reg);
			\draw[flow] (reg) -- (src);
			\draw[flow] (src) -- (corpus);
			\draw[flow] (corpus) -- (exp);
			
			\draw[flow] (tp.south) -- (reg.north);
			\draw[flow] (bridge.south) -- (src.north);
			
			\draw[flow] (exp.south west)
			.. controls +(-0.1,-1.25) and +(1.45,1.00) ..
			(res.north east);
			
			\draw[flow] (res.east) -- (notes.west);
			\draw[flow] (notes.east) -- (pkg.west);
			
			\node[draw, dotted, rounded corners=2pt,
			fit=(refs)(reg)(src)(corpus)(exp),
			inner sep=3.5mm] (core) {};
			
			\node[draw, dotted, rounded corners=2pt,
			fit=(res)(notes)(pkg),
			inner sep=3.5mm] (release) {};
			
			\node[font=\footnotesize] at ([yshift=-5.5mm]core.south)
			{executable artifact core};
			
			\node[font=\footnotesize] at ([yshift=-5.5mm]release.south)
			{evidence, locks, and release packaging};
			
		\end{tikzpicture}%
	}
	\caption{Artifact realization and evidence flow. Project claims are first fixed in the registry, corpus, replay scripts, and packaging discipline; the paper sits downstream of that evidence flow.}
	\label{fig:artifact-pipeline}
\end{figure}
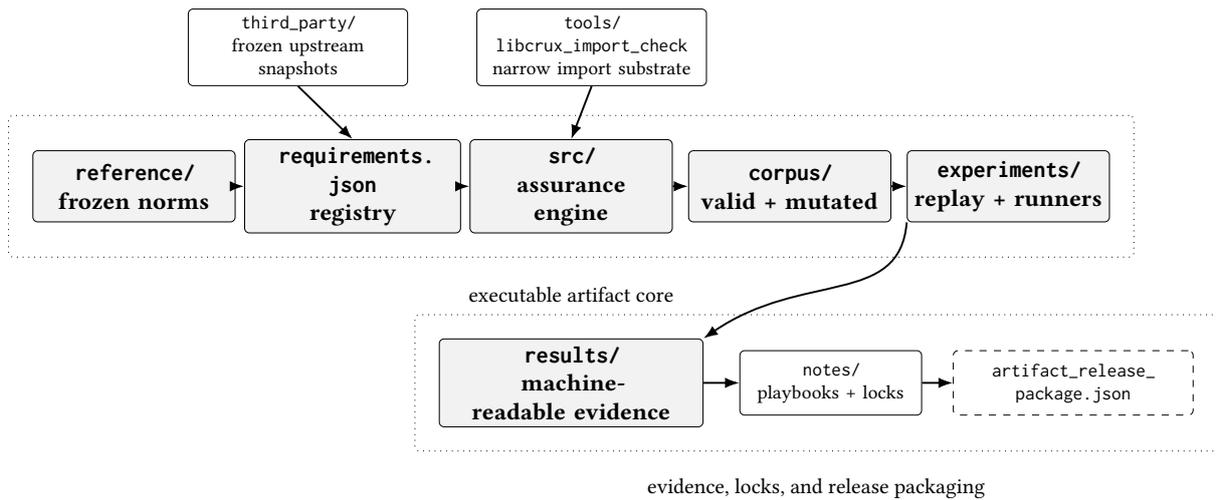

\minorhead{Assurance engine}
\label{subsec:assurance-engine}

The principal runner loads the registry, loads the corpus manifest, resolves
artifact paths, and dispatches on artifact type. Certificates are evaluated
through a certificate-specific DER and X.509 extractor; raw SubjectPublicKeyInfo
artifacts are evaluated through an SPKI extractor; and private-key containers are
evaluated through a dedicated PKCS\#8 / OneAsymmetricKey parser. Each detector
produces requirement-tagged findings. The policy layer then computes the
applicable requirement set, interprets findings under the selected mode, and
emits a final disposition of \texttt{pass}, \texttt{warn}, or \texttt{block}
together with first-hit, redundancy, and missing-expected-detection summaries.

Detector code determines whether a given artifact violates a given condition.
Policy code determines the artifact's operational fate under the active profile
and mode. The same underlying findings can therefore be rendered differently in
\strictmode and \deployablemode without creating two separate detector
universes. The engine also emits CSV and JSON views of the registry itself,
which makes it possible to audit the evaluation outputs alongside the policy
topology that produced them.

The engine does not assume that manifest paths are anchored at one hard-coded
root. It resolves the manifest root from the actual file layout before running
the corpus, which keeps replay stable across packaged and unpackaged
environments.

\minorhead{Detector classes}
\label{subsec:detector-classes}

The implemented detector inventory currently spans three classes. \emph{Structural}
detectors capture exact container and representation constraints: absent
AlgorithmIdentifier parameters, OID-to-length consistency, public-key payload
size, signatureAlgorithm parameter absence, canonical encode/decode identity,
and private-key seed or expanded-key length. \emph{Policy} detectors capture
profile semantics that are not exhausted by representation alone: \mlkem
\texttt{keyUsage}, \mldsa positive signing semantics, prohibited
encipherment/agreement bits, and the PKIX-specific HashML-DSA exclusion.
Finally, \emph{import-crypto} detectors perform checks whose meaning depends on
the semantics of the imported key material itself, such as seed-to-expanded
consistency and the \mlkem expanded-key hash condition.

All 17 active requirements in the current registry map to implemented detector
metadata. Quantitatively, the current artifact contains 10 structural
requirements, 4 policy requirements, and 3 import-crypto requirements.

\newpage

\minorhead{Import-validation bridge}
\label{subsec:import-validation-bridge}

The importer-owned slice is where a certificate-centric design would be most
likely to fail. The artifact therefore uses a narrow import-validation bridge
backed by a frozen \texttt{libcrux} snapshot \cite{libcrux_snapshot}. The bridge
is a small, measured import-validation substrate that exposes the operation
needed by the current profile for executable consistency checks: whether a given
seed and a given expanded private key correspond for a stated parameter set.

Concretely, the local wrapper invokes a dedicated binary under
\texttt{tools/libcrux\_import\_check}, passing the parameter set and the two
private-key components as hexadecimal arguments. A zero exit code indicates
consistency, an exit code of one indicates a mismatch, and any other exit code
is treated as a bridge failure. The private-key parser calls this bridge only
after the relevant container forms and lengths have already been validated,
which keeps the import-crypto checks narrow, interpretable, and subordinate to
the local parsing logic. Within this paper, \texttt{libcrux} is a precise
import-validation substrate, not a general runtime oracle.

This design records which cryptographic import semantics are delegated to a
frozen external substrate while still covering checks that go beyond container
lengths and address seed/expanded coherence.

\minorhead{Corpus generation and mutation pipeline}
\label{subsec:corpus-generation-mutation}

The corpus pipeline preserves provenance at the level of individual artifacts.
Valid artifacts are generated locally through explicit corpus scripts and then
recorded in \texttt{corpus/manifest.jsonl} with stage, algorithm, parameter set,
source note, and SHA-256 hash. Invalid artifacts are produced in three ways:
scripted OpenSSL-side generation for certificate-profile cases, deterministic
DER mutation for fine-grained container and SPKI faults, and private-key-container
reconstruction for importer-side failures. Every invalid artifact carries its
own \texttt{expected\_detection} set and its mutation-family labels, which lets
the evaluation layer distinguish missed expectations from redundant or
unexpected findings.

The corpus therefore functions as a ledger, not as a loose directory of
``good'' and ``bad'' files. The current controlled corpus contains 48 artifacts
in total, with 21 valid and 27 invalid cases, balanced across \mlkem and \mldsa,
and spread across the three active stages. The invalid side spans encoding,
size/shape, inter-field consistency, policy, field-domain, and import-validation
fault families. Because the mutation layer is deterministic, the same mutation
family can be regenerated and replayed without ambiguity once the frozen valid
input artifacts are fixed.

The valid OpenSSL-generated corpus is frozen by hash on first generation and
then reused on later replays.

\minorhead{Freeze/replay pipeline}
\label{subsec:freeze-replay-pipeline}

The canonical executable entrypoint is \texttt{./experiments/replay\_freeze.sh}.
Its job is to rerun detectors and regenerate the entire paper-facing evidence
surface. In its current form, the replay pipeline checks the execution
environment; prepares frozen third-party material; builds the import-validation
bridge; reuses or regenerates the controlled corpus and DER mutations; reruns
the extended suite in both modes; reruns stage-specific coverage for the
certificate/SPKI and private-key slices; regenerates policy summaries,
owner-stage summaries, workflow files, and operator gate packs; reruns the
certificate-level baseline comparison; refreshes the bounded public appendix and
the cross-tool layer; rewrites claim locks, visual locks, and positioning locks;
refreshes the artifact release package; records an explicit upgrade decision;
and finally executes the local smoke tests.

\section{Requirement Registry and Policy Model}
\label{sec:requirement-registry}

\minorhead{Global inventory of 17 active requirements}
\label{subsec:global-inventory}

The active registry currently contains 17 requirements, all marked
\texttt{covered} in the constructibility field and all mapped to implemented
detector metadata. The inventory is nearly balanced by algorithm family, with 8
\mlkem requirements and 9 \mldsa requirements, but its main balance is by
workflow semantics: 10 requirements belong to CA pre-issuance and 7 belong to
the artifact importer. The resulting distribution is 5 requirements in
\emph{certificate/profile}, 5 in \emph{SPKI/public-key}, and 7 in
\emph{private-key-container/import}.

\Cref{tab:registry-inventory} presents the current inventory in full.

\begingroup
\scriptsize
\begin{xltabular}{\textwidth}{@{}>{\RaggedRight\arraybackslash}p{2.15cm}>{\RaggedRight\arraybackslash}p{3.15cm}Y>{\RaggedRight\arraybackslash}p{1.15cm}>{\RaggedRight\arraybackslash}p{0.9cm}cc@{}}
	\caption{Active registry inventory in \pkixcore. The current artifact is not built around one family of homogeneous checks, but around a mixed topology of structural, policy, and import-crypto requirements with explicit mode actions.}
	\label{tab:registry-inventory}\\
	\toprule
	Gate pack & Requirement ID & Operational intent & Detector kind & Norm. & \strictmode & \deployablemode \\
	\midrule
	\endfirsthead
	
	\toprule
	Gate pack & Requirement ID & Operational intent & Detector kind & Norm. & \strictmode & \deployablemode \\
	\midrule
	\endhead
	
	\midrule
	\multicolumn{7}{r@{}}{\footnotesize Continued on next page} \\
	\endfoot
	
	\bottomrule
	\endlastfoot
	
	\multicolumn{7}{@{}l}{\textbf{CA certificate/profile}} \\\\
	\gatepack{ca-certificate-profile} & \reqtab{MLKEM-CERT-KU-KEYENCIPHERMENT-ONLY} & Enforce that an \mlkem certificate uses \texttt{keyEncipherment} as the sole active \texttt{keyUsage} bit when the extension is present. & policy & must & block & block \\
	\gatepack{ca-certificate-profile} & \reqtab{MLDSA-CERT-SIGNATURE-AID-PARAMS-ABSENT} & Require absent parameters in \mldsa certificate signature AlgorithmIdentifiers at both TBS and outer layers. & structural & must & block & block \\
	\gatepack{ca-certificate-profile} & \reqtab{MLDSA-CERT-KU-AT-LEAST-ONE-SIGNING-BIT} & Enforce the positive PKIX condition that an \mldsa certificate expose at least one signing-related \texttt{keyUsage} bit. & policy & must & block & block \\
	\gatepack{ca-certificate-profile} & \reqtab{MLDSA-CERT-KU-NO-ENCIPHERMENT-OR-AGREEMENT} & Forbid encipherment and agreement semantics for \mldsa certificates. & policy & must & block & block \\
	\gatepack{ca-certificate-profile} & \reqtab{MLDSA-PKIX-HASHML-FORBIDDEN} & Block HashML-DSA certificate signatures inside the covered PKIX profile. & policy & must & block & block \\
	\midrule
	
	\multicolumn{7}{@{}l}{\textbf{CA SPKI/public-key}} \\\\
	\gatepack{ca-spki-public-key} & \reqtab{MLKEM-SPKI-AID-PARAMS-ABSENT} & Require an \mlkem SPKI AlgorithmIdentifier with absent parameters. & structural & must & block & block \\
	\gatepack{ca-spki-public-key} & \reqtab{MLKEM-SPKI-PUBLIC-KEY-LENGTH} & Enforce parameter-set-specific \mlkem public-key payload lengths. & structural & must & block & block \\
	\gatepack{ca-spki-public-key} & \reqtab{MLKEM-SPKI-ENCODE-DECODE-IDENTITY} & Require decode/re-encode identity for \mlkem public-key canonicality. & structural & must & block & warn \\
	\gatepack{ca-spki-public-key} & \reqtab{MLDSA-SPKI-AID-PARAMS-ABSENT} & Require an \mldsa SPKI AlgorithmIdentifier with absent parameters. & structural & must & block & block \\
	\gatepack{ca-spki-public-key} & \reqtab{MLDSA-SPKI-PUBLIC-KEY-LENGTH} & Enforce parameter-set-specific \mldsa public-key payload lengths. & structural & must & block & block \\
	\midrule
	
	\\\\
	
	\multicolumn{7}{@{}l}{\textbf{Artifact importer private-key/import}} \\\\
	\gatepack{import-private-key} & \reqtab{MLKEM-PRIVATE-SEED-LENGTH} & Enforce 64-byte \mlkem seed-form private-key containers. & structural & must & block & block \\
	\gatepack{import-private-key} & \reqtab{MLKEM-PRIVATE-EXPANDED-LENGTH} & Enforce parameter-set-specific \mlkem expanded private-key lengths. & structural & must & block & block \\
	\gatepack{import-private-key} & \reqtab{MLKEM-PRIVATE-BOTH-CONSISTENCY} & Check that the seed and expanded \mlkem representations in \texttt{both} form are mutually consistent. & import-crypto & should & block & block \\
	\gatepack{import-private-key} & \reqtab{MLKEM-PRIVATE-EXPANDED-HASH-CHECK} & Validate the \mlkem expanded-key hash relation required by the final container layout. & import-crypto & must & block & block \\
	\gatepack{import-private-key} & \reqtab{MLDSA-PRIVATE-SEED-LENGTH} & Enforce 32-byte \mldsa seed-form private-key containers. & structural & must & block & block \\
	\gatepack{import-private-key} & \reqtab{MLDSA-PRIVATE-EXPANDED-LENGTH} & Enforce parameter-set-specific \mldsa expanded private-key lengths. & structural & must & block & block \\
	\gatepack{import-private-key} & \reqtab{MLDSA-PRIVATE-BOTH-CONSISTENCY} & Check that the seed and expanded \mldsa representations in \texttt{both} form are mutually consistent. & import-crypto & should & block & block \\
\end{xltabular}
\endgroup

\minorhead{CA certificate/profile requirements}
\label{subsec:ca-certificate-profile-reqs}

The certificate/profile slice contains 5 requirements, all owned by the CA and
all blocking in both modes. Their common property is that they express
\emph{issuance semantics} beyond DER well-formedness. The \mlkem certificate
requirement is a positive and negative \texttt{keyUsage} rule: if the extension
is present, \texttt{keyEncipherment} must be the only active bit. The \mldsa
slice contains two \texttt{keyUsage} rules, one positive and one negative. The
positive rule requires at least one signing-related bit; the negative rule
forbids encipherment and agreement semantics.

The remaining certificate/profile requirements close two final-standards
AlgorithmIdentifier seams. One requires absent parameters for \mldsa signature
identifiers in certificates. The other blocks HashML-DSA certificate signatures
inside the covered PKIX profile \cite{rfc9881}. Together, these five
requirements define a CA-side profile gate: they tell the issuance path whether
the certificate satisfies the final PKIX profile beyond basic parse success.

\minorhead{CA SPKI/public-key requirements}
\label{subsec:ca-spki-public-key-reqs}

The SPKI/public-key slice also contains 5 requirements, all CA-owned. Unlike the
certificate/profile slice, the SPKI slice is entirely structural in the current
profile. It requires absent parameters in SPKI AlgorithmIdentifiers for both
algorithms, enforces parameter-set-specific payload lengths for both algorithms,
and adds a canonicality check for \mlkem public-key encodings
\cite{fips203,rfc9935}. These checks are structurally motivated, but they are
still issuance-side checks because the CA is the last party that can prevent a
malformed or semantically suspicious public key from being embedded into a
certificate and pushed downstream.

This slice can be exercised both on raw SPKI artifacts and on certificates that
carry SPKI defects inside a larger X.509 container.

\minorhead{Artifact-importer private-key requirements}
\label{subsec:artifact-importer-private-key-reqs}

The importer slice contains 7 requirements and marks the extension beyond
certificate-only linting. Four requirements belong to \mlkem and three to
\mldsa. Both algorithm families require checks for seed-form length,
expanded-form length, and seed/expanded consistency in the \texttt{both}
representation. \mlkem adds a further expanded-key hash check, which follows
from the final container and key-layout semantics in the standards corpus
\cite{fips203,rfc9935}. All 7 requirements block in both modes.

The two consistency checks are labeled \texttt{should} rather than
\texttt{must}, but they still block in the current profile because the importer
boundary is a high-consequence boundary and because executable evidence for
those checks exists through the narrow bridge. This example shows why the
registry keeps \emph{normative strength} and \emph{mode action} as distinct
fields. A requirement can be derived from a \texttt{should}-level standards
statement and still receive a blocking operational action in a profile that
prioritizes safe import behavior.

\minorhead{Detector-kind distribution and normative strength}
\label{subsec:detector-kind-normative-strength}

The current registry contains 10 structural requirements, 4 policy requirements,
and 3 import-crypto requirements. The normative-strength distribution is also
heterogeneous: 15 requirements are labeled \texttt{must} and 2 are labeled
\texttt{should}. The two \texttt{should}-level rows are active importer checks
with implemented detectors and blocking action.

Policy requirements bring PKIX semantics into the registry. Import-crypto
requirements bring the importer boundary into the registry. The assurance claim
rests on the coexistence of these classes in one executable profile, not on a
larger structural linter alone.

\minorhead{Single exercised mode split}
\label{subsec:single-exercised-mode-split}

The single exercised mode split illustrates how the policy model is used. The requirement \\
\req{MLKEM-SPKI-ENCODE-DECODE-IDENTITY} is implemented locally and triggered by
a controlled SPKI mutation that preserves outer container shape while violating
public-key canonicality. In \strictmode the artifact blocks the case. In
\deployablemode it warns. The underlying detector evidence is the same, and the
coverage summary still records the expected detection as met. Only the operator
consequence changes.

This avoids maintaining separate detector sets or using a deployable mode that
suppresses evidence. The warning path is recorded in policy summaries, workflow
outputs, and coverage reports.

\minorhead{Policy outputs and operator-ready matrices}
\label{subsec:policy-outputs-operator-ready}

The policy model is made inspectable through a family of generated outputs. The
pair of policy summaries provides the mode-specific top-line view. The policy
matrix freezes row-level actions across the registry. The stage-owner summary
collapses the inventory by workflow boundary. The operator-gate matrix
re-expresses the same policy in pack form. The workflow files bind these pieces
into an owner/stage recipe, and the operator-readiness summary freezes the
current default mode, reference mode, pack purposes, command lines, and files to
inspect.

This output family answers a question that standards alone do not answer and
that raw detector code answers only indirectly: what should an operator run, in
which mode, at which stage, and how should the result be read. Because all of
these outputs are generated from the same registry, the artifact makes policy as
data, workflow as data, and paper claims about both inspectable from the same
source.

\partline{Evaluation}
\section{Corpus Design, Mutation Families, and Evaluation Protocol}
\label{sec:evaluation-protocol}

\minorhead{Controlled corpus composition}
\label{subsec:controlled-corpus-composition}

The primary empirical evidence in the paper is a frozen controlled corpus of 48 artifacts. The corpus is balanced by algorithm family, with 24 \mlkem artifacts and 24 \mldsa artifacts; nearly balanced by validity, with 21 valid and 27 invalid artifacts; and topologically balanced across the three active workflow surfaces fixed in Parts~I and~II, namely 14 certificate/profile artifacts, 20 SPKI/public-key artifacts, and 14 private-key-container/import artifacts. Because the corpus is indexed by stage, fault family, and expected detection, it provides executable evidence for the operational model rather than an undifferentiated collection of ``test files.''

\Cref{tab:controlled-corpus-design} fixes the composition of the primary evidence layer. The current artifact is evaluated against all three operationally relevant surfaces, and the SPKI slice is split between raw representation artifacts and certificate-carried SPKI defects.

\begin{table}[H]
	\caption{Controlled corpus design. The table records the empirical geometry of
		the primary evidence layer: every active owner-stage row is exercised, every
		stage contains both valid and invalid material, and the SPKI slice is tested
		both on raw SPKI objects and on certificate-carried defects.}
	\label{tab:controlled-corpus-design}
	\centering
	\scriptsize
	\begin{tabularx}{\textwidth}{@{}>{\raggedright\arraybackslash}p{2.75cm}>{\raggedright\arraybackslash}p{2.0cm}>{\raggedright\arraybackslash}p{2.8cm}c c c>{\raggedright\arraybackslash}X@{}}
		\toprule
		Stage & Owner & Artifact forms & Valid & Invalid & Total & Representative exercised faults \\
		\midrule
		certificate/profile
		& CA pre-issuance
		& 14 certificates
		& 7 & 7 & 14
		& PKIX \texttt{keyUsage} semantics, signature \texttt{AlgorithmIdentifier} parameter defects, HashML-DSA policy prohibition \\
		
		SPKI/public-key
		& CA pre-issuance
		& 17 raw SPKI objects + 3 certificate-carried SPKI defects
		& 7 & 13 & 20
		& absent-parameters violations, payload-length defects, OID/payload inconsistency, \mlkem canonicality mutation \\
		
		private-key-container/import
		& Artifact importer
		& 14 private-key containers
		& 7 & 7 & 14
		& seed-form and expanded-form length defects, \texttt{both} inconsistency, \mlkem expanded-key hash mismatch \\
		\midrule
		Total
		& ---
		& 17 certificates, 17 raw SPKI objects, 14 private-key containers
		& 21 & 27 & 48
		& 7 fault families across certificate, SPKI, and importer boundaries \\
		\bottomrule
	\end{tabularx}
\end{table}

The type distribution is as follows: among the 48 artifacts, 17 are raw
certificates, 17 are raw SPKI objects, and 14 are private-key containers.

\minorhead{Valid artifact generation}
\label{subsec:valid-artifact-generation}

The 21 valid artifacts are generated locally through a deterministic OpenSSL path and then frozen by manifest and hash. They are not scraped from the public web, borrowed from the baseline, or post-edited by hand. In the current corpus, there are 7 valid artifacts for each active stage: 7 valid certificates, 7 valid SPKI/public-key artifacts, and 7 valid private-key containers. Every active gate is exercised on a non-trivial valid set and produces no findings on that set.

\minorhead{Invalid mutation families}
\label{subsec:invalid-mutation-families}

The 27 invalid artifacts are arranged into 7 labeled fault families: \emph{encoding/container},
\emph{size/shape}, \emph{inter-field consistency},
\emph{profile/usage-policy}, \emph{field-domain}, \emph{algorithm-policy},
and \emph{import-validation}. Each invalid artifact has a declared fault identity and a declared detection expectation before any detector is run.

\Cref{tab:mutation-family-catalog} summarizes the current mutation catalogue. For example, absent-parameters obligations are exercised separately with \texttt{NULL} and non-\texttt{NULL} parameter values; the ML-KEM canonicality rule is exercised by an unreduced ByteEncode12 coefficient mutation that preserves the outer container shape; and the private-key importer slice includes a dedicated \mlkem expanded-key hash mismatch, not only trivial truncation cases.

\begin{table}[H]
	\caption{Invalid mutation-family catalogue. Each family has an explicit
		operational purpose and a traceable relation to one or more requirement rows
		in the registry.}
	\label{tab:mutation-family-catalog}
	\centering
	\scriptsize
	\begin{tabularx}{\textwidth}{@{}>{\raggedright\arraybackslash}p{2.3cm}c>{\raggedright\arraybackslash}p{2.4cm}>{\raggedright\arraybackslash}p{4.1cm}>{\raggedright\arraybackslash}X@{}}
		\toprule
		Fault family & Count & Principal stage(s) & Representative artifacts & Assurance purpose \\
		\midrule
		encoding/container & 8 & SPKI/public-key, certificate/profile & \path{der-mut-mlkem768-spki-aid-null-pub}, \path{der-mut-mldsa44-cert-signature-aid-octet-params} & Exercise absent-parameters obligations for SPKI and certificate signature \texttt{AlgorithmIdentifier} fields \\
		
		size/shape & 7 & SPKI/public-key, private-key/import & \path{der-mut-mlkem768-spki-payload-truncated-pub}, \path{der-mut-mldsa44-key-expanded-short} & Exercise parameter-set-specific payload, seed, and expanded-key length checks \\
		
		inter-field consistency & 5 & SPKI/public-key, private-key/import & \path{der-mut-mlkem512-spki-oid-swapped-to-mlkem768-pub}, \path{der-mut-mldsa44-key-both-mismatch} & Exercise OID/payload agreement and \texttt{both}-form seed/expanded consistency \\
		
		profile/usage-policy & 4 & certificate/profile & \path{openssl-mut-mlkem768-keyusage-digital-signature-cert}, \path{der-mut-mldsa65-cert-keyusage-empty} & Exercise positive and negative PKIX \texttt{keyUsage} semantics \\
		
		field-domain & 1 & SPKI/public-key & \path{der-mut-mlkem768-spki-unreduced-byteencode12-pub} & Exercise ML-KEM decode/re-encode canonicality rather than mere container shape \\
		
		algorithm-policy & 1 & certificate/profile & \path{der-mut-mldsa44-cert-signature-hashmldsa44} & Exercise final-standards PKIX prohibition of HashML-DSA certificate signatures \\
		
		import-validation & 1 & private-key/import & \path{der-mut-mlkem512-key-hash-mismatch} & Exercise import-side cryptographic validation that cannot be reduced to certificate linting \\
		\bottomrule
	\end{tabularx}
\end{table}

\minorhead{Expected-detection labeling}
\label{subsec:expected-detection-labeling}

Every invalid artifact in the controlled corpus carries an explicit
\texttt{expected\_detection} label in the manifest. The label is requirement-centric, so an artifact counts as successfully covered when the local run recovers the requirement identifiers that the corpus states should fire.

The certificate mutations \path{der-mut-mldsa44-cert-signature-aid-null} and
\path{der-mut-mldsa44-cert-signature-aid-octet-params}, for example, trigger redundant error instances because the same absent-parameters rule is exercised at more than one certificate locus. Their first useful hit remains the same registry row, \path{MLDSA-CERT-SIGNATURE-AID-PARAMS-ABSENT}, and the evaluation should not overweight them merely because the certificate offers more than one place for the defect to surface.

\minorhead{Evaluation protocol and success criteria}
\label{subsec:evaluation-protocol-success}

The local artifact is executed twice over the full controlled corpus, once in
\strictmode and once in \deployablemode. The success criteria are:
\begin{enumerate}
	\item every invalid artifact should meet its expected detection set;
	\item no valid artifact should be blocked or warned;
	\item no unexpected requirement should become outcome-critical; and
	\item stage-local coverage summaries should show no open detector gaps inside
	the declared scope.
\end{enumerate}

\begin{figure}[H]
	\centering
	\resizebox{\linewidth}{!}{%
		\begin{tikzpicture}[x=1cm,y=1cm,>=Latex,
			block/.style={
				draw, rounded corners=2pt,
				minimum height=1.0cm,
				text width=2.75cm,
				inner sep=3pt,
				align=center,
				font=\small\bfseries,
				fill=black!5
			},
			box/.style={
				draw, rounded corners=2pt,
				minimum height=1.05cm,
				text width=3.10cm,
				inner sep=3pt,
				align=center,
				font=\small
			},
			smallbox/.style={
				draw, rounded corners=2pt,
				minimum height=0.95cm,
				text width=3.15cm,
				inner sep=3pt,
				align=center,
				font=\footnotesize
			},
			outbox/.style={
				draw, rounded corners=2pt,
				minimum height=1.55cm,
				text width=4.25cm,
				inner sep=4pt,
				align=center,
				font=\footnotesize
			},
			wideout/.style={
				draw, rounded corners=2pt,
				minimum height=1.80cm,
				text width=11.20cm,
				inner sep=4pt,
				align=center,
				font=\footnotesize
			},
			flow/.style={
				-{Latex[length=2.1mm]},
				thick,
				rounded corners=2pt
			}
			]
			
			\node[block] (registry) at (1.7,7.0)
			{Requirement\\registry\\17 active rows};
			
			\node[box] (valid) at (5.6,8.9)
			{Deterministic\\valid generation\\21 valid artifacts};
			
			\node[box] (mut) at (5.6,5.1)
			{Deterministic\\mutation families\\27 invalid artifacts};
			
			\node[box, text width=3.55cm] (manifest) at (10.4,7.0)
			{Manifest labeling\\stage, validity,\\fault family, expected\\detection, SHA-256};
			
			\node[block, text width=3.25cm] (eval) at (15.1,7.0)
			{Dual-mode\\local evaluation\\\strictmode\ and\\\deployablemode};
			
			\node[outbox] (primary) at (20.0,8.95)
			{Primary outputs\\
				\texttt{extended\_registry\_}\\
				\texttt{summary*}\\
				\texttt{certificate\_spki\_}\\
				\texttt{coverage*}\\
				\texttt{private\_key\_}\\
				\texttt{coverage*}};
			
			\node[smallbox] (baseline) at (8.0,2.8)
			{Comparable certificate\\subset\\17 certificates = 7\\valid + 10 invalid};
			
			\node[smallbox] (appendix) at (11.8,2.8)
			{Public\\appendix\\26 public artifacts};
			
			\node[smallbox] (cross) at (15.6,2.8)
			{Cross-tool behavior\\layer\\57 artifacts,\\4 tool rows};
			
			\node[wideout] (support) at (11.8,-0.10)
			{Supporting outputs\\
				certificate-only baseline compare\\
				bounded public appendix\\
				cross-tool behavior matrix};
			
			\coordinate (rstem) at (3.40,7.00);
			\coordinate (ventry) at (4.00,8.90);
			\coordinate (mentry) at (4.00,5.10);
			
			\coordinate (bmid) at (8.00,4.55);
			\coordinate (amid) at (11.80,4.55);
			\coordinate (cmid) at (15.60,4.55);
			
			\path (eval.north east) ++(0.70,0) coordinate (psplit);
			\path (eval.south east) ++(0.70,0) coordinate (esplit);
			\coordinate (outerlane) at (22.40,6.20);
			
			\path (support.north) ++(-3.80,0) coordinate (supL);
			\path (support.north) ++(0,0)      coordinate (supC);
			\path (support.north) ++(3.80,0)   coordinate (supR);
			
			\draw[flow] (registry.east) -- (rstem);
			\draw[flow] (rstem) |- (ventry) -- (valid.west);
			\draw[flow] (rstem) |- (mentry) -- (mut.west);
			
			\draw[flow] (valid.east) -- ++(0.55,0) |- (manifest.north west);
			\draw[flow] (mut.east)   -- ++(0.55,0) |- (manifest.south west);
			
			\draw[flow] (manifest.east) -- (eval.west);
			
			\draw[flow] (eval.north east) -- (psplit) |- (primary.west);
			\draw[flow] (eval.south east) -- (esplit) -- (outerlane) |- (support.east);
			
			\draw[flow] (manifest.south west) |- (bmid) -- (baseline.north);
			\draw[flow] (manifest.south)      |- (amid) -- (appendix.north);
			\draw[flow] (manifest.south east) |- (cmid) -- (cross.north);
			
			\draw[flow] (baseline.south) -- (supL);
			\draw[flow] (appendix.south) -- (supC);
			\draw[flow] (cross.south)    -- (supR);
			
		\end{tikzpicture}%
	}
	\caption{Evaluation design with primary and supporting evidence lanes. The controlled corpus is the paper's primary evidence layer; the baseline, appendix, and cross-tool layers remain supporting evidence.}
	\label{fig:evaluation-protocol}
\end{figure}
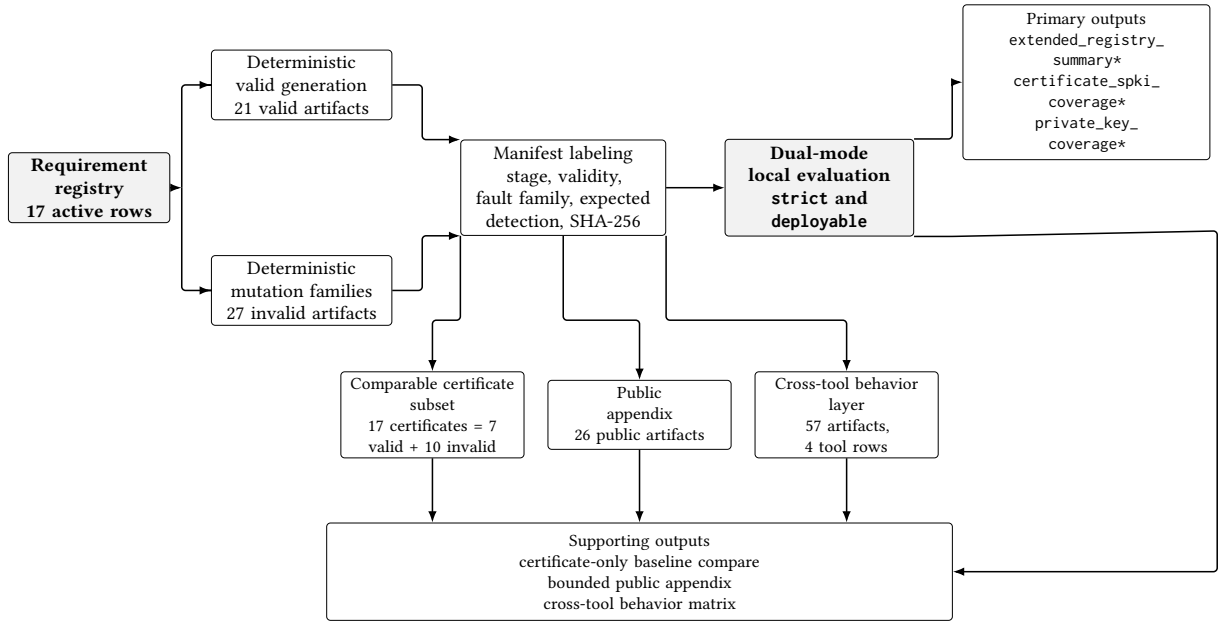

\Cref{fig:evaluation-protocol} shows how the primary and supporting evidence lanes fit together. The controlled corpus is the primary lane. The certificate-only baseline comparison, the bounded public appendix, and the cross-tool behavior matrix are supporting lanes.

\minorhead{Baseline comparison protocol}
\label{subsec:baseline-comparison-protocol}

The frozen baseline comparison is certificate-only. This is the fair slice, because the baseline can be executed on certificates in a way that is not directly matched for raw SPKI objects and private-key containers in the same CLI path \cite{jzlint_snapshot}. The comparable slice therefore contains 17 certificates: 7 valid and 10 invalid. The 10 invalid cases include pure certificate-profile defects and certificate-carried SPKI defects. The local artifact is evaluated on that same slice, and the comparison is normalized in terms of useful detection, miss, pass, and fatal runtime failure.

\minorhead{Bounded public appendix protocol}
\label{subsec:bounded-public-appendix-protocol}

The public appendix is a bounded support layer. It currently contains 26 public artifacts, split into 8 certificates and 18 private-key containers, across two provider paths, \texttt{ossl35} and \texttt{bc}. The appendix contains only valid material; its purpose is to support external-validity discussion by showing that the operator-facing workflow is not confined to one local generation path.

\minorhead{Cross-tool behavior protocol}
\label{subsec:crosstool-protocol}

The cross-tool layer is behavioral. It covers 57 artifacts in total, comprising 31 controlled artifacts and the 26 appendix artifacts, for a total of 25 certificates and 32 private-key containers. Each row records tool behavior in one of five coarse classes: \emph{accepted},
\emph{rejected-semantic}, \emph{rejected-structural},
\emph{runtime-failure}, or \emph{not-applicable}. The purpose is to make visible the difference between parse/import acceptance and policy conformance.

That distinction is especially relevant for \tool{OpenSSL}. In this matrix,
\tool{OpenSSL} is read only as a parse/import signal.

\minorhead{Reproducibility conditions}
\label{subsec:reproducibility-conditions}

All three evidence lanes are pinned to frozen local state. The replay entrypoint rebuilds or reuses the frozen third-party snapshots, reruns corpus generation and mutation, replays strict and deployable evaluation, regenerates coverage and policy outputs, refreshes baseline comparison and cross-tool behavior, and refreshes release and claim-lock metadata. The importer slice depends on a narrow local bridge over a frozen \tool{libcrux} substrate; the baseline lane depends on a frozen \tool{JZLint} snapshot with explicitly recorded runtime behavior on this host \cite{libcrux_snapshot,jzlint_snapshot}.

\section{Controlled Results}
\label{sec:controlled-results}

\minorhead{Overall \strictmode/\deployablemode outcomes}
\label{subsec:overall-strict-deployable-outcomes}

In \strictmode, the 48-artifact corpus yields 27 invalid blocks and 21 passes. In \deployablemode, it yields 26 invalid blocks, 1 invalid warning, and the same 21 passes. In both modes, all 27 invalid artifacts meet their expected detection labels, no valid artifact is blocked or warned, and no unexpected errors become outcome-critical. The deployable warning preserves detection and changes only the operator action on a single known condition.

\Cref{tab:controlled-results-by-stage} records these results at the stage level. All three stages meet the declared success criteria without valid-set findings.

\begin{table}[H]
	\caption{Controlled results by stage and mode. The only mode split in the corpus is a single \mlkem SPKI canonicality artifact that blocks in \strictmode and warns in \deployablemode.}
	\label{tab:controlled-results-by-stage}
	\centering
	{\scriptsize
		\begin{tabularx}{\textwidth}{@{}>{\raggedright\arraybackslash}X *{7}{>{\centering\arraybackslash}m{1.45cm}}@{}}
			\toprule
			Stage
			& \shortstack[c]{Valid\\pass}
			& \shortstack[c]{Strict\\block}
			& \shortstack[c]{Strict\\warn}
			& \shortstack[c]{Deployable\\block}
			& \shortstack[c]{Deployable\\warn}
			& \shortstack[c]{Expected invalid\\detected}
			& \shortstack[c]{False\\positives} \\
			\midrule
			certificate/profile & 7 & 7 & 0 & 7 & 0 & 7/7 & 0 \\
			SPKI/public-key & 7 & 13 & 0 & 12 & 1 & 13/13 & 0 \\
			private-key-container/import & 7 & 7 & 0 & 7 & 0 & 7/7 & 0 \\
			\midrule
			Total & 21 & 27 & 0 & 26 & 1 & 27/27 & 0 \\
			\bottomrule
		\end{tabularx}
	}
\end{table}

\minorhead{Per-stage outcomes}
\label{subsec:per-stage-outcomes}

The certificate/profile slice contributes 7 valid passes and 7 invalid blocks in both modes. The private-key/import slice has the same shape. The SPKI slice is larger, with 7 valid passes and 13 invalid artifacts, and it is the only stage where mode action changes: one invalid artifact, the unreduced ML-KEM ByteEncode12 case, blocks in \strictmode and warns in \deployablemode.

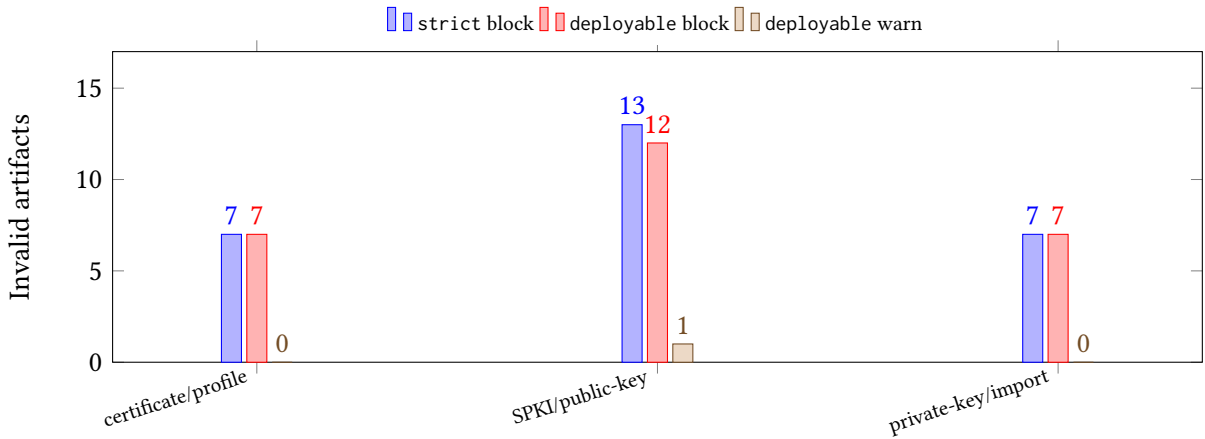
\begin{figure}[H]
	\centering
	\begin{tikzpicture}
		\begin{axis}[
			width=\columnwidth,
			height=5.7cm,
			ybar,
			bar width=7.5pt,
			symbolic x coords={certificate/profile,SPKI/public-key,private-key/import},
			xtick=data,
			xticklabel style={font=\scriptsize, rotate=18, anchor=east},
			ymin=0,
			ymax=17,
			ylabel={Invalid artifacts},
			legend style={at={(0.5,1.02)}, anchor=south, legend columns=3, draw=none, font=\scriptsize},
			nodes near coords,
			nodes near coords align={vertical},
			enlarge x limits=0.18
			]
			\addplot coordinates {(certificate/profile,7) (SPKI/public-key,13) (private-key/import,7)};
			\addplot coordinates {(certificate/profile,7) (SPKI/public-key,12) (private-key/import,7)};
			\addplot coordinates {(certificate/profile,0) (SPKI/public-key,1) (private-key/import,0)};
			\legend{\strictmode block,\deployablemode block,\deployablemode warn}
		\end{axis}
	\end{tikzpicture}
	\caption{Mode action by stage over the invalid side of the controlled corpus.
		Each stage also has 7 valid
		passes in both modes, omitted here for readability.}
	\label{fig:stage-mode-outcomes}
\end{figure}

\minorhead{Certificate/profile results}
\label{subsec:certificate-profile-results}

The certificate/profile stage contains 7 invalid artifacts spanning PKIX usage semantics, signature \texttt{AlgorithmIdentifier} conformance, and final-policy closure. All 7 are blocked in both modes. The stage exercises all 5 active certificate/profile requirements, and its qualitative point is that it covers positive semantics as well as negative bit forbiddance. Two invalid artifacts exercise \path{MLKEM-CERT-KU-KEYENCIPHERMENT-ONLY}; two exercise the positive \mldsa signing-bit requirement
\path{MLDSA-CERT-KU-AT-LEAST-ONE-SIGNING-BIT}; two exercise the \mldsa
certificate-signature absent-parameters rule; and one exercises the HashML-DSA
PKIX prohibition \cite{rfc9881,rfc9935}.

A certificate stage limited to malformed \texttt{AlgorithmIdentifier} parameters would still be useful, but this stage also catches semantic emptiness and policy violations that matter directly to a CA operator. An empty \texttt{keyUsage} extension in an \mlkem or \mldsa certificate is a misprofiled issuance artifact. The local results treat it as such.

\minorhead{SPKI/public-key results}
\label{subsec:spki-public-key-results}

The SPKI/public-key stage is the largest slice of the controlled corpus. Its 13 invalid artifacts comprise 10 raw SPKI defects and 3 certificate-carried SPKI defects. Four cases exercise \path{MLKEM-SPKI-AID-PARAMS-ABSENT}, two exercise
\path{MLDSA-SPKI-AID-PARAMS-ABSENT}, four exercise
\path{MLKEM-SPKI-PUBLIC-KEY-LENGTH}, two exercise
\path{MLDSA-SPKI-PUBLIC-KEY-LENGTH}, and one exercises the canonicality rule
\path{MLKEM-SPKI-ENCODE-DECODE-IDENTITY}. In \strictmode all 13 are blocked. In
\deployablemode 12 are blocked and the canonicality case warns.

Some defects are pure representation failures on raw SPKI material. Others are inter-field consistency failures where OID and payload geometry disagree. Others are carried through a certificate shell but still belong conceptually to the SPKI boundary. The stage therefore tests the declared scope: assurance across representation and profile boundaries, not only at the certificate surface \cite{fips203,rfc9935}.

\minorhead{Private-key/import results}
\label{subsec:private-key-import-results}

The importer slice matches the CA slices on the reported metrics, and this is one of the central empirical results of the paper. All 7 active private-key/import requirements are covered, all 7 corresponding invalid artifacts are detected in both modes, and the stage carries no false positives. Four invalid artifacts belong to the \mlkem importer slice and three to the \mldsa importer slice. Each invalid artifact yields the expected first useful requirement hit, which means the stage is complete in aggregate and interpretable artifact by artifact.

\begin{table}[H]
	\caption{Importer-stage detection grid. The private-key slice is fully exercised and detected in both modes.}
	\label{tab:importer-detection-grid}
	\centering
	\scriptsize
	\begin{tabularx}{\columnwidth}{@{}>{\raggedright\arraybackslash}p{1.15cm}>{\raggedright\arraybackslash}X c c@{}}
		\toprule
		Family & Exercised importer requirements & Invalid artifacts & Outcome \\
		\midrule
		\mlkem & seed length, expanded length, \texttt{both} consistency, expanded-key hash check & 4 & 4/4 detected in both modes \\
		\mldsa & seed length, expanded length, \texttt{both} consistency & 3 & 3/3 detected in both modes \\
		\midrule
		Total & all 7 active importer requirements covered & 7 & 7/7 detected in both modes \\
		\bottomrule
	\end{tabularx}
\end{table}

\minorhead{Redundant detections and first-hit interpretation}
\label{subsec:redundant-detections-first-hit}

Two artifacts in the controlled corpus deliberately produce redundant error instances: \path{der-mut-mldsa44-cert-signature-aid-null} and
\path{der-mut-mldsa44-cert-signature-aid-octet-params}. Both exercise the same certificate-signature absent-parameters rule at more than one certificate location. The local artifact reports two error instances but only one unique requirement.

\minorhead{Error-free valid set and false-positive analysis}
\label{subsec:error-free-valid-set}

All 21 valid artifacts pass in both modes. There are no blocking valid artifacts, no warning valid artifacts, and no unexpected errors on the valid side. The valid set is evenly distributed across stages, with 7 valid certificate/profile artifacts, 7 valid SPKI/public-key artifacts, and 7 valid private-key containers. This is the practical complement to the 27/27 invalid coverage result.

\section{Baseline Insufficiency Analysis}
\label{sec:baseline-insufficiency}

\minorhead{Why certificate-only is the fair slice}
\label{subsec:why-certificate-only-fair-slice}

The frozen baseline comparison must be read through a narrow lens. The lens is certificate-only, and that is the fairest lens available. The baseline is a real current-state artifact and remains useful as secondary evidence, but it is not directly executable over raw SPKI objects and private-key containers in the same workflow shape as the local artifact \cite{jzlint_snapshot}. For that reason, the comparison slice is fixed to 17 certificates: 7 valid and 10 invalid. The invalid side includes both direct certificate-profile defects and certificate-carried SPKI defects.

\minorhead{Headline result: 5/10 versus 10/10}
\label{subsec:headline-5-10-vs-10-10}

Within the fair certificate slice, the gap is clear. The frozen baseline meets 5 of the 10 expected invalid detections. The local artifact meets all 10 of 10. At the same time, the baseline fatally rejects 3 valid \mlkem certificates, while the local artifact fatally rejects none. These numbers leave the baseline informative, but insufficient as an operator assurance path for the covered profile.

\begin{figure}[H]
	\centering
	\begin{tikzpicture}
		\begin{axis}[
			width=\columnwidth,
			height=5.8cm,
			ybar,
			bar width=8pt,
			symbolic x coords={Detected invalid,Missed invalid,Fatal valid,Valid pass},
			xtick=data,
			xticklabel style={font=\scriptsize, rotate=15, anchor=east},
			ymin=0,
			ymax=13,
			ylabel={Artifacts},
			legend style={at={(0.5,1.02)}, anchor=south, legend columns=2, draw=none, font=\scriptsize},
			nodes near coords,
			nodes near coords align={vertical},
			enlarge x limits=0.16
			]
			\addplot coordinates {(Detected invalid,5) (Missed invalid,5) (Fatal valid,3) (Valid pass,4)};
			\addplot coordinates {(Detected invalid,10) (Missed invalid,0) (Fatal valid,0) (Valid pass,7)};
			\legend{Frozen baseline,Extended local artifact}
		\end{axis}
	\end{tikzpicture}
	\caption{Certificate-only comparison on the fair slice.}
	\label{fig:baseline-contrast}
\end{figure}
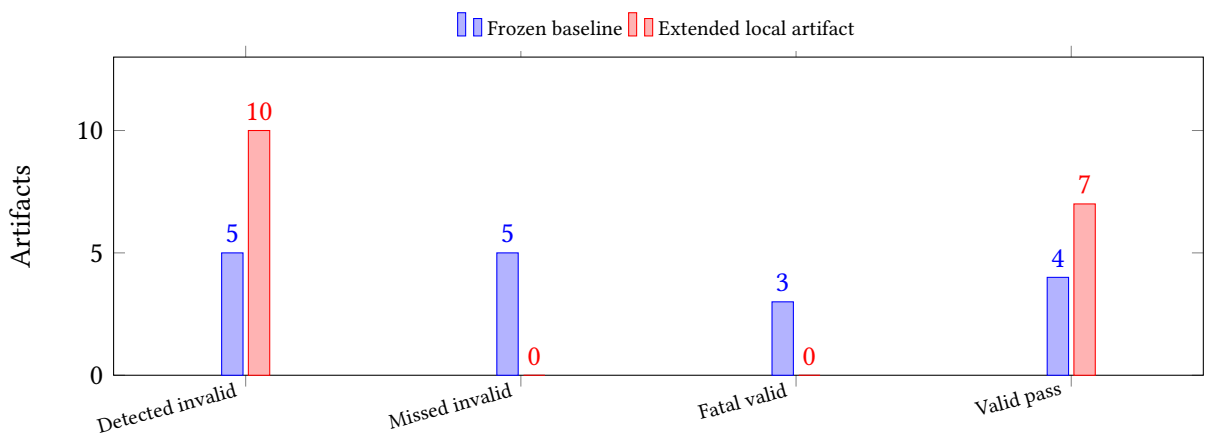

The gap is split between two operator-relevant failure modes: incomplete invalid detection and runtime fragility on valid material.

\minorhead{Fatal valid-certificate fragility}
\label{subsec:fatal-valid-certificate-fragility}

The three fatal valid failures all occur on valid \mlkem end-entity certificates for parameter sets 512, 768, and 1024. In the local evidence frame, this is evidence about the frozen baseline path on this host, not a universal theorem about every future environment or every possible build of the baseline. A certificate-level assurance path that can fatally fail on valid PQ certificates is operationally brittle.

The local artifact avoids that brittleness on the same slice. For an operator, a fatal rejection on valid input is worse than a quiet miss on some invalid cases because it breaks trust in the gate itself.

\minorhead{Positive semantics versus prohibited-bit checking}
\label{subsec:positive-semantics-vs-prohibited-bits}

One of the most informative requirement-level patterns is the contrast between positive semantics and prohibited-bit checking. The baseline recovers the encipherment/agreement prohibition on the \mldsa certificate mutation
\path{openssl-mut-mldsa65-keyusage-key-encipherment-cert}; however, it does not recover the positive requirement that at least one signing-related bit be set. Consequently, it misses the empty-\texttt{keyUsage} case
\path{der-mut-mldsa65-cert-keyusage-empty} and does not fully model the operational semantics of a valid \mldsa certificate profile.

A parallel pattern appears on the \mlkem side. The baseline catches the explicitly prohibited \\ \texttt{digitalSignature} mutation but misses the empty
\texttt{keyUsage} case. The local artifact detects both, because
\path{MLKEM-CERT-KU-KEYENCIPHERMENT-ONLY} is implemented as a profile rule rather than as a partial prohibited-bit test. These two patterns show that final PKIX assurance cannot be reduced to looking for the wrong bits; it also has to reason about the required semantics of the right ones \cite{rfc9881,rfc9935}.

\minorhead{Requirement-level gap analysis}
\label{subsec:requirement-level-gap-analysis}

\Cref{tab:baseline-requirement-gap} gives the requirement-level decomposition of the certificate-only comparison. Part~III specifies which requirement families are fully recovered, which are only partially recovered, which are absent, and which are blocked by runtime fragility rather than by a true semantic detector.

\begin{table}[H]
	\caption{Requirement-level decomposition of the fair certificate-only
		comparison. ``Baseline useful detections'' counts semantic or structural hits;
		``blocked by fatal'' counts cases where the baseline never reached a useful
		decision because the runtime path failed.}
	\label{tab:baseline-requirement-gap}
	\centering
	\scriptsize
	\begin{tabularx}{\textwidth}{@{}>{\raggedright\arraybackslash}p{3.2cm}>{\raggedright\arraybackslash}p{2.0cm}c c c c>{\raggedright\arraybackslash}X@{}}
		\toprule
		Requirement & Stage & \makecell{Expected\\invalids} & \makecell{Baseline useful\\detections} & \makecell{Blocked by\\fatal} & \makecell{Extended\\detections} & Baseline status \\
		\midrule
		\path{MLDSA-CERT-KU-AT-LEAST-ONE-SIGNING-BIT} & certificate/profile & 2 & 0 & 0 & 2 & incomplete \\
		\path{MLDSA-CERT-KU-NO-ENCIPHERMENT-OR-AGREEMENT} & certificate/profile & 1 & 1 & 0 & 1 & covered-for-prohibited-bits \\
		\path{MLDSA-CERT-SIGNATURE-AID-PARAMS-ABSENT} & certificate/profile & 2 & 2 & 0 & 2 & partial \\
		\path{MLDSA-PKIX-HASHML-FORBIDDEN} & certificate/profile & 1 & 0 & 0 & 1 & gap \\
		\path{MLDSA-SPKI-AID-PARAMS-ABSENT} & SPKI/public-key & 0 & 0 & 0 & 0 & partial \\
		\path{MLDSA-SPKI-PUBLIC-KEY-LENGTH} & SPKI/public-key & 0 & 0 & 0 & 0 & gap-or-unverified \\
		\path{MLKEM-CERT-KU-KEYENCIPHERMENT-ONLY} & certificate/profile & 2 & 1 & 0 & 2 & incomplete \\
		\path{MLKEM-SPKI-AID-PARAMS-ABSENT} & SPKI/public-key & 2 & 2 & 0 & 2 & partial \\
		\path{MLKEM-SPKI-ENCODE-DECODE-IDENTITY} & SPKI/public-key & 0 & 0 & 0 & 0 & covered-by-external-lint-path \\
		\path{MLKEM-SPKI-PUBLIC-KEY-LENGTH} & SPKI/public-key & 1 & 0 & 1 & 1 & covered-for-encoded-key \\
		\bottomrule
	\end{tabularx}
\end{table}

Three patterns stand out. First, positive certificate semantics are the clearest coverage gap, especially for \mldsa signing-bit presence and the empty
\texttt{keyUsage} cases. Second, final-policy closure on HashML-DSA remains a true gap rather than merely a runtime casualty. Third, some \mlkem structural coverage is entangled with the baseline's external encoded-key path, which means that even where there is conceptual coverage, there may still be unusable runtime behavior on the way to that coverage.

\minorhead{What the baseline comparison does and does not prove}
\label{subsec:what-baseline-proves}

The comparison proves a bounded and useful proposition: on the fair certificate-only slice, the frozen baseline is insufficient as an assurance path for the covered profile because it both misses expected invalid artifacts and shows runtime fragility on valid material. That result supports the paper's architectural claim that workflow-centric PQ assurance has not been solved by simply adding some PQ lints to an existing certificate path.

The comparison does \emph{not} prove that the baseline is worthless, that no other environment could improve its behavior, or that raw-SPKI and private-key surfaces are directly comparable to its current CLI.

\section{Policy Closure Case Studies}
\label{sec:policy-closure-case-studies}

\minorhead{HashML-DSA prohibition}
\label{subsec:hashmldsa-prohibition}

A direct example of policy closure in the current corpus is the HashML-DSA prohibition for the covered PKIX certificate profile. RFC~9881 fixes the final ML-DSA PKIX conventions and excludes HashML-DSA from the covered certificate-signature path \cite{rfc9881}. The local artifact turns that clause into the blocking requirement \path{MLDSA-PKIX-HASHML-FORBIDDEN}. The mutation
\path{der-mut-mldsa44-cert-signature-hashmldsa44} then exercises that requirement directly. The clause becomes an executable gate: the artifact blocks the case, the expected detection is met, and the baseline leaves the case uncovered.

\minorhead{ML-DSA positive signing semantics}
\label{subsec:mldsa-positive-signing-semantics}

The \mldsa \texttt{keyUsage} rules illustrate why operational assurance needs positive semantics as well as prohibitions. The mutation
\path{openssl-mut-mldsa65-keyusage-key-encipherment-cert} exercises both the positive requirement that at least one signing-related bit be present and the negative requirement that encipherment/agreement semantics be absent. The mutation \path{der-mut-mldsa65-cert-keyusage-empty} isolates the positive side by presenting an empty \texttt{keyUsage} extension. The local artifact blocks both cases. The baseline, by contrast, recovers only the negative encipherment/agreement semantics and leaves the positive requirement incomplete.

Operationally, an empty or signing-less \texttt{keyUsage} extension is still unacceptable even if it contains no forbidden encryption bit. The registry model encodes ``must include a signing semantic'' as a first-class requirement.

\minorhead{ML-KEM \texttt{keyUsage} semantics}
\label{subsec:mlkem-keyusage-semantics}

The \mlkem certificate rule has the same structural lesson. RFC~9935 says that if \texttt{keyUsage} is present, \texttt{keyEncipherment} must be the only active bit \cite{rfc9935}. The local artifact therefore treats the mutations
\path{openssl-mut-mlkem768-keyusage-digital-signature-cert} and
\path{der-mut-mlkem768-cert-keyusage-empty} as two faces of the same profile rule. The baseline catches the explicit prohibited-bit case and misses the empty case.

\minorhead{ML-KEM encode/decode identity as deployable warning}
\label{subsec:mlkem-encode-decode-warning}

The only exercised mode split in the current artifact occurs on
\path{MLKEM-SPKI-ENCODE-DECODE-IDENTITY}. The mutation
\path{der-mut-mlkem768-spki-unreduced-byteencode12-pub} preserves the outer
SPKI shape and payload length while injecting a coefficient value that decodes but does not round-trip canonically under the ML-KEM encode/decode relation
\cite{fips203,rfc9935}. In \strictmode the artifact blocks the case. In
\deployablemode it warns. The expected detection remains met in both modes.

\minorhead{General pattern: from clause to gate}
\label{subsec:general-pattern-clause-to-gate}

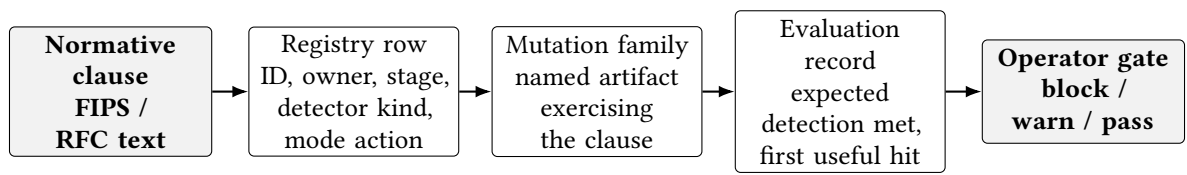
\begin{figure}[H]
	\centering
	\begin{tikzpicture}[x=1cm,y=1cm,>=Latex,
		headbox/.style={
			draw, rounded corners=2pt,
			minimum height=1.15cm,
			text width=2.45cm,
			inner sep=3pt,
			align=center,
			font=\small\bfseries,
			fill=black!5
		},
		nodebox/.style={
			draw, rounded corners=2pt,
			minimum height=1.15cm,
			text width=2.55cm,
			inner sep=3pt,
			align=center,
			font=\small
		},
		flow/.style={-{Latex[length=2.1mm]}, thick}
		]
		
		\def\xA{0.0}
		\def\xB{3.2}
		\def\xC{6.4}
		\def\xD{9.6}
		\def\xE{12.8}
		
		\node[headbox] (clause)   at (\xA,2.1) {Normative clause\\FIPS / RFC text};
		\node[nodebox] (registry) at (\xB,2.1) {Registry row\\ID, owner, stage,\\detector kind,\\mode action};
		\node[nodebox] (mutation) at (\xC,2.1) {Mutation family\\named artifact\\exercising the clause};
		\node[nodebox] (eval)     at (\xD,2.1) {Evaluation record\\expected detection met,\\first useful hit};
		\node[headbox] (gate)     at (\xE,2.1) {Operator gate\\block / warn / pass};
		
		\draw[flow] (clause) -- (registry);
		\draw[flow] (registry) -- (mutation);
		\draw[flow] (mutation) -- (eval);
		\draw[flow] (eval) -- (gate);
		
	\end{tikzpicture}
	\caption{From standards clause to operator gate. The case studies in
		\Cref{tab:policy-closure-cases} are
		instances of the same reproducible translation pattern.}
	\label{fig:clause-to-gate}
\end{figure}

\Cref{tab:policy-closure-cases,fig:clause-to-gate} summarize the general pattern. The reusable contribution lies in the method that turns final standards text into registry rows, mutation families, expected detections, and operator actions.

\begin{table}[H]
	\caption{Policy-closure case studies. Each row follows the same pattern: a normative clause is translated into a requirement, exercised by a named mutation, and surfaced as an operator-visible gate decision.}
	\label{tab:policy-closure-cases}
	\centering
	\scriptsize
	\begin{tabularx}{\textwidth}{@{}>{\raggedright\arraybackslash}p{2.55cm}>{\raggedright\arraybackslash}p{3.25cm}>{\raggedright\arraybackslash}p{3.35cm}cc>{\raggedright\arraybackslash}X@{}}
		\toprule
		Source clause & Requirement & Exercising mutation(s) & \strictmode & \deployablemode & Operational reading \\
		\midrule
		RFC~9881 HashML-DSA exclusion for the covered certificate profile & \path{MLDSA-PKIX-HASHML-FORBIDDEN} & \path{der-mut-mldsa44-cert-signature-hashmldsa44} & block & block & Final-policy closure becomes an executable certificate gate; the baseline leaves the case uncovered \\
		
		RFC~9881 signing-semantics requirement for \mldsa certificates & \path{MLDSA-CERT-KU-AT-LEAST-ONE-SIGNING-BIT} & \path{openssl-mut-mldsa65-keyusage-key-encipherment-cert}; \path{der-mut-mldsa65-cert-keyusage-empty} & block & block & Positive semantics matter; ``no forbidden bit'' is not enough for a valid signing profile \\
		
		RFC~9935 \texttt{keyEncipherment}-only rule for \mlkem certificates & \path{MLKEM-CERT-KU-KEYENCIPHERMENT-ONLY} & \path{openssl-mut-mlkem768-keyusage-digital-signature-cert}; \path{der-mut-mlkem768-cert-keyusage-empty} & block & block & Profile correctness requires both the right bit and the absence of the wrong ones \\
		
		FIPS~203 / RFC~9935 canonicality at the \mlkem SPKI boundary & \path{MLKEM-SPKI-ENCODE-DECODE-IDENTITY} & \path{der-mut-mlkem768-spki-unreduced-byteencode12-pub} & block & warn & The mode split preserves evidence: the same defect is detected in both modes, but operator action changes \\
		\bottomrule
	\end{tabularx}
\end{table}

\partline{Significance}
\section{Operator Workflow and Deployment Guidance}
\label{sec:operator-workflow}

\minorhead{What the CA must run}
\label{subsec:ca-must-run}

The controlled results in \Cref{sec:controlled-results} are translated here into an operating recipe for the parties that issue or import artifacts. In the current \pkixcore release line, the certification authority owns two active pre-issuance gates:
\gatepack{ca-certificate-profile} and \gatepack{ca-spki-public-key}. Together,
they cover the full CA-side requirement inventory of 10 active requirements:
5 at the certificate/profile surface and 5 at the SPKI/public-key surface.

In \deployablemode, the operator runs
\path{./experiments/run_extended.sh}\ \texttt{-{}-mode deployable}
and
\path{./experiments/run_coverage.sh}\ \texttt{-{}-mode deployable}.
In \strictmode, the same path is replayed with
\texttt{-{}-mode strict}. The same artifact therefore supports two operational
postures: a low-noise issuance posture and a maximal-assurance audit posture.

The outputs inspected by the CA are equally important. For the certificate/profile
surface, the decisive outputs are
\path{results/extended_registry_summary_deployable.json},
\path{results/policy_summary_deployable.json},
\path{results/certificate_spki_coverage_deployable.json}, and
\path{results/operator_readiness_summary.json}, with the strict analogues used
for audit replay. For the SPKI/public-key surface, the same summary and policy
artifacts apply, with \path{results/operator_gate_matrix.json} added when the
operator needs a direct gate-pack view. The CA therefore runs a named gate, in
a named mode, with named outputs that support an issuance decision.

\minorhead{What the artifact importer must run}
\label{subsec:importer-must-run}

The importer owns the \gatepack{import-private-key} gate and, in the current
profile, all 7 private-key-container/import requirements. This path is a separate assurance frontier
whose inputs, detector kinds, and failure semantics differ materially from the
CA side.

The importer recipe therefore includes one step that the CA recipe does not:
building the narrow import-validation bridge with
\path{./experiments/build_libcrux_import_check.sh}. After that, the importer
runs the same registry-driven evaluator in either \deployablemode or
\strictmode, followed by
\path{./experiments/run_private_key_coverage.sh}\ \texttt{-{}-mode <mode>}. The outputs of
interest are
\path{results/private_key_coverage.json} or
\path{results/private_key_coverage_deployable.json},
together with the shared policy summaries and gate matrices. The operator then
decides whether a private-key container can be admitted to import, must be
rejected outright, or should be escalated for deeper audit.

The distinction between importer and CA keeps certificate parsing and private-key import as separate assurance surfaces. The importer needs a stable answer to a narrower question: ``Can this container be imported into the covered implementation substrate under the declared profile?''

\minorhead{Default deployable mode and strict audit mode}
\label{subsec:default-deployable-strict-audit}

The two-mode policy model reflects distinct operational postures. Some checks must always block. Others are treated as evidence-preserving warnings in a deployable issuance context while
remaining blocking in a strict conformance context.

In the current release line, \deployablemode is the default CA-facing posture.
It keeps the same detection surface as \strictmode for the controlled corpus,
but downgrades exactly one exercised requirement,
\req{MLKEM-SPKI-ENCODE-DECODE-IDENTITY}, from block to warn. All other active
requirements remain blocking. The practical effect is a different operator
consequence for a single canonicality condition at the ML-KEM SPKI boundary.

By contrast, \strictmode remains the reference assurance posture. It should be
used for pre-release audits, claims reproduction, corpus-extension work, and any
situation where the operator wants the artifact to present the strictest policy
line. The same evidence model is used for both postures.

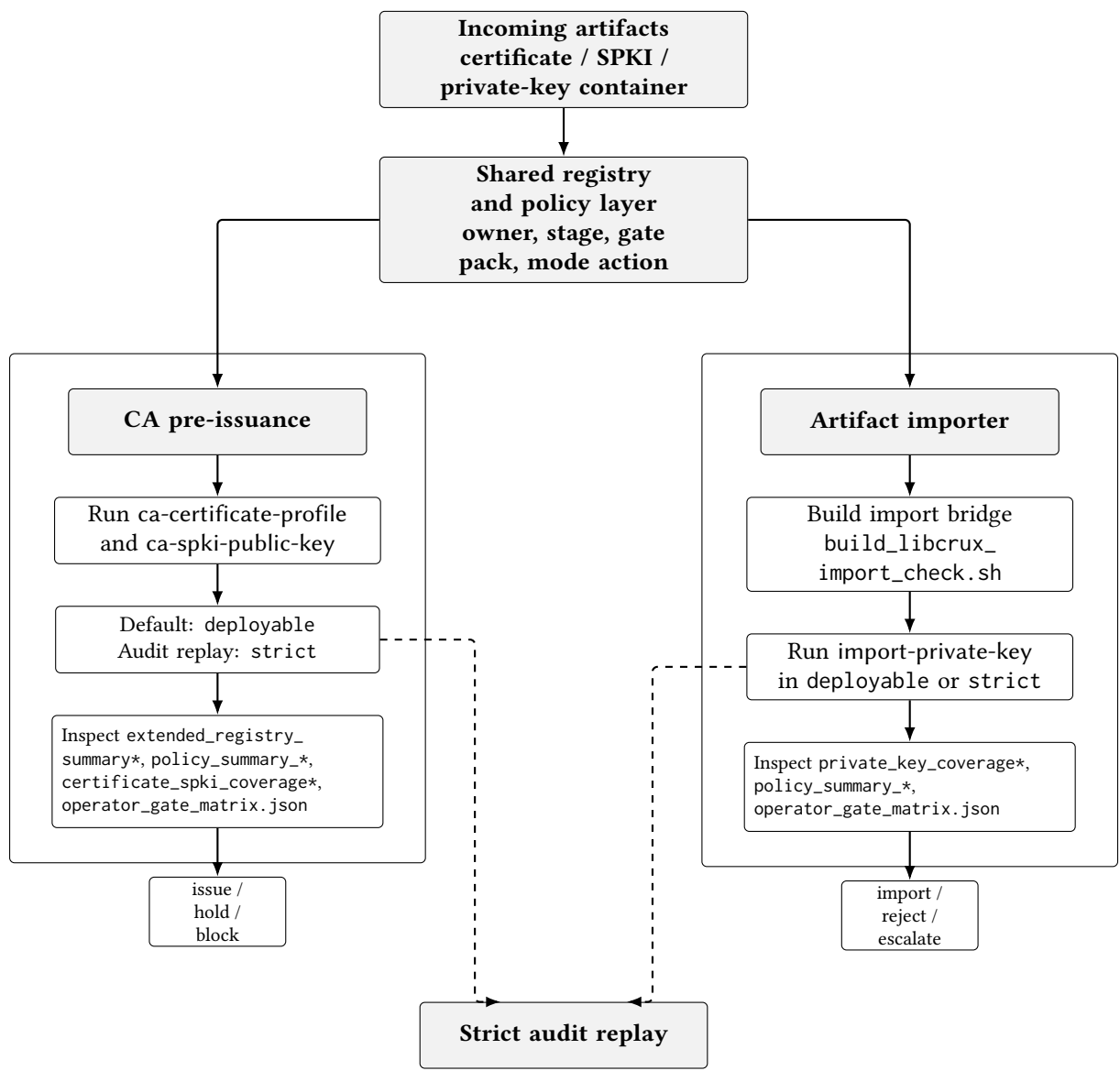
\begin{figure}[H]
	\centering
	\resizebox{\linewidth}{!}{%
		\begin{tikzpicture}[
			>=Latex,
			node distance=6mm,
			headbox/.style={
				draw, rounded corners=2pt,
				minimum height=0.95cm,
				text width=4.30cm,
				inner sep=3pt,
				align=center,
				font=\small\bfseries,
				fill=black!5
			},
			lane/.style={
				draw, rounded corners=2pt,
				inner xsep=6mm,
				inner ysep=5mm
			},
			box/.style={
				draw, rounded corners=2pt,
				minimum height=0.95cm,
				text width=4.45cm,
				inner sep=3pt,
				align=center,
				font=\small
			},
			modebox/.style={
				draw, rounded corners=2pt,
				text width=4.35cm,
				inner sep=4pt,
				align=center,
				font=\footnotesize
			},
			inspectbox/.style={
				draw, rounded corners=2pt,
				text width=4.45cm,
				inner sep=4pt,
				align=left,
				font=\scriptsize
			},
			decision/.style={
				draw, rounded corners=2pt,
				minimum width=1.95cm,
				minimum height=0.95cm,
				inner sep=2pt,
				align=center,
				font=\scriptsize
			},
			flow/.style={
				-{Latex[length=2.1mm]},
				thick,
				rounded corners=2pt
			},
			dashedflow/.style={
				-{Latex[length=2.1mm]},
				thick,
				dashed,
				rounded corners=2pt
			}
			]
			
			\node[headbox, text width=5.05cm] (input)
			{Incoming artifacts\\certificate / SPKI / private-key container};
			
			\node[headbox, text width=5.05cm, below=7mm of input] (policy)
			{Shared registry and policy layer\\owner, stage, gate pack, mode action};
			
			\node[headbox, text width=4.05cm]
			(caHead) at ([xshift=-4.95cm,yshift=-2.00cm]policy.south)
			{CA pre-issuance};
			
			\node[box, below=6mm of caHead] (caRun)
			{Run \gatepack{ca-certificate-profile}\\and \gatepack{ca-spki-public-key}};
			
			\node[modebox, below=6mm of caRun] (caMode)
			{Default: \deployablemode\\Audit replay: \strictmode};
			
			\node[inspectbox, below=6mm of caMode] (caOut)
			{\begin{tabular}{@{}l@{}}
					Inspect \texttt{extended\_registry\_}\\
					\texttt{summary*}, \texttt{policy\_summary\_*},\\
					\texttt{certificate\_spki\_coverage*},\\
					\texttt{operator\_gate\_matrix.json}
			\end{tabular}};
			
			\node[decision, below=7mm of caOut] (caDec)
			{issue /\\hold /\\block};
			
			\node[headbox, text width=4.05cm]
			(impHead) at ([xshift=4.95cm,yshift=-2.00cm]policy.south)
			{Artifact importer};
			
			\node[box, below=6mm of impHead] (impBuild)
			{Build import bridge\\\texttt{build\_libcrux\_}\\\texttt{import\_check.sh}};
			
			\node[box, below=6mm of impBuild] (impRun)
			{Run \gatepack{import-private-key}\\in \deployablemode\ or \strictmode};
			
			\node[inspectbox, below=6mm of impRun] (impOut)
			{\begin{tabular}{@{}l@{}}
					Inspect \texttt{private\_key\_coverage*},\\
					\texttt{policy\_summary\_*},\\
					\texttt{operator\_gate\_matrix.json}
			\end{tabular}};
			
			\node[decision, below=7mm of impOut] (impDec)
			{import /\\reject /\\escalate};
			
			\coordinate (decMid) at ($(caDec.south)!0.5!(impDec.south)$);
			\node[headbox, text width=3.90cm]
			(audit) at ([yshift=-1.25cm]decMid)
			{Strict audit replay};
			
			\begin{scope}[on background layer]
				\node[lane, fit=(caHead)(caRun)(caMode)(caOut)] (caLane) {};
				\node[lane, fit=(impHead)(impBuild)(impRun)(impOut)] (impLane) {};
			\end{scope}
			
			\draw[flow] (input) -- (policy);
			
			\draw[flow] (policy.west) -| (caHead.north);
			\draw[flow] (policy.east) -| (impHead.north);
			
			\draw[flow] (caHead) -- (caRun);
			\draw[flow] (caRun) -- (caMode);
			\draw[flow] (caMode) -- (caOut);
			\draw[flow] (caOut) -- (caDec);
			
			\draw[flow] (impHead) -- (impBuild);
			\draw[flow] (impBuild) -- (impRun);
			\draw[flow] (impRun) -- (impOut);
			\draw[flow] (impOut) -- (impDec);
			
			\coordinate (auditL) at ([xshift=-0.90cm]audit.north);
			\coordinate (auditR) at ([xshift= 0.90cm]audit.north);
			
			\path (caLane.east |- caMode.south) ++(0.70cm,0) coordinate (caDash);
			\path (impLane.west |- impRun.south) ++(-0.70cm,0) coordinate (impDash);
			
			\draw[dashedflow] (caMode.east) -| (caDash) |- (auditL);
			\draw[dashedflow] (impRun.west) -| (impDash) |- (auditR);
			
		\end{tikzpicture}%
	}
	\caption{Operator execution workflow in the current \pkixcore release line.}
	\label{fig:operator-execution-workflow}
\end{figure}

\minorhead{Outputs to inspect and decisions to make}
\label{subsec:outputs-inspect-decisions}

The workflow is expressed through generated outputs. Some are shared across the
artifact regardless of stage, such as \path{results/policy_matrix.csv},
\path{results/stage_owner_summary.json}, and
\path{results/reference_workflow.json}. Others are mode-specific and stage-aware
summaries that tell the operator which requirements block, which warn, and which
surfaces were exercised without findings.

\Cref{tab:operator-execution-matrix} gives the operator-facing view and records the commands, outputs, and decisions for each active gate pack.

\begin{table}[H]
	\caption{Operator execution matrix. The table gives the shortest faithful recipe for each active gate pack, including the commands to run, the outputs to inspect, and the decision supported by those outputs.}
	\label{tab:operator-execution-matrix}
	\centering
	{\scriptsize
		\setlength{\tabcolsep}{3.5pt}
		\begin{tabularx}{\textwidth}{@{}>{\raggedright\arraybackslash}p{2.25cm}>{\raggedright\arraybackslash}p{0.95cm}>{\raggedright\arraybackslash}p{1.50cm}>{\raggedright\arraybackslash}p{2.95cm}>{\raggedright\arraybackslash}p{2.85cm}>{\raggedright\arraybackslash}p{3.75cm}@{}}
			\toprule
			Gate pack & Owner & Default mode & Commands & Outputs to inspect & Operator decision \\
			\midrule
			\gatepack{ca-certificate-profile}
			& CA
			& \deployablemode
			& \path{run_extended.sh}\ \texttt{-{}-mode deployable}; \path{run_coverage.sh}\ \texttt{-{}-mode deployable} \newline
			audit replay: same commands with \texttt{-{}-mode strict}
			& \path{extended_registry_summary_*}, \path{policy_summary_*}, \path{certificate_spki_coverage*}, \path{operator_readiness_summary.json}
			& Issue, hold, or block on certificate-profile semantics before issuance \\
			
			\gatepack{ca-spki-public-key}
			& CA
			& \deployablemode
			& \path{run_extended.sh}\ \texttt{-{}-mode deployable}; \path{run_coverage.sh}\ \texttt{-{}-mode deployable} \newline
			audit replay: same commands with \texttt{-{}-mode strict}
			& \path{extended_registry_summary_*}, \path{policy_summary_*}, \path{certificate_spki_coverage*}, \path{operator_gate_matrix.json}
			& Issue, hold, or block on SPKI structure, parameters, and key-material conditions before issuance \\
			
			\gatepack{import-private-key}
			& Importer
			& \deployablemode or \strictmode
			& \path{build_libcrux_import_check.sh}; \path{run_extended.sh}\ \texttt{-{}-mode <mode>}; \path{run_private_key_coverage.sh}\ \texttt{-{}-mode <mode>}
			& \path{private_key_coverage*}, \path{extended_registry_summary_*}, \path{policy_summary_*}, \path{operator_gate_matrix.json}
			& Import, reject, or escalate based on container form, length, and consistency checks before use \\
			\bottomrule
		\end{tabularx}
	}
\end{table}

\section{Supporting External Evidence}
\label{sec:supporting-external-evidence}

\Cref{tab:external-evidence-layers} summarizes the support layer.

\begin{table}[H]
	\caption{Supporting external evidence layers.}
	\label{tab:external-evidence-layers}
	\centering
	\scriptsize
	\begin{tabularx}{\textwidth}{@{}>{\raggedright\arraybackslash}p{2.35cm}c>{\raggedright\arraybackslash}p{2.25cm}>{\raggedright\arraybackslash}p{2.7cm}>{\raggedright\arraybackslash}p{3.75cm}>{\raggedright\arraybackslash}X@{}}
		\toprule
		Evidence layer & Count & Surface & Coverage highlights & What it supports & What it does not support \\
		\midrule
		Controlled corpus
		& 48
		& certificate, SPKI, private-key
		& 21 valid, 27 invalid; all 17 active requirements exercised; strict and deployable views
		& Main performance claims, operator workflow claims, requirement-level closure
		& Internet-scale prevalence, universal toolchain claims \\
		
		Bounded public appendix
		& 26
		& certificate, private-key
		& 2 providers; 8 certificates; 18 private-key containers; all 6 parameter sets
		& Bounded external-validity support for valid certificate and private-key handling
		& Census claims, invalid-coverage generalization, ecosystem-rate inference \\
		
		Cross-tool behavior matrix
		& 57
		& certificate, private-key
		& 4 tool rows; 31 controlled artifacts and 26 appendix artifacts
		& Parse-versus-policy divergence and baseline/runtime fragility patterns
		& Proof that any given external tool is policy-conformant or normatively complete \\
		\bottomrule
	\end{tabularx}
\end{table}

\minorhead{Appendix behavior against baseline}
\label{subsec:appendix-behavior-against-baseline}

The appendix certificate slice contains 8 valid certificates, all of which the local artifact accepts without findings. The frozen baseline, by contrast, fatally rejects 4 of those 8 valid certificates, all of them in the \mlkem family. In this appendix slice there are no invalid certificates, so the comparison concerns whether externally sourced but in-scope valid certificates survive contact with the baseline.

\minorhead{Cross-tool behavior and parse-versus-policy divergence}
\label{subsec:cross-tool-parse-policy-divergence}

In the current frozen run, the cross-tool layer covers 57 artifacts in total, including 25 certificates and 32 private-key containers, drawn from both the controlled corpus and the public appendix. Four tool rows are recorded: the local extended artifact, the frozen \tool{JZLint} baseline, the \tool{OpenSSL} CLI path, and \tool{pkilint} in its available-or-insufficient state.

The local artifact exhibits structured policy behavior across the full matrix, with 40 acceptances, 8 semantic rejections, and 9 structural rejections. The baseline is certificate-only and is therefore not applicable to all 32 private-key-container artifacts; on the applicable certificate slice it still exhibits 9 runtime failures. The \tool{OpenSSL} path accepts 50 artifacts and rejects 7 structurally. Finally, \tool{pkilint} appears as unavailable-or-insufficient.

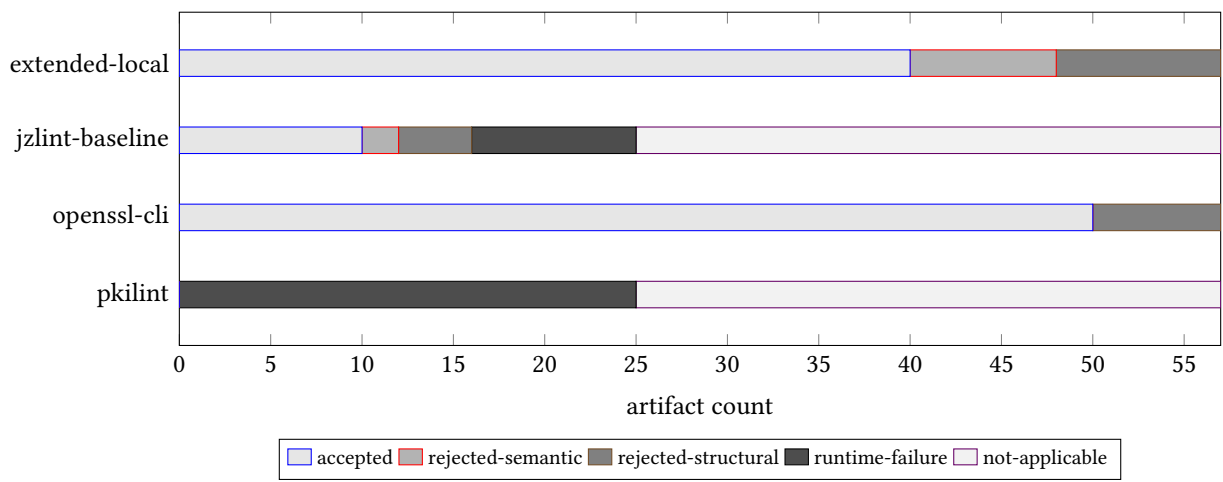
\begin{figure}[H]
	\centering
	\begin{tikzpicture}
		\begin{axis}[
			xbar stacked,
			width=0.96\textwidth,
			height=6.0cm,
			xmin=0, xmax=57,
			bar width=10pt,
			enlarge y limits=0.22,
			xlabel={artifact count},
			symbolic y coords={extended-local,jzlint-baseline,openssl-cli,pkilint},
			ytick=data,
			y dir=reverse,
			legend style={font=\scriptsize, at={(0.5,-0.28)}, anchor=north, legend columns=5},
			every axis plot/.append style={draw=black},
			tick label style={font=\small},
			label style={font=\small}
			]
			\addplot+[fill=black!10] coordinates {(40,extended-local) (10,jzlint-baseline) (50,openssl-cli) (0,pkilint)};
			\addplot+[fill=black!30] coordinates {(8,extended-local) (2,jzlint-baseline) (0,openssl-cli) (0,pkilint)};
			\addplot+[fill=black!50] coordinates {(9,extended-local) (4,jzlint-baseline) (7,openssl-cli) (0,pkilint)};
			\addplot+[fill=black!70] coordinates {(0,extended-local) (9,jzlint-baseline) (0,openssl-cli) (25,pkilint)};
			\addplot+[fill=black!5]  coordinates {(0,extended-local) (32,jzlint-baseline) (0,openssl-cli) (32,pkilint)};
			\legend{accepted,rejected-semantic,rejected-structural,runtime-failure,not-applicable}
		\end{axis}
	\end{tikzpicture}
	\caption{Cross-tool behavior is evidence of divergence. Parse acceptance and policy conformance separate materially across the current tool rows.}
	\label{fig:cross-tool-behavior}
\end{figure}

\section{Limitations and Threats to Validity}
\label{sec:limitations-threats-validity}

\Cref{tab:threats-to-validity-matrix} condenses all five threat classes to validity.

\begin{table}[H]
	\caption{Threats-to-validity matrix. Each threat is paired with the concrete mitigation already built into the artifact or the manuscript framing.}
	\label{tab:threats-to-validity-matrix}
	\centering
	\scriptsize
	\begin{tabularx}{\textwidth}{@{}>{\raggedright\arraybackslash}p{2.15cm}>{\raggedright\arraybackslash}p{4.0cm}>{\raggedright\arraybackslash}p{4.15cm}>{\raggedright\arraybackslash}X@{}}
		\toprule
		Threat class & Limiting fact & Mitigation in this paper & Claims protected by the mitigation \\
		\midrule
		Construct validity & Only \pkixcore and 17 active requirements are executable; runtime-consumer remains out of scope
		& State owner/stage boundaries explicitly and avoid runtime claims that the artifact does not execute
		& Workflow-centric assurance claim remains narrow and defensible \\
		
		Comparison validity & Baseline comparison is certificate-only and baseline runtime fragility appears on the frozen host
		& Frame baseline evidence as certificate-level insufficiency rather than universal head-to-head superiority
		& Comparative claims remain fair and stage-bounded \\
		
		External validity & Public appendix is small, curated, and valid-heavy
		& Use appendix as bounded support only; exclude prevalence or ecosystem-rate rhetoric
		& External-support claim remains modest and honest \\
		
		Toolchain validity & External rows mix policy-aware evaluation, parse-only signals, and unavailable paths
		& Interpret the matrix behaviorally and distinguish acceptance from conformance
		& Cross-tool discussion supports divergence claims without overreach \\
		
		Normative drift & Future profile documents or errata may alter the best registry encoding
		& Anchor claims to the current final normative floor and the current profile version
		& Final-standards framing remains temporally explicit rather than timelessly asserted \\
		\bottomrule
	\end{tabularx}
\end{table}

\section{Conclusion}
\label{sec:conclusion}

Final standards settle the normative floor for ML-KEM and ML-DSA in X.509/PKIX, but they do not by themselves yield operational assurance. This paper shows that post-quantum assurance becomes executable when final normative clauses are translated into owner-assigned, stage-specific, and mode-aware workflow decisions. In the narrow but executable \pkixcore{} profile, that translation is realized as a registry-driven assurance artifact spanning certificate/profile, SPKI/public-key, and private-key-container/import surfaces, organized into operator-facing gate packs for CA pre-issuance and artifact import.

Within that profile, the paper shows that a reproducible assurance workflow can be built and evaluated with disciplined scope. The current artifact reifies 17 active requirements from the final standards corpus, assigns them by owner and stage, and exercises them against a frozen mutation-based corpus containing 48 artifacts across valid and invalid cases. On that primary evidence layer, the artifact meets all expected invalid detections, produces no false positives on the valid set, and preserves the same underlying detection coverage across strict and deployable modes, with a single explicit warning downgrade on an ML-KEM SPKI canonicality condition. The importer-owned private-key boundary is treated as a first-class assurance surface, and all 7 active importer requirements are covered in the present release line.

The comparative and supporting evidence clarify the operational meaning of these results. On the fair certificate-only slice, the frozen baseline misses expected invalid cases and exhibits fatal fragility on some valid ML-KEM certificates, whereas the local artifact closes the same slice without fatal valid rejections. The public appendix and the cross-tool matrix further show that parse acceptance, structural validity, profile conformance, and import readiness are distinct judgments. This distinction explains why post-quantum X.509 assurance cannot be reduced to parser success, certificate-only linting, or isolated algorithm-aware checks.

The paper's contribution is methodological as well as empirical. It gives a reproducible pattern for turning standards prose into executable assurance: extract a requirement, normalize it into a checkable clause, assign it to an owner and stage, bind it to mutation-backed evidence, and expose the result through an operator-facing policy surface. In that sense, the paper extends prior PQ public-key linting work from detector catalogs toward accountable operational workflow. The objective is to make assurance decisions reviewable, replayable, and attributable before issuance and before import.

The work does not claim runtime-consumer coverage, ecosystem prevalence, universal tool behavior, or full PQ-family breadth beyond ML-KEM and ML-DSA. The next steps are to extend the registry discipline to runtime-consumer boundaries, widen artifact-family coverage where executable evidence can be maintained, broaden bounded external support without turning it into prevalence rhetoric, and continue refining mode-aware policy for deployment-facing use. The present result establishes that once final post-quantum standards exist, the essential assurance task is no longer to ask only whether an artifact parses, but whether the right owner can justify the right decision at the right boundary with replayable evidence.

\clearpage
\printbibliography

\clearpage
\appendix

\newcolumntype{L}[1]{>{\raggedright\arraybackslash}p{#1}}

\section{Full Requirement Catalogue}
\label{app:full-requirement-catalogue}

The body text presents the registry as an operational object. This appendix
makes the registry inspectable row by row. The catalogue is extracted from the
final normative floor given by FIPS~203, FIPS~204, RFC~9881, and RFC~9935
\cite{fips203,fips204,rfc9881,rfc9935}, and is frozen into a machine-readable
schema whose records can be audited, mutated, replayed, and cited without
rewriting the paper each time a result table is rebuilt.

\minorhead{Requirement schema}
\label{app:req-schema}

The executable registry keeps more structure than a flat list of lint names.
Each record carries provenance, operational ownership, fault-family alignment,
detector expectations, and per-mode actions. \Cref{tab:app-registry-schema}
summarizes the fields used for review and replay.

{\small
\setlength{\LTleft}{0pt}
\setlength{\LTright}{0pt}
\begin{longtable}{@{}>{\raggedright\arraybackslash}p{0.24\textwidth}>{\raggedright\arraybackslash}p{0.72\textwidth}@{}}
\caption{Registry schema fields used by the executable requirement catalogue.}\label{tab:app-registry-schema}\\
\toprule
Field & Role in the executable registry \\
\midrule
\endfirsthead
\caption[]{Registry schema fields used by the executable requirement catalogue. (continued)}\\
\toprule
Field & Role in the executable registry \\
\midrule
\endhead
\midrule
\multicolumn{2}{r}{\emph{Continued on next page}}\\
\endfoot
\bottomrule
\endlastfoot
\texttt{id} & Stable requirement identifier used by the registry, policy summaries, coverage reports, and expected-detection labels. \\
\texttt{algorithm} & Covered algorithm family, either ML-KEM or ML-DSA. \\
\texttt{artifact\_type} & Concrete artifact surface: certificate, spki, or private-key-container. \\
\texttt{stage} & Operational stage at which the check is owned and executed. \\
\texttt{owner} & Responsible workflow owner, namely CA pre-issuance or artifact importer. \\
\texttt{gate\_pack} & Operator-facing gate grouping exposed by the workflow. \\
\texttt{fault\_family} & Fault class exercised by the mutation catalogue and used for coverage accounting. \\
\texttt{requirement} & Natural-language executable clause distilled from the normative sources. \\
\texttt{expected\_detector} & Detector or validator expected to fire when the requirement is exercised. \\
\texttt{detector\_kind} & Structural, policy, or import-crypto detector category. \\
\texttt{normative\_strength} & Normative level as recorded from the source text, currently must or should. \\
\texttt{baseline\_status} & Position of the frozen baseline relative to this requirement: covered, partial, incomplete, gap, or external-lint only. \\
\texttt{constructibility} & Whether the project can construct and exercise the requirement in the frozen corpus. \\
\texttt{mode\_action} & Per-mode operator consequence, recorded separately for strict and deployable. \\
\texttt{source / source\_locators} & Normative source anchors used to justify the clause and audit the extraction. \\
\end{longtable}
}

\minorhead{ML-KEM requirements}
\label{app:req-mlkem}

The current \pkixcore profile contains eight active \mlkem requirements: two at
the certificate and SPKI issuance boundary and four at the importer boundary,
plus the exercised encode/decode identity rule that becomes the only
\deployablemode warning.

{\scriptsize
	\setlength{\LTleft}{0pt}
	\setlength{\LTright}{0pt}
	\setlength{\tabcolsep}{3pt}
	\renewcommand{\arraystretch}{1.05}
	\newcommand{\idcell}[1]{{\ttfamily\scriptsize\nolinkurl{#1}}}
	
	\begin{longtable}{@{}>{\raggedright\arraybackslash}p{2.8cm}>{\raggedright\arraybackslash}p{1.8cm}>{\raggedright\arraybackslash}p{4.6cm}>{\raggedright\arraybackslash}p{1.45cm}>{\centering\arraybackslash}p{0.8cm}>{\centering\arraybackslash}p{0.8cm}>{\raggedright\arraybackslash}p{1.6cm}@{}}
		\caption{ML-KEM requirement catalogue in the current \pkixcore registry. The \textbf{S} and \textbf{D} columns denote strict and deployable actions.}
		\label{tab:app-mlkem-reqs}\\
		\toprule
		ID & Surface / owner & Executable clause & Detector & S & D & Base \\
		\midrule
		\endfirsthead
		
		\caption[]{ML-KEM requirement catalogue in the current \pkixcore registry. The \textbf{S} and \textbf{D} columns denote strict and deployable actions. (continued)}\\
		\toprule
		ID & Surface / owner & Executable clause & Detector & S & D & Base \\
		\midrule
		\endhead
		
		\midrule
		\multicolumn{7}{@{}r@{}}{\emph{Continued on next page}}\\
		\endfoot
		
		\bottomrule
		\endlastfoot
		
		\idcell{MLKEM-SPKI-AID-PARAMS-ABSENT} & SPKI / CA & SPKI AlgorithmIdentifier uses the correct ML-KEM OID and absent parameters. & structural & block & block & partial \\
		\idcell{MLKEM-CERT-KU-KEYENCIPHERMENT-ONLY} & cert/profile / CA & if keyUsage is present, only keyEncipherment may be active. & policy & block & block & incomplete \\
		\idcell{MLKEM-SPKI-PUBLIC-KEY-LENGTH} & SPKI / CA & SPKI public-key payload length matches the parameter set. & structural & block & block & covered-for-encoded-key \\
		\idcell{MLKEM-SPKI-ENCODE-DECODE-IDENTITY} & SPKI / CA & ML-KEM decode/re-encode is identity on the encapsulation key. & structural & block & warn & covered-by-external-lint-path \\
		\idcell{MLKEM-PRIVATE-SEED-LENGTH} & priv/import / importer & seed-form private key length is 64 bytes. & structural & block & block & gap \\
		\idcell{MLKEM-PRIVATE-EXPANDED-LENGTH} & priv/import / importer & expanded-form private key length matches the parameter set. & structural & block & block & gap \\
		\idcell{MLKEM-PRIVATE-BOTH-CONSISTENCY} & priv/import / importer & both-form ML-KEM import rejects seed/expanded inconsistency. & import-crypto & block & block & gap \\
		\idcell{MLKEM-PRIVATE-EXPANDED-HASH-CHECK} & priv/import / importer & expanded-form ML-KEM import performs the hash check before acceptance. & import-crypto & block & block & gap \\
	\end{longtable}
}

\minorhead{ML-DSA requirements}
\label{app:req-mldsa}

The current \pkixcore profile contains nine active \mldsa requirements: four at
the certificate boundary, two at the SPKI boundary, and three at the importer
boundary. The distinguishing policy closure point is that positive
certificate-profile semantics are encoded as first-class requirements alongside
prohibited-bit checking.

{\scriptsize
	\setlength{\LTleft}{0pt}
	\setlength{\LTright}{0pt}
	\setlength{\tabcolsep}{3pt}
	\renewcommand{\arraystretch}{1.05}
	\newcommand{\idcell}[1]{{\ttfamily\scriptsize\nolinkurl{#1}}}
	
	\begin{longtable}{@{}>{\raggedright\arraybackslash}p{2.8cm}>{\raggedright\arraybackslash}p{1.8cm}>{\raggedright\arraybackslash}p{4.6cm}>{\raggedright\arraybackslash}p{1.45cm}>{\centering\arraybackslash}p{0.8cm}>{\centering\arraybackslash}p{0.8cm}>{\raggedright\arraybackslash}p{1.6cm}@{}}
		\caption{ML-DSA requirement catalogue in the current \pkixcore registry. The \textbf{S} and \textbf{D} columns denote strict and deployable actions.}
		\label{tab:app-mldsa-reqs}\\
		\toprule
		ID & Surface / owner & Executable clause & Detector & S & D & Base \\
		\midrule
		\endfirsthead
		
		\caption[]{ML-DSA requirement catalogue in the current \pkixcore registry. The \textbf{S} and \textbf{D} columns denote strict and deployable actions. (continued)}\\
		\toprule
		ID & Surface / owner & Executable clause & Detector & S & D & Base \\
		\midrule
		\endhead
		
		\midrule
		\multicolumn{7}{@{}r@{}}{\emph{Continued on next page}}\\
		\endfoot
		
		\bottomrule
		\endlastfoot
		
		\idcell{MLDSA-SPKI-AID-PARAMS-ABSENT} & SPKI / CA & SPKI AlgorithmIdentifier uses the correct ML-DSA OID and absent parameters. & structural & block & block & partial \\
		\idcell{MLDSA-CERT-SIGNATURE-AID-PARAMS-ABSENT} & cert/profile / CA & certificate signatureAlgorithm uses an ML-DSA OID with absent parameters. & structural & block & block & partial \\
		\idcell{MLDSA-CERT-KU-AT-LEAST-ONE-SIGNING-BIT} & cert/profile / CA & if keyUsage is present, at least one signing bit must be active. & policy & block & block & incomplete \\
		\idcell{MLDSA-CERT-KU-NO-ENCIPHERMENT-OR-AGREEMENT} & cert/profile / CA & keyUsage forbids encipherment and agreement bits for ML-DSA. & policy & block & block & covered-for-prohibited-bits \\
		\idcell{MLDSA-SPKI-PUBLIC-KEY-LENGTH} & SPKI / CA & SPKI public-key payload length matches the parameter set. & structural & block & block & gap-or-unverified \\
		\idcell{MLDSA-PKIX-HASHML-FORBIDDEN} & cert/profile / CA & HashML-DSA is forbidden in the covered PKIX certificate profile. & policy & block & block & gap \\
		\idcell{MLDSA-PRIVATE-SEED-LENGTH} & priv/import / importer & seed-form private key length is 32 bytes. & structural & block & block & gap \\
		\idcell{MLDSA-PRIVATE-EXPANDED-LENGTH} & priv/import / importer & expanded-form private key length matches the parameter set. & structural & block & block & gap \\
		\idcell{MLDSA-PRIVATE-BOTH-CONSISTENCY} & priv/import / importer & both-form ML-DSA import rejects seed/expanded inconsistency. & import-crypto & block & block & gap \\
	\end{longtable}
}

\minorhead{Gate-pack index}
\label{app:gate-pack-index}

The requirement rows above are operator-facing only when bundled into named
gate packs. \Cref{tab:app-gate-pack-index} gives the compact index used by the
CA and importer workflows.

{\scriptsize
	\setlength{\LTleft}{0pt}
	\setlength{\LTright}{0pt}
	\setlength{\tabcolsep}{3pt}
	\renewcommand{\arraystretch}{1.08}
	
	\begin{longtable}{@{}L{0.12\textwidth}L{0.10\textwidth}L{0.13\textwidth}L{0.27\textwidth}L{0.11\textwidth}L{0.19\textwidth}@{}}
		\caption{Operator gate-pack index for the frozen release line.}
		\label{tab:app-gate-pack-index}\\
		\toprule
		Gate pack & Surface / owner & Title & Requirement IDs & Mode note & Strict commands \\
		\midrule
		\endfirsthead
		
		\caption[]{Operator gate-pack index for the frozen release line. (continued)}\\
		\toprule
		Gate pack & Surface / owner & Title & Requirement IDs & Mode note & Strict commands \\
		\midrule
		\endhead
		
		\midrule
		\multicolumn{6}{@{}r@{}}{\emph{Continued on next page}}\\
		\endfoot
		
		\bottomrule
		\endlastfoot
		
		\code{ca-certificate-profile}
		& cert/profile / CA
		& CA certificate/profile gate
		& \code{MLDSA-CERT-KU-AT-LEAST-ONE-SIGNING-BIT}, \code{MLDSA-CERT-KU-NO-ENCIPHERMENT-OR-AGREEMENT},\par
		\code{MLDSA-CERT-SIGNATURE-AID-PARAMS-ABSENT}, \code{MLDSA-PKIX-HASHML-FORBIDDEN},\par
		\code{MLKEM-CERT-KU-KEYENCIPHERMENT-ONLY}
		& strict 5 block /\par deployable 5 block,\par 0 warn
		& \code{experiments/run_extended.sh --mode strict},\par
		\code{experiments/run_coverage.sh --mode strict}
		\\
		
		\code{ca-spki-public-key}
		& SPKI / CA
		& CA SPKI/public-key gate
		& \code{MLDSA-SPKI-AID-PARAMS-ABSENT}, \code{MLDSA-SPKI-PUBLIC-KEY-LENGTH},\par
		\code{MLKEM-SPKI-AID-PARAMS-ABSENT}, \code{MLKEM-SPKI-ENCODE-DECODE-IDENTITY},\par
		\code{MLKEM-SPKI-PUBLIC-KEY-LENGTH}
		& strict 5 block /\par deployable 4 block,\par 1 warn
		& \code{experiments/run_extended.sh --mode strict},\par
		\code{experiments/run_coverage.sh --mode strict}
		\\
		
		\code{import-private-key}
		& priv/import / importer
		& Importer private-key gate
		& \code{MLDSA-PRIVATE-BOTH-CONSISTENCY}, \code{MLDSA-PRIVATE-EXPANDED-LENGTH},\par
		\code{MLDSA-PRIVATE-SEED-LENGTH}, \code{MLKEM-PRIVATE-BOTH-CONSISTENCY},\par
		\code{MLKEM-PRIVATE-EXPANDED-HASH-CHECK}, \code{MLKEM-PRIVATE-EXPANDED-LENGTH},\par
		\code{MLKEM-PRIVATE-SEED-LENGTH}
		& strict 7 block /\par deployable 7 block,\par 0 warn
		& \code{experiments/build_libcrux_import_check.sh},\par
		\code{experiments/run_extended.sh --mode strict},\par
		\code{experiments/run_private_key_coverage.sh --mode strict}
		\\
		
	\end{longtable}
}

\section{Controlled Corpus Inventory and Mutation Catalogue}
\label{app:controlled-corpus-inventory}

The controlled corpus is the paper's primary evidence layer. It exercises every active requirement on valid and mutated artifacts, and all identifiers listed below appear in the frozen manifest for exact replay from the packaged artifact.

\minorhead{Valid artifacts}
\label{app:valid-artifacts}

The valid side of the controlled corpus contains 21 locally generated artifacts:
7 certificates, 7 SPKI/public-key artifacts, and 7 private-key containers. The
This stage symmetry ensures that every active assurance frontier is exercised on a
non-trivial valid set before invalid mutations are considered.

{\scriptsize
	\setlength{\LTleft}{0pt}
	\setlength{\LTright}{0pt}
	\setlength{\tabcolsep}{3pt}
	\renewcommand{\arraystretch}{1.05}
	\newcommand{\artifactcell}[1]{{\ttfamily\scriptsize\nolinkurl{#1}}}
	\newcommand{\pathcell}[1]{{\ttfamily\scriptsize\nolinkurl{#1}}}
	
	\begin{longtable}{@{}>{\raggedright\arraybackslash}p{3.2cm}>{\raggedright\arraybackslash}p{1.7cm}>{\raggedright\arraybackslash}p{1.9cm}>{\raggedright\arraybackslash}p{1.7cm}>{\raggedright\arraybackslash}p{5.0cm}@{}}
		\caption{Valid artifacts in the frozen controlled corpus.}
		\label{tab:app-valid-artifacts}\\
		\toprule
		Artifact ID & Type & Parameter set & Surface & Path \\
		\midrule
		\endfirsthead
		
		\caption[]{Valid artifacts in the frozen controlled corpus. (continued)}\\
		\toprule
		Artifact ID & Type & Parameter set & Surface & Path \\
		\midrule
		\endhead
		
		\midrule
		\multicolumn{5}{@{}r@{}}{\emph{Continued on next page}}\\
		\endfoot
		
		\bottomrule
		\endlastfoot
		
		\artifactcell{openssl-mldsa44-ee-pub} & spki & ML-DSA-44 & SPKI & \pathcell{corpus/valid/openssl/openssl_mldsa44_ee_pub.pem} \\
		\artifactcell{openssl-mldsa65-ca-pub} & spki & ML-DSA-65 & SPKI & \pathcell{corpus/valid/openssl/openssl_mldsa65_ca_pub.pem} \\
		\artifactcell{openssl-mldsa65-ee-pub} & spki & ML-DSA-65 & SPKI & \pathcell{corpus/valid/openssl/openssl_mldsa65_ee_pub.pem} \\
		\artifactcell{openssl-mldsa87-ee-pub} & spki & ML-DSA-87 & SPKI & \pathcell{corpus/valid/openssl/openssl_mldsa87_ee_pub.pem} \\
		\artifactcell{openssl-mlkem1024-ee-pub} & spki & ML-KEM-1024 & SPKI & \pathcell{corpus/valid/openssl/openssl_mlkem1024_ee_pub.pem} \\
		\artifactcell{openssl-mlkem512-ee-pub} & spki & ML-KEM-512 & SPKI & \pathcell{corpus/valid/openssl/openssl_mlkem512_ee_pub.pem} \\
		\artifactcell{openssl-mlkem768-ee-pub} & spki & ML-KEM-768 & SPKI & \pathcell{corpus/valid/openssl/openssl_mlkem768_ee_pub.pem} \\
		\artifactcell{openssl-mldsa44-ee-cert} & certificate & ML-DSA-44 & cert/profile & \pathcell{corpus/valid/openssl/openssl_mldsa44_ee_cert.pem} \\
		\artifactcell{openssl-mldsa65-ca-cert} & certificate & ML-DSA-65 & cert/profile & \pathcell{corpus/valid/openssl/openssl_mldsa65_ca_cert.pem} \\
		\artifactcell{openssl-mldsa65-ee-cert} & certificate & ML-DSA-65 & cert/profile & \pathcell{corpus/valid/openssl/openssl_mldsa65_ee_cert.pem} \\
		\artifactcell{openssl-mldsa87-ee-cert} & certificate & ML-DSA-87 & cert/profile & \pathcell{corpus/valid/openssl/openssl_mldsa87_ee_cert.pem} \\
		\artifactcell{openssl-mlkem1024-ee-cert} & certificate & ML-KEM-1024 & cert/profile & \pathcell{corpus/valid/openssl/openssl_mlkem1024_ee_cert.pem} \\
		\artifactcell{openssl-mlkem512-ee-cert} & certificate & ML-KEM-512 & cert/profile & \pathcell{corpus/valid/openssl/openssl_mlkem512_ee_cert.pem} \\
		\artifactcell{openssl-mlkem768-ee-cert} & certificate & ML-KEM-768 & cert/profile & \pathcell{corpus/valid/openssl/openssl_mlkem768_ee_cert.pem} \\
		\artifactcell{openssl-mldsa44-ee-key} & private-key-container & ML-DSA-44 & priv/import & \pathcell{corpus/valid/openssl/openssl_mldsa44_ee_key.pem} \\
		\artifactcell{openssl-mldsa65-ca-key} & private-key-container & ML-DSA-65 & priv/import & \pathcell{corpus/valid/openssl/openssl_mldsa65_ca_key.pem} \\
		\artifactcell{openssl-mldsa65-ee-key} & private-key-container & ML-DSA-65 & priv/import & \pathcell{corpus/valid/openssl/openssl_mldsa65_ee_key.pem} \\
		\artifactcell{openssl-mldsa87-ee-key} & private-key-container & ML-DSA-87 & priv/import & \pathcell{corpus/valid/openssl/openssl_mldsa87_ee_key.pem} \\
		\artifactcell{openssl-mlkem1024-ee-key} & private-key-container & ML-KEM-1024 & priv/import & \pathcell{corpus/valid/openssl/openssl_mlkem1024_ee_key.pem} \\
		\artifactcell{openssl-mlkem512-ee-key} & private-key-container & ML-KEM-512 & priv/import & \pathcell{corpus/valid/openssl/openssl_mlkem512_ee_key.pem} \\
		\artifactcell{openssl-mlkem768-ee-key} & private-key-container & ML-KEM-768 & priv/import & \pathcell{corpus/valid/openssl/openssl_mlkem768_ee_key.pem} \\
	\end{longtable}
}

\minorhead{Invalid mutation families}
\label{app:invalid-mutation-families}

The invalid side of the corpus is organized first by fault family and then by
artifact-level mutation.

{\scriptsize
	\setlength{\LTleft}{0pt}
	\setlength{\LTright}{0pt}
	\setlength{\tabcolsep}{3pt}
	\renewcommand{\arraystretch}{1.08}
	
	\begin{longtable}{@{}L{0.16\textwidth}L{0.08\textwidth}L{0.24\textwidth}L{0.08\textwidth}L{0.40\textwidth}@{}}
		\caption{Mutation families exercised by the 27 invalid controlled artifacts.}
		\label{tab:app-mutation-families}\\
		\toprule
		Fault family & Invalid artifacts & Representative mutation tokens & Covered requirements & Requirement IDs \\
		\midrule
		\endfirsthead
		
		\caption[]{Mutation families exercised by the 27 invalid controlled artifacts. (continued)}\\
		\toprule
		Fault family & Invalid artifacts & Representative mutation tokens & Covered requirements & Requirement IDs \\
		\midrule
		\endhead
		
		\midrule
		\multicolumn{5}{@{}r@{}}{\emph{Continued on next page}}\\
		\endfoot
		
		\bottomrule
		\endlastfoot
		
		encoding/container
		& 8
		& \code{aid-parameters-null}, \code{aid-parameters-present-non-null},\par
		\code{signature-aid-parameters-null}
		& 3
		& \code{MLKEM-SPKI-AID-PARAMS-ABSENT}, \code{MLDSA-SPKI-AID-PARAMS-ABSENT},\par
		\code{MLDSA-CERT-SIGNATURE-AID-PARAMS-ABSENT}
		\\
		
		size/shape
		& 7
		& \code{spki-public-key-truncate}, \code{private-key-seed-length-short},\par
		\code{private-key-expanded-length-short}
		& 6
		& \code{MLKEM-SPKI-PUBLIC-KEY-LENGTH}, \code{MLKEM-PRIVATE-SEED-LENGTH},\par
		\code{MLKEM-PRIVATE-EXPANDED-LENGTH}, \code{MLDSA-SPKI-PUBLIC-KEY-LENGTH},\par
		\code{MLDSA-PRIVATE-SEED-LENGTH}, \code{MLDSA-PRIVATE-EXPANDED-LENGTH}
		\\
		
		inter-field-consistency
		& 5
		& \code{spki-oid-length-mismatch}, \code{private-key-both-seed-expanded-mismatch},\par
		\code{aid-oid-family-swap}
		& 2
		& \code{MLKEM-PRIVATE-BOTH-CONSISTENCY},\par
		\code{MLDSA-PRIVATE-BOTH-CONSISTENCY}
		\\
		
		profile/usage-policy
		& 4
		& \code{keyusage-empty}, \code{keyusage-missing-key-encipherment},\par
		\code{keyusage-extra-prohibited-bit}
		& 3
		& \code{MLKEM-CERT-KU-KEYENCIPHERMENT-ONLY},\par
		\code{MLDSA-CERT-KU-AT-LEAST-ONE-SIGNING-BIT},\par
		\code{MLDSA-CERT-KU-NO-ENCIPHERMENT-OR-AGREEMENT}
		\\
		
		field-domain
		& 1
		& \code{mlkem-unreduced-byteencode12-value}
		& 1
		& \code{MLKEM-SPKI-ENCODE-DECODE-IDENTITY}
		\\
		
		algorithm-policy
		& 1
		& \code{hashml-dsa-signature-oid-in-pkix-cert},\par
		\code{hashml-dsa-pkix-context}
		& 1
		& \code{MLDSA-PKIX-HASHML-FORBIDDEN}
		\\
		
		import-validation
		& 1
		& \code{mlkem-expanded-key-hash-mismatch}
		& 1
		& \code{MLKEM-PRIVATE-EXPANDED-HASH-CHECK}
		\\
		
	\end{longtable}
}

\minorhead{Expected-detection map}
\label{app:expected-detection-map}

Each invalid artifact has a declared expected-detection set in the manifest
before the evaluator is run. The map below therefore records both what failed
and what the artifact was expected to fail \emph{for}. The only
strict/deployable action split in the current corpus is the unreduced
\mlkem ByteEncode12 mutation, which remains detected in both modes while
becoming a warning in \deployablemode.

{\scriptsize
	\setlength{\LTleft}{0pt}
	\setlength{\LTright}{0pt}
	\setlength{\tabcolsep}{2.5pt}
	\renewcommand{\arraystretch}{1.04}
	\newcommand{\idcell}[1]{{\ttfamily\scriptsize\nolinkurl{#1}}}
	
	\begin{longtable}{@{}>{\raggedright\arraybackslash}p{3.1cm}>{\raggedright\arraybackslash}p{1.5cm}>{\raggedright\arraybackslash}p{3.0cm}>{\raggedright\arraybackslash}p{3.2cm}>{\centering\arraybackslash}p{0.9cm}>{\centering\arraybackslash}p{1.1cm}@{}}
		\caption{Expected-detection ledger for all invalid controlled artifacts.}
		\label{tab:app-expected-detection-map}\\
		\toprule
		Artifact ID & Surface & Mutation token(s) & Expected requirement(s) & Strict & Deployable \\
		\midrule
		\endfirsthead
		
		\caption[]{Expected-detection ledger for all invalid controlled artifacts. (continued)}\\
		\toprule
		Artifact ID & Surface & Mutation token(s) & Expected requirement(s) & Strict & Deployable \\
		\midrule
		\endhead
		
		\midrule
		\multicolumn{6}{@{}r@{}}{\emph{Continued on next page}}\\
		\endfoot
		
		\bottomrule
		\endlastfoot
		
		\idcell{der-mut-mldsa44-spki-oid-swapped-to-mldsa65-pub} & SPKI & spki-oid-length-mismatch & \idcell{MLDSA-SPKI-PUBLIC-KEY-LENGTH} & block & block \\
		\idcell{der-mut-mldsa65-spki-aid-null-pub} & SPKI & aid-parameters-null & \idcell{MLDSA-SPKI-AID-PARAMS-ABSENT} & block & block \\
		\idcell{der-mut-mldsa65-spki-aid-octet-params-pub} & SPKI & aid-parameters-present-non-null & \idcell{MLDSA-SPKI-AID-PARAMS-ABSENT} & block & block \\
		\idcell{der-mut-mldsa87-spki-payload-2602-pub} & SPKI & spki-public-key-extend, rfc9881-appendix-size-transcription-2602 & \idcell{MLDSA-SPKI-PUBLIC-KEY-LENGTH} & block & block \\
		\idcell{der-mut-mldsa65-spki-oid-swapped-to-mlkem768-pub} & SPKI & aid-oid-family-swap, spki-oid-length-mismatch & \idcell{MLKEM-SPKI-PUBLIC-KEY-LENGTH} & block & block \\
		\idcell{der-mut-mlkem512-spki-oid-swapped-to-mlkem768-pub} & SPKI & spki-oid-length-mismatch & \idcell{MLKEM-SPKI-PUBLIC-KEY-LENGTH} & block & block \\
		\idcell{der-mut-mlkem768-cert-spki-aid-null} & SPKI & aid-parameters-null & \idcell{MLKEM-SPKI-AID-PARAMS-ABSENT} & block & block \\
		\idcell{der-mut-mlkem768-cert-spki-aid-octet-params} & SPKI & aid-parameters-present-non-null & \idcell{MLKEM-SPKI-AID-PARAMS-ABSENT} & block & block \\
		\idcell{der-mut-mlkem768-cert-spki-payload-truncated} & SPKI & spki-public-key-truncate & \idcell{MLKEM-SPKI-PUBLIC-KEY-LENGTH} & block & block \\
		\idcell{der-mut-mlkem768-spki-aid-null-pub} & SPKI & aid-parameters-null & \idcell{MLKEM-SPKI-AID-PARAMS-ABSENT} & block & block \\
		\idcell{der-mut-mlkem768-spki-aid-octet-params-pub} & SPKI & aid-parameters-present-non-null & \idcell{MLKEM-SPKI-AID-PARAMS-ABSENT} & block & block \\
		\idcell{der-mut-mlkem768-spki-payload-truncated-pub} & SPKI & spki-public-key-truncate & \idcell{MLKEM-SPKI-PUBLIC-KEY-LENGTH} & block & block \\
		\idcell{der-mut-mlkem768-spki-unreduced-byteencode12-pub} & SPKI & mlkem-unreduced-byteencode12-value & \idcell{MLKEM-SPKI-ENCODE-DECODE-IDENTITY} & block & warn \\
		\idcell{der-mut-mldsa44-cert-signature-aid-null} & cert/profile & signature-aid-parameters-null & \idcell{MLDSA-CERT-SIGNATURE-AID-PARAMS-ABSENT} & block & block \\
		\idcell{der-mut-mldsa44-cert-signature-aid-octet-params} & cert/profile & signature-aid-parameters-present-non-null & \idcell{MLDSA-CERT-SIGNATURE-AID-PARAMS-ABSENT} & block & block \\
		\idcell{der-mut-mldsa44-cert-signature-hashmldsa44} & cert/profile & hashml-dsa-signature-oid-in-pkix-cert, hashml-dsa-pkix-context & \idcell{MLDSA-PKIX-HASHML-FORBIDDEN} & block & block \\
		\idcell{der-mut-mldsa65-cert-keyusage-empty} & cert/profile & keyusage-empty & \idcell{MLDSA-CERT-KU-AT-LEAST-ONE-SIGNING-BIT} & block & block \\
		\idcell{openssl-mut-mldsa65-keyusage-key-encipherment-cert} & cert/profile & keyusage-missing-signature-bit, keyusage-key-encipherment & \begin{tabular}[t]{@{}l@{}}\idcell{MLDSA-CERT-KU-AT-LEAST-ONE-SIGNING-BIT},\\ \idcell{MLDSA-CERT-KU-NO-ENCIPHERMENT-OR-AGREEMENT}\end{tabular} & block & block \\
		\idcell{der-mut-mlkem768-cert-keyusage-empty} & cert/profile & keyusage-empty & \idcell{MLKEM-CERT-KU-KEYENCIPHERMENT-ONLY} & block & block \\
		\idcell{openssl-mut-mlkem768-keyusage-digital-signature-cert} & cert/profile & keyusage-missing-key-encipherment, keyusage-extra-prohibited-bit & \idcell{MLKEM-CERT-KU-KEYENCIPHERMENT-ONLY} & block & block \\
		\idcell{der-mut-mldsa44-key-both-mismatch} & priv/import & private-key-both-seed-expanded-mismatch & \idcell{MLDSA-PRIVATE-BOTH-CONSISTENCY} & block & block \\
		\idcell{der-mut-mldsa44-key-expanded-short} & priv/import & private-key-expanded-length-short & \idcell{MLDSA-PRIVATE-EXPANDED-LENGTH} & block & block \\
		\idcell{der-mut-mldsa44-key-seed-short} & priv/import & private-key-seed-length-short & \idcell{MLDSA-PRIVATE-SEED-LENGTH} & block & block \\
		\idcell{der-mut-mlkem512-key-both-mismatch} & priv/import & private-key-both-seed-expanded-mismatch & \idcell{MLKEM-PRIVATE-BOTH-CONSISTENCY} & block & block \\
		\idcell{der-mut-mlkem512-key-expanded-short} & priv/import & private-key-expanded-length-short & \idcell{MLKEM-PRIVATE-EXPANDED-LENGTH} & block & block \\
		\idcell{der-mut-mlkem512-key-hash-mismatch} & priv/import & mlkem-expanded-key-hash-mismatch & \begin{tabular}[t]{@{}l@{}}\idcell{MLKEM-PRIVATE-EXPANDED-HASH-CHECK},\\ \idcell{MLKEM-PRIVATE-BOTH-CONSISTENCY}\end{tabular} & block & block \\
		\idcell{der-mut-mlkem512-key-seed-short} & priv/import & private-key-seed-length-short & \idcell{MLKEM-PRIVATE-SEED-LENGTH} & block & block \\
	\end{longtable}
}

\section{Detailed Baseline Comparison Tables}
\label{app:detailed-baseline-comparison}

The baseline comparison is certificate-only because the frozen
JZLint CLI baseline does not natively cover raw SPKI artifacts or private-key
containers. Within that fair slice, the baseline remains useful as it catches some defects, misses
others, and exhibits fatal fragility on some otherwise valid \mlkem
certificates \cite{jzlint_snapshot}.

\minorhead{Per-artifact outcomes}
\label{app:baseline-per-artifact}

{\scriptsize
	\setlength{\LTleft}{0pt}
	\setlength{\LTright}{0pt}
	\setlength{\tabcolsep}{3pt}
	\renewcommand{\arraystretch}{1.08}
	
	\begin{longtable}{@{}L{0.24\textwidth}L{0.08\textwidth}L{0.19\textwidth}L{0.07\textwidth}L{0.07\textwidth}L{0.29\textwidth}@{}}
		\caption{Per-artifact certificate-level comparison between the frozen baseline and the local extended artifact.}
		\label{tab:app-baseline-per-artifact}\\
		\toprule
		Artifact ID & Validity & Expected requirement & Baseline & Extended & Comparison class \\
		\midrule
		\endfirsthead
		
		\caption[]{Per-artifact certificate-level comparison between the frozen baseline and the local extended artifact. (continued)}\\
		\toprule
		Artifact ID & Validity & Expected requirement & Baseline & Extended & Comparison class \\
		\midrule
		\endhead
		
		\midrule
		\multicolumn{6}{@{}r@{}}{\emph{Continued on next page}}\\
		\endfoot
		
		\bottomrule
		\endlastfoot
		
		\code{der-mut-mldsa44-cert-signature-aid-null}
		& invalid
		& \code{MLDSA-CERT-SIGNATURE-AID-PARAMS-ABSENT}
		& error
		& error
		& \code{same-detection}
		\\
		
		\code{der-mut-mldsa44-cert-signature-aid-octet-params}
		& invalid
		& \code{MLDSA-CERT-SIGNATURE-AID-PARAMS-ABSENT}
		& error
		& error
		& \code{same-detection}
		\\
		
		\code{der-mut-mldsa44-cert-signature-hashmldsa44}
		& invalid
		& \code{MLDSA-PKIX-HASHML-FORBIDDEN}
		& pass
		& error
		& \code{extended-recovers-baseline-miss}
		\\
		
		\code{der-mut-mldsa65-cert-keyusage-empty}
		& invalid
		& \code{MLDSA-CERT-KU-AT-LEAST-ONE-SIGNING-BIT}
		& pass
		& error
		& \code{extended-recovers-baseline-miss}
		\\
		
		\code{openssl-mut-mldsa65-keyusage-key-encipherment-cert}
		& invalid
		& \code{MLDSA-CERT-KU-AT-LEAST-ONE-SIGNING-BIT}\par
		\texttt{|}\par
		\code{MLDSA-CERT-KU-NO-ENCIPHERMENT-OR-AGREEMENT}
		& error
		& error
		& \code{extended-recovers-baseline-miss}
		\\
		
		\code{der-mut-mlkem768-cert-keyusage-empty}
		& invalid
		& \code{MLKEM-CERT-KU-KEYENCIPHERMENT-ONLY}
		& fatal
		& error
		& \code{extended-recovers-baseline-miss-and-runtime-fragility}
		\\
		
		\code{der-mut-mlkem768-cert-spki-aid-null}
		& invalid
		& \code{MLKEM-SPKI-AID-PARAMS-ABSENT}
		& error
		& error
		& \code{extended-matches-detection-without-baseline-runtime-fragility}
		\\
		
		\code{der-mut-mlkem768-cert-spki-aid-octet-params}
		& invalid
		& \code{MLKEM-SPKI-AID-PARAMS-ABSENT}
		& error
		& error
		& \code{extended-matches-detection-without-baseline-runtime-fragility}
		\\
		
		\code{der-mut-mlkem768-cert-spki-payload-truncated}
		& invalid
		& \code{MLKEM-SPKI-PUBLIC-KEY-LENGTH}
		& fatal
		& error
		& \code{extended-recovers-baseline-miss-and-runtime-fragility}
		\\
		
		\code{openssl-mut-mlkem768-keyusage-digital-signature-cert}
		& invalid
		& \code{MLKEM-CERT-KU-KEYENCIPHERMENT-ONLY}
		& error
		& error
		& \code{extended-matches-detection-without-baseline-runtime-fragility}
		\\
		
		\code{openssl-mldsa44-ee-cert}
		& valid
		& --
		& pass
		& pass
		& \code{same-pass}
		\\
		
		\code{openssl-mldsa65-ca-cert}
		& valid
		& --
		& pass
		& pass
		& \code{same-pass}
		\\
		
		\code{openssl-mldsa65-ee-cert}
		& valid
		& --
		& pass
		& pass
		& \code{same-pass}
		\\
		
		\code{openssl-mldsa87-ee-cert}
		& valid
		& --
		& pass
		& pass
		& \code{same-pass}
		\\
		
		\code{openssl-mlkem1024-ee-cert}
		& valid
		& --
		& fatal
		& pass
		& \code{baseline-runtime-fragile}
		\\
		
		\code{openssl-mlkem512-ee-cert}
		& valid
		& --
		& fatal
		& pass
		& \code{baseline-runtime-fragile}
		\\
		
		\code{openssl-mlkem768-ee-cert}
		& valid
		& --
		& fatal
		& pass
		& \code{baseline-runtime-fragile}
		\\
		
	\end{longtable}
}

\minorhead{Requirement-level comparison}
\label{app:baseline-requirement-level}

The requirement-level view is often more revealing than the artifact-level view
because it separates ``baseline missed the expected defect'' from ``baseline
never had a fair path to the covered artifact type.'' Only the ten certificate
requirements relevant to the baseline slice appear here.

{\scriptsize
\setlength{\LTleft}{0pt}
\setlength{\LTright}{0pt}
\begin{longtable}{@{}>{\raggedright\arraybackslash}p{0.28\textwidth}>{\raggedright\arraybackslash}p{0.10\textwidth}>{\raggedright\arraybackslash}p{0.07\textwidth}>{\raggedright\arraybackslash}p{0.08\textwidth}>{\raggedright\arraybackslash}p{0.08\textwidth}>{\raggedright\arraybackslash}p{0.15\textwidth}>{\raggedright\arraybackslash}p{0.08\textwidth}@{}}
\caption{Requirement-level comparison for the certificate-only baseline slice.}\label{tab:app-baseline-requirement-level}\\
\toprule
Requirement ID & Stage & Expected & Baseline met & Blocked by fatal & Baseline status & Extended met \\
\midrule
\endfirsthead
\caption[]{Requirement-level comparison for the certificate-only baseline slice. (continued)}\\
\toprule
Requirement ID & Stage & Expected & Baseline met & Blocked by fatal & Baseline status & Extended met \\
\midrule
\endhead
\midrule
\multicolumn{7}{r}{\emph{Continued on next page}}\\
\endfoot
\bottomrule
\endlastfoot
\path{MLDSA-CERT-KU-AT-LEAST-ONE-SIGNING-BIT} & cert/profile & 2 & 0 & 0 & incomplete & 2 \\
\path{MLDSA-CERT-KU-NO-ENCIPHERMENT-OR-AGREEMENT} & cert/profile & 1 & 1 & 0 & covered-for-prohibited-bits & 1 \\
\path{MLDSA-CERT-SIGNATURE-AID-PARAMS-ABSENT} & cert/profile & 2 & 2 & 0 & partial & 2 \\
\path{MLDSA-PKIX-HASHML-FORBIDDEN} & cert/profile & 1 & 0 & 0 & gap & 1 \\
\path{MLDSA-SPKI-AID-PARAMS-ABSENT} & SPKI & 0 & 0 & 0 & partial & 0 \\
\path{MLDSA-SPKI-PUBLIC-KEY-LENGTH} & SPKI & 0 & 0 & 0 & gap-or-unverified & 0 \\
\path{MLKEM-CERT-KU-KEYENCIPHERMENT-ONLY} & cert/profile & 2 & 1 & 0 & incomplete & 2 \\
\path{MLKEM-SPKI-AID-PARAMS-ABSENT} & SPKI & 2 & 2 & 0 & partial & 2 \\
\path{MLKEM-SPKI-ENCODE-DECODE-IDENTITY} & SPKI & 0 & 0 & 0 & covered-by-external-lint-path & 0 \\
\path{MLKEM-SPKI-PUBLIC-KEY-LENGTH} & SPKI & 1 & 0 & 1 & covered-for-encoded-key & 1 \\
\end{longtable}
}

\minorhead{Runtime fragility notes}
\label{app:baseline-runtime-fragility}

Fatal baseline behavior matters even when the baseline eventually ``rejects'' an
invalid artifact, because the rejection can happen for the wrong reason and can
also spill onto valid artifacts. The table below records the fatal cases inside
the controlled certificate slice.

{\scriptsize
	\setlength{\LTleft}{0pt}
	\setlength{\LTright}{0pt}
	\setlength{\tabcolsep}{3pt}
	\renewcommand{\arraystretch}{1.08}
	
	\begin{longtable}{@{}L{0.23\textwidth}L{0.07\textwidth}L{0.18\textwidth}L{0.21\textwidth}L{0.27\textwidth}@{}}
		\caption{Fatal baseline cases in the controlled certificate comparison.}
		\label{tab:app-baseline-fragility}\\
		\toprule
		Artifact ID & Validity & Fatal baseline lints & Comparison class & Why the case matters \\
		\midrule
		\endfirsthead
		
		\caption[]{Fatal baseline cases in the controlled certificate comparison. (continued)}\\
		\toprule
		Artifact ID & Validity & Fatal baseline lints & Comparison class & Why the case matters \\
		\midrule
		\endhead
		
		\midrule
		\multicolumn{5}{@{}r@{}}{\emph{Continued on next page}}\\
		\endfoot
		
		\bottomrule
		\endlastfoot
		
		\code{der-mut-mlkem768-cert-keyusage-empty}
		& invalid
		& \code{e_ml_kem_ek_encoding};\par \code{e_ml_kem_ek_length}
		& \code{extended-recovers-baseline-miss-and-runtime-fragility}
		& invalid certificate blocked by fatal baseline behavior before the expected requirement could be credited
		\\
		
		\code{der-mut-mlkem768-cert-spki-payload-truncated}
		& invalid
		& \code{e_ml_kem_ek_encoding};\par \code{e_ml_kem_ek_length}
		& \code{extended-recovers-baseline-miss-and-runtime-fragility}
		& invalid certificate blocked by fatal baseline behavior before the expected requirement could be credited
		\\
		
		\code{openssl-mlkem1024-ee-cert}
		& valid
		& \code{e_ml_kem_ek_encoding};\par \code{e_ml_kem_ek_length}
		& \code{baseline-runtime-fragile}
		& valid certificate rejected by baseline with fatal ML-KEM encoded-key/runtime lints
		\\
		
		\code{openssl-mlkem512-ee-cert}
		& valid
		& \code{e_ml_kem_ek_encoding};\par \code{e_ml_kem_ek_length}
		& \code{baseline-runtime-fragile}
		& valid certificate rejected by baseline with fatal ML-KEM encoded-key/runtime lints
		\\
		
		\code{openssl-mlkem768-ee-cert}
		& valid
		& \code{e_ml_kem_ek_encoding};\par \code{e_ml_kem_ek_length}
		& \code{baseline-runtime-fragile}
		& valid certificate rejected by baseline with fatal ML-KEM encoded-key/runtime lints
		\\
		
	\end{longtable}
}

\section{Public Appendix Ledger}
\label{app:public-appendix-ledger}

The public appendix spans 26 valid public artifacts
across two providers, a certificate slice, and private-key containers covering
all six \mlkem/\mldsa parameter sets.

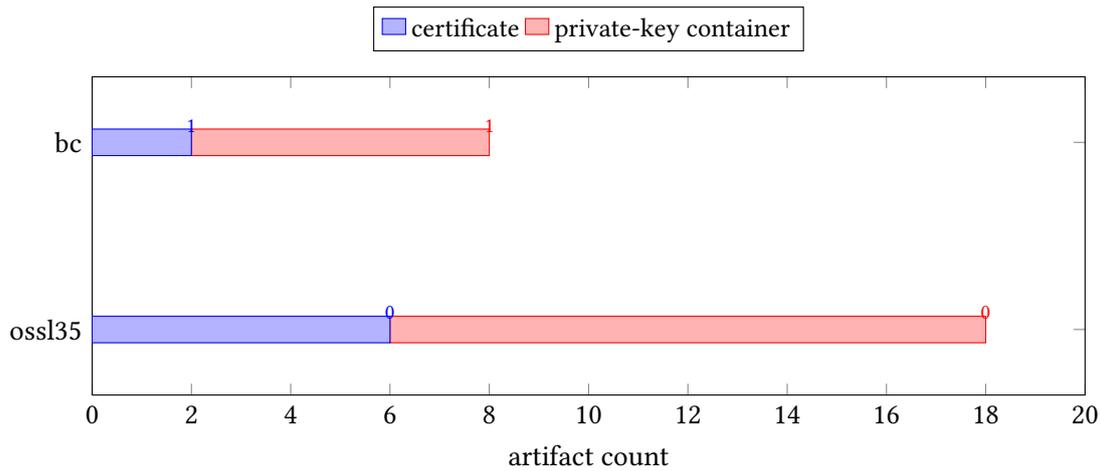
\begin{figure}[t]
  \centering
  \begin{tikzpicture}
    \begin{axis}[
      width=0.92\linewidth,
      height=5.8cm,
      xbar stacked,
      xmin=0, xmax=20,
      symbolic y coords={ossl35,bc},
      ytick=data,
      xlabel={artifact count},
      legend style={at={(0.5,1.08)},anchor=south,legend columns=2,font=\small},
      nodes near coords,
      every node near coord/.append style={font=\scriptsize},
      bar width=10pt,
      enlarge y limits=0.35
    ]
      \addplot coordinates {(6,ossl35) (2,bc)};
      \addplot coordinates {(12,ossl35) (6,bc)};
      \legend{certificate, private-key container}
    \end{axis}
  \end{tikzpicture}
  \caption{Public appendix coverage by provider and artifact type.}
  \label{fig:app-provider-coverage}
\end{figure}

\minorhead{Provider inventories}
\label{app:provider-inventories}

{\scriptsize
	\setlength{\LTleft}{0pt}
	\setlength{\LTright}{0pt}
	\setlength{\tabcolsep}{3pt}
	\renewcommand{\arraystretch}{1.08}
	
	\begin{longtable}{@{}L{0.08\textwidth}L{0.08\textwidth}L{0.09\textwidth}L{0.14\textwidth}L{0.16\textwidth}L{0.39\textwidth}@{}}
		\caption{Provider inventory for the bounded public appendix.}
		\label{tab:app-provider-inventories}\\
		\toprule
		Provider & Artifacts & Certificates & Private-key containers & Algorithms & Parameter sets \\
		\midrule
		\endfirsthead
		
		\caption[]{Provider inventory for the bounded public appendix. (continued)}\\
		\toprule
		Provider & Artifacts & Certificates & Private-key containers & Algorithms & Parameter sets \\
		\midrule
		\endhead
		
		\midrule
		\multicolumn{6}{@{}r@{}}{\emph{Continued on next page}}\\
		\endfoot
		
		\bottomrule
		\endlastfoot
		
		bc
		& 8
		& 2
		& 6
		& ML-DSA, ML-KEM
		& ML-DSA-65, ML-KEM-768
		\\
		
		ossl35
		& 18
		& 6
		& 12
		& ML-DSA, ML-KEM
		& ML-DSA-44, ML-DSA-65, ML-DSA-87, ML-KEM-1024, ML-KEM-512, ML-KEM-768
		\\
		
	\end{longtable}
}

\minorhead{Parameter-set coverage}
\label{app:parameter-set-coverage}

{\scriptsize
\setlength{\LTleft}{0pt}
\setlength{\LTright}{0pt}
\begin{longtable}{@{}>{\raggedright\arraybackslash}p{0.18\textwidth}>{\raggedright\arraybackslash}p{0.22\textwidth}>{\raggedright\arraybackslash}p{0.22\textwidth}>{\raggedright\arraybackslash}p{0.28\textwidth}@{}}
\caption{Parameter-set coverage.}\label{tab:app-parameter-set-coverage}\\
\toprule
Parameter set & Certificate providers & Private-key providers & Covered stages \\
\midrule
\endfirsthead
\caption[]{Parameter-set coverage in the bounded public appendix. (continued)}\\
\toprule
Parameter set & Certificate providers & Private-key providers & Covered stages \\
\midrule
\endhead
\midrule
\multicolumn{4}{r}{\emph{Continued on next page}}\\
\endfoot
\bottomrule
\endlastfoot
ML-DSA-44 & ossl35 & ossl35 & cert/profile, priv/import \\
ML-DSA-65 & bc, ossl35 & bc & cert/profile, priv/import \\
ML-DSA-87 & ossl35 & ossl35 & cert/profile, priv/import \\
ML-KEM-1024 & ossl35 & ossl35 & cert/profile, priv/import \\
ML-KEM-512 & ossl35 & ossl35 & cert/profile, priv/import \\
ML-KEM-768 & bc, ossl35 & bc & cert/profile, priv/import \\
\end{longtable}
}

For the certificate subset of the public appendix, the frozen baseline still
fatally rejects four valid \mlkem certificates while the local artifact passes
all eight appendix certificates without findings.
\minorhead{Certificate slice}
\label{app:certificate-slice}

{\scriptsize
\setlength{\LTleft}{0pt}
\setlength{\LTright}{0pt}
\begin{longtable}{@{}>{\raggedright\arraybackslash}p{0.26\textwidth}>{\raggedright\arraybackslash}p{0.08\textwidth}>{\raggedright\arraybackslash}p{0.12\textwidth}>{\raggedright\arraybackslash}p{0.10\textwidth}>{\raggedright\arraybackslash}p{0.34\textwidth}@{}}
\caption{Certificate slice of the bounded public appendix.}\label{tab:app-certificate-slice}\\
\toprule
Artifact ID & Provider & Parameter set & Surface & Selection rationale \\
\midrule
\endfirsthead
\caption[]{Certificate slice of the bounded public appendix. (continued)}\\
\toprule
Artifact ID & Provider & Parameter set & Surface & Selection rationale \\
\midrule
\endhead
\midrule
\multicolumn{5}{r}{\emph{Continued on next page}}\\
\endfoot
\bottomrule
\endlastfoot
\path{appendix-bc-mldsa65-ta-cert} & bc & ML-DSA-65 & cert/profile & cross-provider pure ML-DSA certificate matching the private-key container variants selected for import-validation \\
\path{appendix-bc-mlkem768-ee-cert} & bc & ML-KEM-768 & cert/profile & cross-provider pure ML-KEM certificate matching the private-key container variants selected for import-validation \\
\path{appendix-ossl35-mldsa44-ta-cert} & ossl35 & ML-DSA-44 & cert/profile & final-R5 pure ML-DSA trust anchor from a widely used public implementation \\
\path{appendix-ossl35-mldsa65-ta-cert} & ossl35 & ML-DSA-65 & cert/profile & final-R5 pure ML-DSA trust anchor from a widely used public implementation \\
\path{appendix-ossl35-mldsa87-ta-cert} & ossl35 & ML-DSA-87 & cert/profile & final-R5 pure ML-DSA trust anchor from a widely used public implementation \\
\path{appendix-ossl35-mlkem1024-ee-cert} & ossl35 & ML-KEM-1024 & cert/profile & final-R5 pure ML-KEM end-entity certificate covering the highest NIST parameter set \\
\path{appendix-ossl35-mlkem512-ee-cert} & ossl35 & ML-KEM-512 & cert/profile & final-R5 pure ML-KEM end-entity certificate covering the lowest NIST parameter set \\
\path{appendix-ossl35-mlkem768-ee-cert} & ossl35 & ML-KEM-768 & cert/profile & final-R5 pure ML-KEM end-entity certificate covering the practical default security level \\
\end{longtable}
}

\minorhead{Private-key slice}
\label{app:private-key-slice}

{\scriptsize
\setlength{\LTleft}{0pt}
\setlength{\LTright}{0pt}
\begin{longtable}{@{}>{\raggedright\arraybackslash}p{0.25\textwidth}>{\raggedright\arraybackslash}p{0.08\textwidth}>{\raggedright\arraybackslash}p{0.12\textwidth}>{\raggedright\arraybackslash}p{0.10\textwidth}>{\raggedright\arraybackslash}p{0.35\textwidth}@{}}
\caption{Private-key-container slice.}\label{tab:app-private-key-slice}\\
\toprule
Artifact ID & Provider & Parameter set & Surface & Selection rationale \\
\midrule
\endfirsthead
\caption[]{Private-key-container slice of the bounded public appendix. (continued)}\\
\toprule
Artifact ID & Provider & Parameter set & Surface & Selection rationale \\
\midrule
\endhead
\midrule
\multicolumn{5}{r}{\emph{Continued on next page}}\\
\endfoot
\bottomrule
\endlastfoot
\path{appendix-bc-mldsa65-both-key} & bc & ML-DSA-65 & priv/import & cross-provider both-form ML-DSA private key exercising seed-expanded consistency parsing \\
\path{appendix-bc-mldsa65-expanded-key} & bc & ML-DSA-65 & priv/import & cross-provider expanded-key ML-DSA private key for import-format external validity \\
\path{appendix-bc-mldsa65-seed-key} & bc & ML-DSA-65 & priv/import & cross-provider seed-only ML-DSA private key for import-format external validity \\
\path{appendix-bc-mlkem768-both-key} & bc & ML-KEM-768 & priv/import & cross-provider both-form ML-KEM private key exercising seed-expanded consistency and hash-check parsing \\
\path{appendix-bc-mlkem768-expanded-key} & bc & ML-KEM-768 & priv/import & cross-provider expanded-key ML-KEM private key exercising hash-check parsing on non-OpenSSL material \\
\path{appendix-bc-mlkem768-seed-key} & bc & ML-KEM-768 & priv/import & cross-provider seed-only ML-KEM private key for import-format external validity \\
\path{appendix-ossl35-mldsa44-both-key} & ossl35 & ML-DSA-44 & priv/import & low-parameter pure ML-DSA both-form private key from ossl35 to cover seed-expanded consistency parsing \\
\path{appendix-ossl35-mldsa44-expanded-key} & ossl35 & ML-DSA-44 & priv/import & low-parameter pure ML-DSA expanded-key private key from ossl35 to complement seed-only import coverage \\
\path{appendix-ossl35-mldsa44-seed-key} & ossl35 & ML-DSA-44 & priv/import & low-parameter pure ML-DSA seed-only private key from ossl35 to cover importer behavior at the low end \\
\path{appendix-ossl35-mldsa87-both-key} & ossl35 & ML-DSA-87 & priv/import & high-parameter pure ML-DSA both-form private key from ossl35 to pair consistency checks with the highest parameter set \\
\path{appendix-ossl35-mldsa87-expanded-key} & ossl35 & ML-DSA-87 & priv/import & high-parameter pure ML-DSA expanded-key private key from ossl35 to close representation coverage at the high end \\
\path{appendix-ossl35-mldsa87-seed-key} & ossl35 & ML-DSA-87 & priv/import & high-parameter pure ML-DSA seed-only private key from ossl35 to show importer coverage at the high end \\
\path{appendix-ossl35-mlkem1024-both-key} & ossl35 & ML-KEM-1024 & priv/import & high-parameter pure ML-KEM both-form private key from ossl35 to combine consistency and hash-check coverage at the high end \\
\path{appendix-ossl35-mlkem1024-expanded-key} & ossl35 & ML-KEM-1024 & priv/import & high-parameter pure ML-KEM expanded-key private key from ossl35 to extend hash-check and sizing coverage to the high end \\
\path{appendix-ossl35-mlkem1024-seed-key} & ossl35 & ML-KEM-1024 & priv/import & high-parameter pure ML-KEM seed-only private key from ossl35 to complete parameter-set coverage for importer behavior \\
\path{appendix-ossl35-mlkem512-both-key} & ossl35 & ML-KEM-512 & priv/import & low-parameter pure ML-KEM both-form private key from ossl35 to cover seed-expanded consistency at the low end \\
\path{appendix-ossl35-mlkem512-expanded-key} & ossl35 & ML-KEM-512 & priv/import & low-parameter pure ML-KEM expanded-key private key from ossl35 to exercise hash-check parsing outside the default set \\
\path{appendix-ossl35-mlkem512-seed-key} & ossl35 & ML-KEM-512 & priv/import & low-parameter pure ML-KEM seed-only private key from ossl35 to widen importer coverage beyond the mid-level parameter set \\
\end{longtable}
}

\section{Cross-Tool Full Matrix}
\label{app:cross-tool-full-matrix}

The cross-tool layer
shows a practical separation that matters to operators: parse acceptance is not
the same thing as policy conformance. 

\minorhead{Behavior taxonomy}
\label{app:behavior-taxonomy}

{\small
\setlength{\LTleft}{0pt}
\setlength{\LTright}{0pt}
\begin{longtable}{@{}>{\raggedright\arraybackslash}p{0.08\textwidth}>{\raggedright\arraybackslash}p{0.86\textwidth}@{}}
\caption{Behavior codes used in the cross-tool matrix.}\label{tab:app-cross-tool-taxonomy}\\
\toprule
Code & Meaning \\
\midrule
\endfirsthead
\caption[]{Behavior codes used in the cross-tool matrix. (continued)}\\
\toprule
Code & Meaning \\
\midrule
\endhead
\midrule
\multicolumn{2}{r}{\emph{Continued on next page}}\\
\endfoot
\bottomrule
\endlastfoot
\textbf{A} & artifact accepted by the named tool in the exercised path \\
\textbf{RS} & artifact rejected for a structural or encoding reason \\
\textbf{RP} & artifact rejected for a profile or semantic-policy reason \\
\textbf{RF} & tool failed or raised a fatal runtime condition on the artifact \\
\textbf{NA} & tool path not applicable for the artifact type or unsupported in the frozen scope \\
\end{longtable}
}

\minorhead{Controlled set}
\label{app:cross-tool-controlled}

The controlled set contains 31 artifacts in the cross-tool layer: 17
certificates and 14 private-key containers. Raw SPKI artifacts are omitted from
this appendix matrix because the external tools exercised here do not offer a
fairly comparable frozen path for them.

{\scriptsize
\setlength{\LTleft}{0pt}
\setlength{\LTright}{0pt}
\begin{longtable}{@{}>{\raggedright\arraybackslash}p{0.36\textwidth}>{\raggedright\arraybackslash}p{0.12\textwidth}>{\raggedright\arraybackslash}p{0.08\textwidth}>{\raggedright\arraybackslash}p{0.08\textwidth}>{\raggedright\arraybackslash}p{0.08\textwidth}>{\raggedright\arraybackslash}p{0.08\textwidth}>{\raggedright\arraybackslash}p{0.08\textwidth}@{}}
\caption{Cross-tool behavior matrix for the controlled set.}\label{tab:app-cross-tool-controlled}\\
\toprule
Artifact ID & Surface & Validity & Local & JZLint & OpenSSL & pkilint \\
\midrule
\endfirsthead
\caption[]{Cross-tool behavior matrix for the controlled set. (continued)}\\
\toprule
Artifact ID & Surface & Validity & Local & JZLint & OpenSSL & pkilint \\
\midrule
\endhead
\midrule
\multicolumn{7}{r}{\emph{Continued on next page}}\\
\endfoot
\bottomrule
\endlastfoot
\path{der-mut-mldsa44-cert-signature-aid-null} & cert/profile & invalid & RS & RS & A & RF \\
\path{der-mut-mldsa44-cert-signature-aid-octet-params} & cert/profile & invalid & RS & RS & A & RF \\
\path{der-mut-mldsa44-cert-signature-hashmldsa44} & cert/profile & invalid & RP & A & A & RF \\
\path{der-mut-mldsa65-cert-keyusage-empty} & cert/profile & invalid & RP & A & A & RF \\
\path{der-mut-mlkem768-cert-keyusage-empty} & cert/profile & invalid & RP & RF & A & RF \\
\path{der-mut-mlkem768-cert-spki-aid-null} & SPKI & invalid & RS & RS & A & RF \\
\path{der-mut-mlkem768-cert-spki-aid-octet-params} & SPKI & invalid & RS & RS & A & RF \\
\path{der-mut-mlkem768-cert-spki-payload-truncated} & SPKI & invalid & RS & RF & A & RF \\
\path{openssl-mldsa44-ee-cert} & cert/profile & valid & A & A & A & RF \\
\path{openssl-mldsa65-ca-cert} & cert/profile & valid & A & A & A & RF \\
\path{openssl-mldsa65-ee-cert} & cert/profile & valid & A & A & A & RF \\
\path{openssl-mldsa87-ee-cert} & cert/profile & valid & A & A & A & RF \\
\path{openssl-mlkem1024-ee-cert} & cert/profile & valid & A & RF & A & RF \\
\path{openssl-mlkem512-ee-cert} & cert/profile & valid & A & RF & A & RF \\
\path{openssl-mlkem768-ee-cert} & cert/profile & valid & A & RF & A & RF \\
\path{openssl-mut-mldsa65-keyusage-key-encipherment-cert} & cert/profile & invalid & RP & RP & A & RF \\
\path{openssl-mut-mlkem768-keyusage-digital-signature-cert} & cert/profile & invalid & RP & RP & A & RF \\
\path{der-mut-mldsa44-key-both-mismatch} & priv/import & invalid & RP & NA & RS & NA \\
\path{der-mut-mldsa44-key-expanded-short} & priv/import & invalid & RS & NA & RS & NA \\
\path{der-mut-mldsa44-key-seed-short} & priv/import & invalid & RS & NA & RS & NA \\
\path{der-mut-mlkem512-key-both-mismatch} & priv/import & invalid & RP & NA & RS & NA \\
\path{der-mut-mlkem512-key-expanded-short} & priv/import & invalid & RS & NA & RS & NA \\
\path{der-mut-mlkem512-key-hash-mismatch} & priv/import & invalid & RP & NA & RS & NA \\
\path{der-mut-mlkem512-key-seed-short} & priv/import & invalid & RS & NA & RS & NA \\
\path{openssl-mldsa44-ee-key} & priv/import & valid & A & NA & A & NA \\
\path{openssl-mldsa65-ca-key} & priv/import & valid & A & NA & A & NA \\
\path{openssl-mldsa65-ee-key} & priv/import & valid & A & NA & A & NA \\
\path{openssl-mldsa87-ee-key} & priv/import & valid & A & NA & A & NA \\
\path{openssl-mlkem1024-ee-key} & priv/import & valid & A & NA & A & NA \\
\path{openssl-mlkem512-ee-key} & priv/import & valid & A & NA & A & NA \\
\path{openssl-mlkem768-ee-key} & priv/import & valid & A & NA & A & NA \\
\end{longtable}
}

\minorhead{Appendix set}
\label{app:cross-tool-appendix}

The appendix set contains 26 valid public artifacts. It is particularly useful
for exposing baseline runtime fragility on otherwise valid \mlkem certificates
and for showing that parse-only acceptance by OpenSSL does not, by itself,
establish profile conformance.

{\scriptsize
\setlength{\LTleft}{0pt}
\setlength{\LTright}{0pt}
\begin{longtable}{@{}>{\raggedright\arraybackslash}p{0.36\textwidth}>{\raggedright\arraybackslash}p{0.12\textwidth}>{\raggedright\arraybackslash}p{0.08\textwidth}>{\raggedright\arraybackslash}p{0.08\textwidth}>{\raggedright\arraybackslash}p{0.08\textwidth}>{\raggedright\arraybackslash}p{0.08\textwidth}>{\raggedright\arraybackslash}p{0.08\textwidth}@{}}
\caption{Cross-tool behavior matrix for the bounded public appendix.}\label{tab:app-cross-tool-appendix}\\
\toprule
Artifact ID & Surface & Validity & Local & JZLint & OpenSSL & pkilint \\
\midrule
\endfirsthead
\caption[]{Cross-tool behavior matrix for the bounded public appendix. (continued)}\\
\toprule
Artifact ID & Surface & Validity & Local & JZLint & OpenSSL & pkilint \\
\midrule
\endhead
\midrule
\multicolumn{7}{r}{\emph{Continued on next page}}\\
\endfoot
\bottomrule
\endlastfoot
\path{appendix-bc-mldsa65-ta-cert} & cert/profile & valid & A & A & A & RF \\
\path{appendix-bc-mlkem768-ee-cert} & cert/profile & valid & A & RF & A & RF \\
\path{appendix-ossl35-mldsa44-ta-cert} & cert/profile & valid & A & A & A & RF \\
\path{appendix-ossl35-mldsa65-ta-cert} & cert/profile & valid & A & A & A & RF \\
\path{appendix-ossl35-mldsa87-ta-cert} & cert/profile & valid & A & A & A & RF \\
\path{appendix-ossl35-mlkem1024-ee-cert} & cert/profile & valid & A & RF & A & RF \\
\path{appendix-ossl35-mlkem512-ee-cert} & cert/profile & valid & A & RF & A & RF \\
\path{appendix-ossl35-mlkem768-ee-cert} & cert/profile & valid & A & RF & A & RF \\
\path{appendix-bc-mldsa65-both-key} & priv/import & valid & A & NA & A & NA \\
\path{appendix-bc-mldsa65-expanded-key} & priv/import & valid & A & NA & A & NA \\
\path{appendix-bc-mldsa65-seed-key} & priv/import & valid & A & NA & A & NA \\
\path{appendix-bc-mlkem768-both-key} & priv/import & valid & A & NA & A & NA \\
\path{appendix-bc-mlkem768-expanded-key} & priv/import & valid & A & NA & A & NA \\
\path{appendix-bc-mlkem768-seed-key} & priv/import & valid & A & NA & A & NA \\
\path{appendix-ossl35-mldsa44-both-key} & priv/import & valid & A & NA & A & NA \\
\path{appendix-ossl35-mldsa44-expanded-key} & priv/import & valid & A & NA & A & NA \\
\path{appendix-ossl35-mldsa44-seed-key} & priv/import & valid & A & NA & A & NA \\
\path{appendix-ossl35-mldsa87-both-key} & priv/import & valid & A & NA & A & NA \\
\path{appendix-ossl35-mldsa87-expanded-key} & priv/import & valid & A & NA & A & NA \\
\path{appendix-ossl35-mldsa87-seed-key} & priv/import & valid & A & NA & A & NA \\
\path{appendix-ossl35-mlkem1024-both-key} & priv/import & valid & A & NA & A & NA \\
\path{appendix-ossl35-mlkem1024-expanded-key} & priv/import & valid & A & NA & A & NA \\
\path{appendix-ossl35-mlkem1024-seed-key} & priv/import & valid & A & NA & A & NA \\
\path{appendix-ossl35-mlkem512-both-key} & priv/import & valid & A & NA & A & NA \\
\path{appendix-ossl35-mlkem512-expanded-key} & priv/import & valid & A & NA & A & NA \\
\path{appendix-ossl35-mlkem512-seed-key} & priv/import & valid & A & NA & A & NA \\
\end{longtable}
}

\section{Reproduction and Release Manifest}
\label{app:reproduction-release-manifest}

The artifact is designed to be rerun. This appendix records the canonical replay
path, the frozen toolchain summary, the reference hashes, the release tiers, and
the reviewer-first reading order.

\minorhead{Replay log}
\label{app:canonical-replay-log}

{\scriptsize
\setlength{\LTleft}{0pt}
\setlength{\LTright}{0pt}
\begin{longtable}{@{}>{\raggedright\arraybackslash}p{0.08\textwidth}>{\raggedright\arraybackslash}p{0.34\textwidth}>{\raggedright\arraybackslash}p{0.54\textwidth}@{}}
\caption{Canonical replay sequence distilled from \texttt{experiments/replay\_freeze.sh}.}\label{tab:app-canonical-replay}\\
\toprule
Step & Command or script & Purpose \\
\midrule
\endfirsthead
\caption[]{Canonical replay sequence distilled from \texttt{experiments/replay\_freeze.sh}. (continued)}\\
\toprule
Step & Command or script & Purpose \\
\midrule
\endhead
\midrule
\multicolumn{3}{r}{\emph{Continued on next page}}\\
\endfoot
\bottomrule
\endlastfoot
STEP-01 & \path{./experiments/check_environment.sh} & capture toolchain state and write environment.txt \\
STEP-02 & \path{./experiments/prepare_third_party.sh} & prepare frozen third-party snapshots needed for baseline replay \\
STEP-03 & \path{./experiments/build_libcrux_import_check.sh} & build narrow import-validation bridge used by the importer slice \\
STEP-04 & \path{./experiments/generate_corpus_openssl.sh} & regenerate valid local corpus artifacts \\
STEP-05 & \path{./experiments/generate_mutations_openssl.sh} & regenerate OpenSSL-driven certificate mutations \\
STEP-06 & \path{./experiments/generate_der_mutations.sh} & regenerate DER-level mutations for SPKI and private keys \\
STEP-07 & \path{./experiments/run_extended.sh --mode strict} & produce strict registry results \\
STEP-08 & \path{./experiments/run_extended.sh --mode deployable} & produce deployable registry results \\
STEP-09 & \path{./experiments/run_coverage.sh --mode strict/deployable} & produce certificate and SPKI coverage summaries in both modes \\
STEP-10 & \path{./experiments/run_private_key_coverage.sh --mode strict/deployable} & produce importer/private-key coverage summaries in both modes \\
STEP-11 & \path{./experiments/run_baseline.sh} & replay certificate-level JZLint baseline; baseline\_exit\_code=1 is frozen as expected upstream fragility \\
STEP-12 & \path{./experiments/run_baseline_compare.sh} & materialize certificate-level baseline-versus-extended comparison \\
STEP-13 & \path{./experiments/run_real_world_appendix.sh} & materialize the bounded public appendix and its summary outputs \\
STEP-14 & \path{./experiments/run_reference_workflow.sh}, \path{./experiments/run_operator_gate_packs.sh} & materialize owner/stage workflow and gate-pack outputs \\
STEP-15 & \path{./experiments/run_cross_tool_behavior.sh} & materialize the secondary parse-vs-policy behavior matrix \\
STEP-16 & \path{./experiments/run_claim_lock.sh}, \path{./experiments/run_visual_plan.sh}, \path{./experiments/run_positioning_lock.sh} & lock claim, visual, and positioning scaffolding \\
STEP-17 & \path{./experiments/run_artifact_packaging.sh}, \path{./experiments/run_upgrade_decision.sh}, \path{./experiments/run_smoke_tests.sh} & package the artifact, freeze the upgrade decision, and run smoke tests \\
\end{longtable}
}

\minorhead{Environment and hashes}
\label{app:environment-hashes}

{\small
\setlength{\LTleft}{0pt}
\setlength{\LTright}{0pt}
\begin{longtable}{@{}>{\raggedright\arraybackslash}p{0.18\textwidth}>{\raggedright\arraybackslash}p{0.78\textwidth}@{}}
\caption{Toolchain summary captured in \texttt{results/environment.txt}.}\label{tab:app-environment}\\
\toprule
Component & Frozen value \\
\midrule
\endfirsthead
\caption[]{Toolchain summary captured in \texttt{results/environment.txt}. (continued)}\\
\toprule
Component & Frozen value \\
\midrule
\endhead
\midrule
\multicolumn{2}{r}{\emph{Continued on next page}}\\
\endfoot
\bottomrule
\endlastfoot
Java & openjdk version "17.0.18" 2026-01-20 \\
Maven & Apache Maven 3.9.14 (996c630dbc656c76214ce58821dcc58be960875b) \\
Rust & rustc 1.82.0 (f6e511eec 2024-10-15); cargo 1.82.0 (8f40fc59f 2024-08-21) \\
Python & Python 3.14.3 \\
OpenSSL & OpenSSL 3.6.1 27 Jan 2026 (Library: OpenSSL 3.6.1 27 Jan 2026) \\
Environment status & ok \\
\end{longtable}
}

{\scriptsize
\setlength{\LTleft}{0pt}
\setlength{\LTright}{0pt}
\begin{longtable}{@{}>{\raggedright\arraybackslash}p{0.14\textwidth}>{\raggedright\arraybackslash}p{0.34\textwidth}>{\raggedright\arraybackslash}p{0.48\textwidth}@{}}
\caption{Frozen reference artifacts.}\label{tab:app-reference-hashes}\\
\toprule
Hash prefix & Reference path & Note \\
\midrule
\endfirsthead
\caption[]{Frozen reference artifacts recorded in \texttt{references/MANIFEST.sha256}. (continued)}\\
\toprule
Hash prefix & Reference path & Note \\
\midrule
\endhead
\midrule
\multicolumn{3}{r}{\emph{Continued on next page}}\\
\endfoot
\bottomrule
\endlastfoot
\texttt{a4852298eebc…} & \path{../reference/2025-1241.pdf} & Public Key Linting for ML-KEM and ML-DSA \\
\texttt{fe1f12f32a7e…} & \path{../reference/NIST.FIPS.203.pdf} & FIPS 203 \\
\texttt{57239b9f84c0…} & \path{../reference/NIST.FIPS.204.pdf} & FIPS 204 \\
\texttt{4b586ec732b2…} & \path{../reference/rfc9881.pdf} & RFC 9881 \\
\texttt{977e6fb42ecd…} & \path{../reference/rfc9935.pdf} & RFC 9935 \\
\texttt{4206e03ad6ff…} & \path{../reference/github/jzlint-main.zip} & JZLint snapshot, internal commit d6fdf02ad31f085e88d252b368f50e9da87debfd \\
\texttt{764f83e3bd00…} & \path{../reference/github/libcrux-main.zip} & libcrux snapshot, internal commit 6b9eca5a7b507e1d8423f85e5711572b4a661e8d \\
\end{longtable}
}

\end{document}